\documentclass[12pt]{article}
\usepackage{youngtab}
\usepackage{amsmath,amssymb,latexsym,cite}
\usepackage[]{hyperref}
\input xy
\xyoption{all}

\begin{document}

\begin{flushright}
UTTG-01-15
\end{flushright}

\vspace*{0.5in}

\begin{center}

{\large\bf Chiral operators in two-dimensional (0,2) theories}
\\
{\large\bf  and a test of
triality}

\vspace*{0.2in}

Jirui Guo$^1$, Bei Jia$^2$, Eric Sharpe$^1$

\vspace*{0.2in}

\begin{tabular}{cc}
{ \begin{tabular}{l}
$^1$ Department of Physics MC 0435 \\
850 West Campus Drive\\
Virginia Tech \\
Blacksburg, VA  24061
\end{tabular}
} &
{ \begin{tabular}{l}
$^2$ Theory Group \\
Physics Department \\
University of Texas \\
Austin, TX  78712
\end{tabular}
}
\end{tabular}

{\tt jrkwok@vt.edu}, {\tt beijia@physics.utexas.edu}, {\tt ersharpe@vt.edu}

\end{center}

In this paper we compute spaces of chiral operators in general two-dimensional
(0,2) nonlinear sigma models, 
both in theories twistable to the A/2 or B/2 model, as well
as in non-twistable theories, and apply them to check recent duality
conjectures.  The fact that in a nonlinear sigma model, the 
Fock vacuum can act as a section of a line bundle on the target space plays
a crucial role in our (0,2) computations, so we begin with a review of
this property.  We also take this opportunity to show how 
even in (2,2) theories,
the Fock vacuum encodes in this way choices of target space spin structures,
and discuss how such choices enter the A and B model topological field theories.
We then compute chiral operators in general (0,2) nonlinear sigma models,
and apply them to the
recent Gadde-Gukov-Putrov triality proposal, which says that
certain triples of
(0,2) GLSMs should RG flow to nontrivial IR fixed points.  We find that
different UV theories in the same proposed universality class do not 
necessarily have the same space of chiral operators -- but, the mismatched
operators do not contribute to elliptic genera and are in non-integrable
representations of the proposed IR affine symmetry groups, suggesting that
the mismatched states become massive along RG flow.  We find this state
matching
in examples not only of different geometric phases of the same GLSMs,
but also in phases of different GLSMs, indirectly
verifying the triality proposal, and giving a clean demonstration that (0,2) 
chiral rings are not topologically protected.
We also check proposals for enhanced IR affine $E_6$ symmetries in one such
model, verifying that (matching) chiral states in phases of corresponding
GLSMs transform as ${\bf 27}$'s, ${\bf \overline{27}}$'s.

\begin{flushleft}
January 2015
\end{flushleft}

\newpage

\tableofcontents

\newpage

\section{Introduction}

Over the last few years we have seen a tremendous surge of development in
two-dimensional GLSMs and theories with (2,2) and (0,2) supersymmetry.
To give just a few examples, we now have examples of 
nonperturbatively-realized geometries 
\cite{horitong,hori2,meron,cdhps,hkm,es-rflat,hori-knapp},
perturbative GLSMs for Pfaffians \cite{hori2,jklmr1,jsw},
non-birational GLSM phases \cite{horitong,meron,cdhps},
examples of closed strings on noncommutative resolutions 
\cite{cdhps,es-rflat,add-ss,ncgw}, 
localization techniques and new computations of, for
example, Gromov-Witten invariants and elliptic genera 
(see for example \cite{bc1,dgfl,jklmr2,beht1,beht2,gg}), 
and progress in heterotic string compactifications,
ranging from nonperturbative correlations to new two-dimensional
dualities and an understanding of non-K\"ahler moduli
(see for example 
\cite{dgks1,dgks2,ggp1,ggp-exact,jsw,ms1,ags,ossa-svanes,ossa-svanes2,svanes,gfrt}).

This paper concerns chiral states and rings in (0,2) theories.
Chiral rings have an extensive history in two-dimensional (2,2) supersymmetric
theories, providing tools such as quantum cohomology to help analyze
quantum field theories.  They also can exist in (0,2) supersymmetric theories
(see for example \cite{ks,ade,es-other,es-snowbird,guffin-katz,guffinrev,mm1,mm2,mcorist1,dgks1,dgks2}),
and are a current research topic.

As one particularly illuminating example, we apply chiral states to
discuss 
triality proposals for two-dimensional (0,2) theories, described in
\cite{ggp1,ggp-exact}.  Specifically, those papers proposed that all points
in phase diagrams of certain triples of (0,2) GLSMs should RG flow to the
same IR fixed point, yielding a `triality' relating three naively different
two-dimensional gauge theories.  Now, many two-dimensional
gauge theory dualities have a simple understanding as different presentations
of the same IR geometry, as discussed in \cite{jsw}, but as also observed
there, triality is different -- of the six total geometric
phases, there are three pairs such that each pair is associated with
the same geometry, but the geometries associated to different pairs are
simply different.  Thus, it would be useful to
have further insight into triality, as it cannot be understood as simply
as other two-dimensional gauge dualities.

We begin in section~\ref{sect:02chiral} with a general overview of
chiral states and rings in two-dimensional (0,2) theories.  These have
been described previously only in the A/2 and B/2 models; here, we describe 
chiral operators in general (0,2) nonlinear sigma models.  The correct
counting of chiral states in two-dimensional (0,2) nonlinear sigma models
utilizes the fact that the Fock vacuum transforms as a section of a line
bundle over the target space, a phenomenon closely related to
the fractional fermion number property.  
This has been observed in a few other papers,
but the basic idea still seems to be somewhat obscure to the community,
so we take this opportunity to review the details.  We also
observe that even in (2,2) theories,
the Fock vacuum can be a section of a nontrivial line bundle, encoding
choices of spin structure on the target space.  
When applied to (0,2) theories, the resulting spectra are shown to satisfy
basic consistency properties, such as invariance under Serre duality
and also invariance under dualizing bundle factors.

In section~\ref{sect:triality} we apply chiral operators to study examples of
triality.  We begin with a brief overview of triality, and how in general terms
one can keep track of all of the global symmetries in chiral operator
computations (as some of the symmetries are realized in a nonobvious fashion,
encoded implicitly in the Bott-Borel-Weil theorem), and also how in principle
states in UV nonlinear sigma models, computed in terms of sheaf cohomology,
should be related to state countings in the IR limit, which should be at least
partially encoded in Lie algebra cohomology.  We also briefly discuss 
pseudo-topological twists of examples of triality, to the A/2 and B/2 models.

Next, in each of three examples of triality, we compute chiral operators
in two geometric phases in each of two GLSMs related by triality, keeping
track of global symmetries, both to better understand triality and also
to gain insight into properties of (0,2) chiral states.  
We find that only a subset of all chiral 
states computed match between phases and GLSMs.  However, (0,2) chiral
states, unlike (2,2) chiral states, are not protected against
nonperturbative corrections, and as RG flow will take the weakly-coupled
nonlinear sigma models to a strong coupling regime, such a state mismatch
should not be unexpected.  
As noted in \cite{ggp1,ggp-exact}, the
global symmetry groups should be enhanced to affine groups of
certain levels in the IR, and chiral states should lie in integrable
representations of those affine symmetry algebras.  We check that
the subset of the states that
match between phases and GLSMs, 
all lie in integrable representations of the
affine symmetry algebras predicted by
\cite{ggp1,ggp-exact}, consistent with the predictions of triality.  
The mismatched states all lie in non-integrable
representations, which one would hope would not survive to the IR.  
Furthermore, the mismatched states all come in pairs whose contributions
to at least leading terms in elliptic genera, refined by any of the computed
global symmetries, cancel out, consistent with the conjecture that the
mismatched states become massive along the RG flow.  (In principle, pairs of
massive states could also become massless along RG flow, but we did not
observe any examples of such in the examples computed.)
Thus, by counting chiral states in examples, we find nontrivial evidence
for triality, as well as extremely clean examples of states varying
along RG flow.

Such behavior along RG flow is of course typical of non-protected operators, but
many workers in the (0,2) community have sometimes implicitly assumed that (0,2)
chiral rings would be protected, hence triality provides clean examples
stress-testing assumptions about (0,2) chiral rings.  (See also
\cite{mcmel} which reached a related conclusion in certain other
special cases.) 

Furthermore, in some examples of triality the global symmetry algebras are
further enhanced, and we find that the matching chiral states can be
recombined into integrable representations of the larger symmetry algebras,
again consistent with the predictions of \cite{ggp1,ggp-exact}.

Finally, as the Bott-Borel-Weil theorem in mathematics plays a crucial
role in our computations, in appendix~\ref{app:bbw} we include a short
overview, so as to make this paper self-contained.

\section{Chiral rings in two-dimensional (0,2) theories}
\label{sect:02chiral}

Part of the purpose of this paper is to compute
chiral operators in general (0,2) nonlinear sigma models, as a means of
constructing tests of the Gadde-Gukov-Putrov triality proposal 
\cite{ggp1,ggp-exact}.

To that end, let us first address chiral rings in gauged linear sigma models.
These are two-dimensional gauge theories, so one might expect that chiral
rings should be given as rings of gauge-invariant operators modulo relations
determined by the superpotential, just as they are computed in four-dimensional
gauge theories.  Unfortunately, even in (2,2) theories in two dimensions,
this gives an incomplete result.

Consider, for example, the (2,2) GLSM describing the quintic hypersurface
in ${\mathbb P}^4$, a $U(1)$ gauge theory with five fields $\phi_i$ of
charge $+1$ and one field $p$ of charge $-5$.  The chiral ring computed as
above includes operators of the form
\begin{displaymath}
p \left( \mbox{degree 5 polynomial in }\phi_i\right)
\end{displaymath}
modulo relations of the form $p d G$, where $G$ is the quintic hypersurface.
Certainly these form part of the chiral ring -- in particular, these
encode complex structure deformations in $H^{2,1}$.  However, it is well-known
that not all complex structure deformations of a hypersurface or complete
intersection can be expressed as polynomials of the form above.
Such non-algebraic complex structure deformations contribute to cohomology
$H^{2,1}$ and to the chiral ring in the nonlinear sigma model.  Since the
(2,2) chiral ring lies in a topological subsector, in principle those
same non-algebraic deformations ought to appear in the complete chiral
ring of the GLSM.  Unfortunately, it is not known at present how to present
those elements of the chiral ring of the GLSM, those non-algebraic complex
structure deformations, in terms of gauge-invariant operators.  

Although we do not know how to build the complete chiral ring in a (2,2)
GLSM, we do know how to build the complete chiral ring in a (2,2) nonlinear
sigma model.  Thus, in this paper we shall focus on chiral rings in
nonlinear sigma models.

\subsection{Review of chiral rings in (2,2) theories and Fock vacuum
subtleties}   \label{sect:rev:Fock}

Let us begin our discussion of (0,2) chiral rings with a brief review
of pertinent aspects of
chiral rings in (2,2) supersymmetric nonlinear sigma models, 
focusing on some subtleties in Fock vacua that do not seem to be
widely appreciated but which will play a crucial 
role in the (0,2) generalization.  

Consider a (2,2) nonlinear sigma model on a complex K\"ahler manifold
$X$.  Let $\phi: \Sigma \rightarrow X$ denote the worldsheet scalars,
and
$\psi_{\pm}^{i,\overline{\imath}}$
worldsheet fermions coupling to the tangent bundle of $X$, in the 
standard fashion.  Then, the chiral ring we will generalize to (0,2)
is the ring of states in (R,R) sectors\footnote{
Usually we compute states in R sectors rather than NS because the low-energy
states can be obtained via a cohomology computation.  For example, from
\cite{lvw}[section 2], in an $N=2$ SCFT, in the R sector
\begin{displaymath}
L_0 \: = \: \{ G_0^-, G_0^+ \} \: + \: c/12,
\end{displaymath}
and so $L_0 - c/12$ lives in cohomology, whereas in the NS sector,
\begin{displaymath}
L_0 \: = \: (1/4) \{ G^-_{-1/2}, G^+_{+1/2} \} \: + \:
(1/4)\{ G^-_{+1/2}, G^-_{-1/2} \} ,
\end{displaymath}
which is not a cohomology computation, but rather an analogue of a harmonic
representative computation.  The two are related by supersymmetry and
should give equivalent results, but the cohomology computation is 
significantly simpler, especially in the absence of an explicit target space
metric.
}, 
which following standard methods (see {\it e.g.} \cite{lvw})
are of the form 
\begin{displaymath}
b_{\overline{\imath}_1, \cdots, \overline{\imath}_n, j_1,
\cdots, j_m} \psi_+^{\overline{\imath}_1} \cdots
\psi_+^{\overline{\imath}_n} \psi_-^{j_1} \cdots
\psi_-^{j_m} | 0 \rangle .
\end{displaymath}

The factor
\begin{displaymath}
b_{\overline{\imath}_1, \cdots, \overline{\imath}_n, j_1,
\cdots, j_m} \psi_+^{\overline{\imath}_1} \cdots
\psi_+^{\overline{\imath}_n} \psi_-^{j_1} \cdots
\psi_-^{j_m} 
\end{displaymath}
has a standard understanding in terms of the cohomology of the space.
What we will need in our discussion of (0,2) chiral rings, and may
be less widely understood, is that the Fock vacuum $|0\rangle$ might also
couple to a nontrivial bundle on the target space\footnote{
We should distinguish this from Fock vacua coupling to nontrivial bundles
over a moduli space of SCFT's, say, which is well-known, see for
example \cite{bcov}.
}
and contribute to the state counting.
This phenomenon has also been discussed in \cite{hetstx,es-rflat,bmp1},
but as this phenomenon may be obscure,
to make this paper self-contained, we will briefly review it here.
We will also take this opportunity to describe how the same phenomenon
arises even in (2,2) theories, in describing spin structures on target
spaces, something which to our knowledge has not previously appeared in 
the literature.  (See also \cite{dfm1,dfm2} for a general discussion
of spinors in strings, and \cite{hori-index}[section 2] for a discussion
of the same phenomenon in one-dimensional theories.)

\subsubsection{Fock vacua coupling to nontrivial bundles on the target}

We can understand this phenomenon as follows.
In a chiral R sector, the Fock vacuum couples to $(K_X)^{\pm 1/2}$ in general.
This follows from the usual multiplicity of Fock vacua in the
presence of periodic fermions.  Schematically,
if we define two vacua $|0\rangle$,
$|0\rangle'$ by
\begin{displaymath}
\psi_0^i |0\rangle' \: = \: 0, \: \: \:
\psi_0^{\overline{\imath}} |0\rangle \: = \: 0 ,
\end{displaymath}
then
\begin{displaymath}
|0 \rangle' \: = \: \left( \prod_i \psi_0^i \right) | 0 \rangle, \: \: \:
|0 \rangle \: = \: \left( \prod_{\overline{\imath}} \psi_0^{\overline{\imath}}
\right) | 0 \rangle' .
\end{displaymath}
Since
\begin{displaymath}
\prod_i \psi_0^i \: \sim \: K_X^{-1}, \: \: \:
\prod_{\overline{\imath}} \psi_0^{\overline{\imath}} \: \sim \:
K_X,
\end{displaymath}
then, just as in fractional charges,
$|0\rangle'$ transforms as a section of $(K_X)^{-1/2}$ and 
$|0\rangle$ transforms as a section of $(K_X)^{+1/2}$.

In the case of a (2,2) theory, the Fock vacuum couples to
\begin{equation}   \label{eq:rr-fock}
(K_X)^{+1/2} \otimes (K_X)^{-1/2},
\end{equation}
one factor for left-movers, the other for right-movers.

If $X$ is simply-connected, the square roots $(K_X)^{\pm 1/2}$, if they
exist, are
uniquely determined, and the Fock vacuum couples to a trivial bundle.
(If they do not exist in a given sector, then that sector is physically
inconsistent.)

More generally, if $X$ is not simply-connected, then there will be
multiple different square roots $(K_X)^{\pm 1}$.  
These different choices of square roots correspond to different
choices of spin structure on the target space $X$, as spinors on
a complex K\"ahler manifold can be expressed in the form
\cite{lawson-mich}[section II.3] 
\begin{displaymath}
\wedge^{\bullet} TX \otimes (K_X)^{+1/2} .
\end{displaymath}
If $X = T^2$, for example, the different square roots $(K_X)^{1/2}$ 
simply correspond to choices of periodic and antiperiodic boundary
conditions around the legs of the torus.

To compute the chiral ring, we must specify two
square roots, one for left-movers, another for right-movers.
Which square roots
should appear, associated to left- and right-movers, is part of the
specification of the nonlinear sigma model.  In other words,
just as one must specify a metric and $B$ field on $X$ in order
to define a nonlinear sigma model, if $X$ is not simply-connected
then in addition one must also specify a spin structure on $X$,
and that choice of spin structure enters worldsheet physics
via Fock vacua, as above.

In this paper, we will work with simply-connected spaces.
However, in the (0,2) case, Fock vacua will couple to bundles of the
form $(\det {\cal E})^{-1/2} \otimes K_X^{+1/2}$, and so even if there
is no ambiguity, the bundle can be nontrivial.
More generally, suppose $|0\rangle \in
\Gamma(L)$ for some line bundle $L$.  If $L$ has no sections at all,
this merely implies that the Fock vacuum is not BRST-closed:  we still
have a Fock vacuum, it merely does not define a state by itself, analogous
to tachyonic states projected out of closed bosonic string spectra.
At a different extreme, if $L$ admits multiple sections, this is merely
another source of multiplicity beyond that provided by Fermi zero modes.

For completeness, and since this fact does not yet seem to be widely
appreciated in the literature,
let us explore some of the implications of the
statement above on the (2,2) locus.

If $X$ is Calabi-Yau, so that $K_X$ is trivial, then there is a 
canonical\footnote{
More generally, the $n$th roots of the structure sheaf form a finite
group which is canonically Hom($\pi_1(X),\mu_n$) where $\mu_n$ is the
group of $n$th roots of unity \cite{tonypriv}.
} trivial square root, specifically $(K_X)^{+1/2} = {\cal O}_X$.
(If $K_X$ is nontrivial, then in general there will not be any
canonical choice of square root, if square roots in fact exist.)  
Only in that canonical trivial spin
structure on a Calabi-Yau does there exist a nowhere-zero covariantly
constant spinor \cite{tonypriv}.  (For example, on $T^2$, only in the
(R,R) spin structure does $K^{1/2}$ have a section.)

Since the anomaly in the left and right $U(1)_R$ symmetries is determined
by $K_X$ and not the spin structure, if $X$ is Calabi-Yau this theory
should flow to a nontrivial (2,2) SCFT, even if one chooses nontrivial
left- or right-spin structures.  However, the target space string theories
are not likely well-defined if either spin structure is nontrivial
\cite{dfm1,dfm2,distlerpriv}.  
Even if the target space string theory is well-defined, spacetime
supersymmetry must surely be broken if either spin structure is nontrivial.
In an SCFT associated to a Calabi-Yau compactification, there is
an isomorphism between R and NS sector states:  spectral flow rotates
one into the other.  If $(K_X)^{1/2}$ is nontrivial, however, that
isomorphism is broken.  After all, the Fock vacuum ambiguity which
leads to this interpretation in terms of $(K_X)^{1/2}$ only exists
in the R sector, not the NS sector, and so the states that arise in 
the R sector are necessarily distinct\footnote{
The spectral flow operator that takes NS to the relevant
R and conversely has
$\theta = 1/2$ in the notation of \cite{lvw}, and charge $\pm c/6$, and
formally would be associated to a square root of the canonical bundle.
That spectral flow operator would be different from the one in \cite{lvw},
which maps NS to the R sector for the canonical spin structure.  The one
relevant here could not be expressed merely as the exponential of a
boson, unlike the one in \cite{lvw}.
}.  Put another way, spectral
flow relates NS sector states to the R sector states associated with
trivial spin structures, instead of the given spin
structures.  In effect, this is a further
condition for spacetime supersymmetry in $N=2$ SCFT's, beyond the familiar
statement that the difference of left and right charges should be integral
\cite{lvw}[equ'n (2.3)].

In orbifolds, an example of
the effect of having $(K_X)^{1/2}$ nontrivial while
$K_X$ is trivial is given by the Scherk-Schwarz mechanism
\cite{scherk-schwarz}, which breaks supersymmetry in orbifolds by 
assigning
different boundary conditions to fermions than to bosons.
(In full string theories, such boundary conditions might also contribute
to {\it e.g.}
failures of level-matching or modular invariance, but to decide the
matter, one would need to specify the rest of the CFT needed for
a critical string.)

\subsubsection{A, B model topological field theories and non-simply-connected
targets}

For completeness, let us also briefly discuss the A and B model topological
field theories for non-simply-connected target spaces.  In both cases,
the Fock vacuum couples to the ratio of square roots in 
equation~(\ref{eq:rr-fock}).  In conventional treatments of the A and B 
models such as
\cite{ed-tft}, the two square roots are assumed identical, so that the bundle is
trivial.  (If $X$ does not admit a spin structure, then $(K_X)^{1/2}$
does not exist as an honest bundle, only a `twisted' bundle; however, by
formally identifying the contributions from left- and right-moving sectors,
we can still make sense of the RR sector, and hence the topological field
theories.)  

More generally, if the square roots are not identical, 
then the Fock vacuum
couples to a nontrivial bundle.  The two topological field theories
appear to still be well-defined, but their interpretations are
slightly different.  The operators in the A model continue
to be counted by
\begin{displaymath}
H^q(\Omega^p_X),
\end{displaymath}
but the states are now counted by
\begin{displaymath}
H^q\left(\Omega^p_X \otimes (K_X)^{+1/2} \otimes (K_X)^{-1/2} \right) . 
\end{displaymath}
Similarly, the operators in the B model continue to be counted by
\begin{displaymath}
H^q(\wedge^p TX), 
\end{displaymath}
but the states are now counted by
\begin{displaymath}
H^q\left( \wedge^p TX \otimes (K_X)^{+1/2} \otimes (K_X)^{-1/2} \right) . 
\end{displaymath}
We do not interpret this as a violation of the state-operator
corresopndence, which refers to the $SL(2,{\mathbb C})$-invariant
NS-NS vacuum, but instead in terms of spectral flow.  For example,
when A model three-point correlation functions are interpreted in a physical
theory, the physical correlation function takes the form 
\begin{displaymath}
\left( \mbox{spacetime spinor} \right)
\left( \mbox{spacetime boson} \right)
\left( \mbox{spacetime spinor} \right) ,
\end{displaymath}
where the spinor structure is encoded in the Fock vacuum.  We interpret
the issue above similarly.

That said, correlation functions of local observables
are unchanged by the choices of target-space spin
structure, as the combination of
$|0\rangle$ and $\langle 0 |$ result in a factor of
\begin{displaymath}
\left( (K_X)^{+1/2} \otimes (K_X)^{-1/2} \right)^2 \: \cong \: K_X \otimes
K_X^{-1} \: \cong \: {\cal O}_X .
\end{displaymath}
It is possible that nonlocal\footnote{
We would like to thank H.~Jockers for observing this possibility.
} observables may be able to detect the
spin structure.

\subsection{(0,2) chiral states and rings}

Briefly, the (0,2) chiral states in which we are interested are the 
``massless'' or zero-energy
elements of the ring of
states annihilated by a right supercharge.  The set of all states annihilated
by a right supercharge,
an infinite tower, forms a ring.  In this paper, we will focus on 
the ``massless'' elements of that ring, which form a finite-dimensional
subset.  

The elements of the right-chiral ring with fixed conformal
dimension need not form a ring.  Surprisingly, however,
under certain circumstances
\cite{ade} it can be shown that the OPE's nevertheless close into
themselves.  Specifically, if the bundle rank is less than eight,
then the massless chiral states in the A/2 model will close into a ring,
at least in patches on the moduli space.

Let us make explicit what we mean by the (0,2) chiral states
in a (0,2) nonlinear sigma model on a space $X$ with holomorphic
vector bundle ${\cal E}$, satisfying the conditions\footnote{
The second condition suffices to define the theory in an (R,R) sector.
In more general sectors, one would need to separately require that
$c_1({\cal E})$ and $c_1(TX)$ vanish mod 2; however, in this paper we
will only be concerned with the RR sector.
}
\begin{displaymath}
{\rm ch}_2({\cal E}) \: = \: {\rm ch}_2(TX), \: \: \:
c_1({\cal E}) \: \equiv \: c_1(TX) \: {\rm mod}\:2 .
\end{displaymath}

In the (R,R) sector, slightly generalizing the old analysis of
\cite{dg}, the states in the worldsheet theory are of the form
\begin{displaymath}
b_{\overline{\imath}_1, \cdots, \overline{\imath}_n, a_1, \cdots, a_m}
\lambda_-^{a_1} \cdots \lambda_-^{a_m} \psi_+^{\overline{\imath}_1} \cdots
\psi_+^{\overline{\imath}_n} | 0 \rangle
\end{displaymath}
for the Fock vacuum defined by 
\begin{displaymath}
\psi_+^i | 0 \rangle \: = \: 0 \: = \: \lambda_-^{\overline{a}} |0 \rangle .
\end{displaymath}
The Fock vacuum defined as above
transforms as a section of the bundle
\begin{displaymath}
(\det {\cal E})^{-1/2}
\otimes K_X^{+1/2} ,
\end{displaymath}
essentially as a consequence of its fractional charges under
global symmetries, as discussed in section~\ref{sect:rev:Fock}.  
Following standard methods (for example \cite{dg}), since the right
supercharge can be identified with $\overline{\partial}$,
the states above realize a Dolbeault representation of
the sheaf cohomology groups
\begin{displaymath}  
H^{n}\left(X, (\wedge^{m} {\cal E})\otimes (\det {\cal E})^{-1/2}
\otimes K_X^{+1/2} \right) .
\end{displaymath}
(This is a special case of the general result for massless spectra of
heterotic strings on stacks described in \cite{hetstx}[appendix A].)

The ratio of square roots will exist whenever\footnote{
In GLSMs, the analogous constraint for a single $U(1)$ would be the
statement
\begin{displaymath}
\sum_{\alpha} q_{L,\alpha} \: \equiv \: 
\sum_{\beta} q_{R,\beta}
\mbox{ mod } 2 ,
\end{displaymath}
relating the sum of charges of left- and right-moving fields.
However, note that since
\begin{displaymath}
\sum_{\alpha} q_{\alpha}^2 \: \equiv \:
\sum_{\alpha} q_{\alpha} \mbox{ mod }2 ,
\end{displaymath}
the anomaly cancellation condition
\begin{displaymath}
\sum_{\alpha} q_{L,\alpha}^2 \: = \: 
\sum_{\beta} q_{R,\beta}^2
\end{displaymath}
implies the statement above.
See also \cite{hetstx}[appendix A.4] for a discussion of this
condition as it appears in orbifolds and related theories.
}
\begin{displaymath}
c_1({\cal E}) \: \equiv \: c_1(TX) \mbox{ mod } 2 ,
\end{displaymath}
which is typically taken as a consistency condition on heterotic
nonlinear sigma models.  (In fact, to make sense of the (R,NS) and (NS,R)
sectors, we must require that $\det {\cal E}$ and $K_X$ separately
admit square roots, which requires$c_1(E) \equiv 0$ mod $2$ and separately
$c_1(TX) \equiv 0$ mod $2$.  For our purposes in this paper, we will focus
on (R,R) sectors, and so the condition above suffices.)
Thus, square roots will exist in cases of interest.

As an aside, in a typical perturbative heterotic compactification,
it is taken that both $K_X$ and $\det {\cal E}$ are trivial.
In this case, each has a canonical trivial square root.

Now, beyond the ambiguities just described,
there are different choices one could make for R sector Fock vacua.
For example, we could instead consider the Fock vacuum
defined by
\begin{displaymath}
\psi_+^i | 0 \rangle \: = \: 0 \: = \: \lambda_-^{a} |0 \rangle ,
\end{displaymath}
which instead couples to
\begin{displaymath}
(\det {\cal E})^{+1/2}
\otimes K_X^{+1/2} .
\end{displaymath}
In this case, states would be enumerated in the form
\begin{displaymath}
b_{\overline{\imath}_1, \cdots, \overline{\imath}_n, \overline{a}_1, 
\cdots, \overline{a}_m}
\lambda_-^{\overline{a}_1} \cdots \lambda_-^{\overline{a}_m} 
\psi_+^{\overline{\imath}_1} \cdots
\psi_+^{\overline{\imath}_n} | 0 \rangle ,
\end{displaymath}
and counted by
\begin{displaymath}
H^n\left(X, (\wedge^m {\cal E}^*) \otimes
(\det {\cal E})^{+1/2}
\otimes K_X^{+1/2} \right) .
\end{displaymath}
However, the choice of Fock vacuum should not change the states,
and that is reflected in mathematical dualities.  For example, using
the fact that
\begin{displaymath}
\wedge^m {\cal E}^* \: = \: ( \wedge^{r-m} {\cal E}) \otimes (\det {\cal E}^*)
\end{displaymath}
(for $r$ the rank of ${\cal E}$), it is easy to check that
\begin{displaymath}
H^n\left(X, (\wedge^m {\cal E}^*) \otimes
(\det {\cal E})^{+1/2}
\otimes K_X^{+1/2} \right)
\: = \:
H^n\left(X, (\wedge^{r-m} {\cal E}) \otimes (\det {\cal E})^{-1/2} \otimes
K_X^{+1/2} \right) ,
\end{displaymath}
and so we see that these are merely two different descriptions of the
same set of states:
\begin{equation}   \label{eq:chiral-ring}
H^{\bullet}\left(X, (\wedge^{\bullet} {\cal E})\otimes (\det {\cal E})^{-1/2}
\otimes K_X^{+1/2} \right)
\: = \:
H^{\bullet}\left(X, (\wedge^{\bullet} {\cal E}^*)\otimes (\det {\cal E})^{+1/2}
\otimes K_X^{+1/2}\right) .
\end{equation}
Thus, the choice of conventions in picking Fock vacua do not alter
the set of states.

Let us list a few consistency checks:
\begin{itemize}
\item In the A/2 model, where $\det {\cal E}^* \cong K_X$,
the states should be counted by $H^{\bullet}(X,\wedge^{\bullet} {\cal E}^*)$,
and in the B/2 model, where $\det {\cal E} \cong K_X$,
the states should be counted by $H^{\bullet}(X,\wedge^{\bullet} {\cal E})$.
\item On the (2,2) locus, the states should be counted by
\begin{displaymath}
H^{\bullet}(X, \wedge^{\bullet} T^*X) \: = \:
H^{\bullet}(X, \wedge^{\bullet} TX \otimes K_X) .
\end{displaymath}
\item The states should be counted by sheaf cohomology groups that
close into themselves under Serre duality.
\item The structure above is compatible with the left Ramond elliptic genus
of (0,2) nonlinear sigma models.  As discussed in {\it e.g.}
\cite{km,lgeg,jsw}, the leading term in the elliptic genus of a sigma
model on $X$ with bundle ${\cal E}$ is
proportional to
\begin{displaymath}
\int_X \hat{A}(TX) \wedge {\rm ch}\left( (\det {\cal E})^{-1/2} \wedge_{-1}
{\cal E} \right)
\: = \:
\int_X {\rm td}(TX) \wedge {\rm ch}\left( K_X^{+1/2} \otimes
(\det {\cal E})^{-1/2} \wedge_{-1}
{\cal E} \right) ,
\end{displaymath}
which is the Hirzebruch-Riemann-Roch index appropriate for the
sheaf cohomology groups above.
\item The states should be invariant under ${\cal E} \mapsto
{\cal E}^*$ (which should swap the A/2, B/2 models) 
(see {\it e.g.} \cite{es-other}).
\item If the bundle ${\cal E}$ is reducible, the states should be
invariant under separately dualizing factors.
We can check this explicitly as follows.  Suppose
${\cal E}$ splits holomorphically, 
${\cal E} = {\cal A} \oplus {\cal B}$, of ranks
$r_1$, $r_2$, respectively.
Using
\begin{displaymath}
\wedge^{\bullet} {\cal E} \: = \: \sum_{i+j=\bullet} \wedge^i {\cal A} \otimes
\wedge^j {\cal B},
\end{displaymath}
we have
\begin{eqnarray*}
\lefteqn{
H^{\bullet}\left(X, (\wedge^i {\cal A}) \otimes (\wedge^j {\cal B})
\otimes (\det {\cal A})^{-1/2} \otimes (\det {\cal B})^{-1/2} \otimes
K_X^{+1/2}\right) 
} \\
& = &
H^{\bullet}\left(X, (\wedge^i {\cal A}) \otimes (\wedge^{r_2-j} {\cal B}^*)
\otimes (\det {\cal B}) \otimes  (\det {\cal A})^{-1/2} \otimes 
(\det {\cal B})^{-1/2} \otimes
K_X^{+1/2}\right) , \\
& = &
H^{\bullet}\left(X, (\wedge^i {\cal A}) \otimes (\wedge^{r_2-j} {\cal B}^*)
\otimes (\det {\cal A})^{-1/2} \otimes 
(\det {\cal B}^*)^{-1/2} \otimes K_X^{+1/2} \right) .
\end{eqnarray*}
Thus, the spectrum remains invariant if we replace ${\cal A} \oplus {\cal B}$
by ${\cal A} \oplus {\cal B}^*$.  At some level, this reflects the fact
that shuffling between Fock vacua does not change the set of states, and so
is a self-consistency test.
\end{itemize}
The sheaf cohomology groups above in~(\ref{eq:chiral-ring}) can
straightforwardly be shown to satisfy all of
the conditions above.

So far we have only discussed the additive structure of the chiral
states; however, in special cases, there are also results on
product structures.  For example, in the special cases
$\det {\cal E} \cong K_X^{*}$, there exists a pseudo-topological
twists known as the A/2 model, for which nonperturbative
corrections to product structures have been computed for $X$ a toric
variety and ${\cal E}$ a deformation of the tangent bundle,
see for example \cite{ks,es-other,ade,dgks1,dgks2,mm1,mm2}.
In this paper we will focus on additive structures only.

Finally, let us discuss the behavior of these states under deformations
and RG flow.  In a (2,2) theory, the chiral states live in a topologically
protected subsector, and so one expects to have the same additive
structure in the chiral rings everywhere along RG flow and 
under deformations.  For example, this is the physics reason why the
Hodge numbers of Calabi-Yau's are the same in different geometric
phases of the same GLSM.  (In mathematics, this result is a consequence
of motivic integration \cite{andreipriv}.)

By contrast, in (0,2) theories, the chiral
ring $Q$-cohomology computation
is protected only against perturbative corrections, for the same
reasons that the (0,2) superpotential is not perturbatively renormalized.
In applications such as \cite{dg}, where RG flow stays in weakly-coupled
regimes, $Q$-cohomology can be reliably used to count states.
By contrast, in this paper we compute $Q$-cohomology in weakly-coupled
UV nonlinear sigma models which RG flow to strong coupling.  As a result,
one should expect that our $Q$-cohomology computations above will not
necessarily give the correct IR spectrum, but rather additional states
could enter or leave along the RG flow, and in fact that is precisely
what we find.

\section{Application to triality}
\label{sect:triality}

\subsection{Overview of triality}

It was proposed in \cite{ggp1} that triples of (0,2) GLSMs might flow to the
same IR fixed point. One starts with a (0,2) $U(k)$ GLSM:
\begin{center}
\begin{tabular}{ccccc}
$ $ & type & multiplicity & $su(k)$ & $u(1)$  \\ \hline 
$\Phi$ & chiral & $n$ & ${\bf k}$ & 1  \\
$P$ & chiral & $B$ & ${\bf \overline{k}}$ & -1  \\
$\Gamma$ & Fermi & $nB$ & ${\bf 1}$ & 0  \\
$\Psi$ & Fermi & $A$ & ${\bf \overline{k}}$ & -1  \\
$\lambda$ & fermion & 1 & $ad$ & 0  \\
$\Omega$ & Fermi & 2 & ${\bf 1}$ & $k$  \\
\end{tabular}
\end{center}
with a (0,2) superpotential $W = \Gamma P \Phi$, where $B = 2k + A -n$. This GLSM was argued to be dual to a (0,2) $U(n-k)$ GLSM:
\begin{center}
\begin{tabular}{ccccc}
$ $ & type & multiplicity & $su(k)$ & $u(1)$  \\ \hline 
$\tilde{\Phi}$ & chiral & $n$ & ${\bf k}$ & 1  \\
$\tilde{P}$ & chiral & $A$ & ${\bf \overline{k}}$ & -1  \\
$\tilde{\Gamma}$ & Fermi & $nA$ & ${\bf 1}$ & 0  \\
$\tilde{\Psi}$ & Fermi & $B$ & ${\bf \overline{k}}$ & -1  \\
$\tilde{\lambda}$ & fermion & 1 & $ad$ & 0  \\
$\tilde{\Omega}$ & Fermi & 2 & ${\bf 1}$ & $k$  \\
\end{tabular}
\end{center}
with a (0,2) superpotential $\tilde{W} = \tilde{\Gamma} \tilde{P} \tilde{\Phi}$. A further step of duality move leads to yet a third (0,2) GLSM with a $U(A-n+k)$ gauge group:
\begin{center}
\begin{tabular}{ccccc}
$ $ & type & multiplicity & $su(k)$ & $u(1)$  \\ \hline 
$\Phi'$ & chiral & $B$ & ${\bf k}$ & 1  \\
$P'$ & chiral & $n$ & ${\bf \overline{k}}$ & -1  \\
$\Gamma'$ & Fermi & $nB$ & ${\bf 1}$ & 0  \\
$\Psi'$ & Fermi & $A$ & ${\bf \overline{k}}$ & -1  \\
$\lambda'$ & fermion & 1 & $ad$ & 0  \\
$\Omega'$ & Fermi & 2 & ${\bf 1}$ & $k$  \\
\end{tabular}
\end{center}
with a (0,2) superpotential $W' = \Gamma' P' \Phi'$.

This phenomenon, labelled ``triality,'' can be
understood as follows.  If we integrate out the gauge field, then the
large-radius limit of the
first (0,2) theory can be understood as a nonlinear sigma model on
\begin{displaymath}
X_1 \: = \: G(k,n)
\end{displaymath}
with bundle 
\begin{displaymath}
{\cal E}_1 \: = \: S^A \oplus (Q^*)^{2k+A-n} \oplus (\det S^*)^2 ,
\end{displaymath}
where $S$ denotes the universal subbundle on $G(k,n)$, and $Q$ the
universal quotient bundle on $G(k,n)$.
This theory has four flavor symmetries, three of which rotate bundle
factors
\begin{displaymath}
SU(A) \times SU(2k+A-n) \times SU(2) ,
\end{displaymath}
and the fourth of which, $SU(n)$, acts on the base.
The other theories related by triality can be obtained by cyclically permuting
\begin{displaymath}
A, \: \: \: 2k+A-n, \: \: \: n, 
\end{displaymath}
and simultaneously replacing $k$ by $n-k$.
For example,
the large-radius limit of the 
second theory is given by a nonlinear sigma model on
\begin{displaymath}
X_2 \: = \: G(n-k,A)
\end{displaymath}
with bundle
\begin{displaymath}
{\cal E}_2 \: = \: S^{2k+A-n} \oplus (Q^*)^n \oplus (\det S^*)^2 ,
\end{displaymath}
and the large-radius limit of the 
third theory is given by a nonlinear sigma model on
\begin{displaymath}
X_3 \: = \: G(A-n+k, 2k+A-n)
\end{displaymath}
with bundle
\begin{displaymath}
{\cal E}_3 \: = \: S^n \oplus (Q^*)^A \oplus (\det S^*)^2 .
\end{displaymath}

Now, in order for the geometric description above to make sense,
the values of $n$, $A$, and $k$ are constrained.  For example, to
make sense of $G(A-n+k,2k+A-n)$, we require
\begin{displaymath}
0 \: < \: A-n+k \: < \: 2k+A-n .
\end{displaymath}
In addition, in order for the triality to be interesting, we would also like
supersymmetry to remain unbroken, which can be checked by {\it e.g.}
computing elliptic genera as refined Witten indices.
Happily, these two requirements -- that the geometric description be sensible,
and that supersymmetry be unbroken -- coincide in these theories.

In the UV, in addition to some nonanomalous $U(1)$'s, the theories above
have nonanomalous
\begin{displaymath}
SU(n) \times SU(A) \times SU(2k+A-n) \times SU(2)
\end{displaymath}
symmetries, as discussed above.  It was proposed in
\cite{ggp-exact} that in the IR, the theories above flow to a common
nontrivial SCFT, in which the global flavor symmetries above are
enhanced to affine symmetries \cite{ggp-exact}[equ'n (3.1)]
\begin{displaymath}
SU(n)_{k+A-n} \times SU(A)_{k} \times SU(2k+A-n)_{n-k} \times SU(2)_{1} .
\end{displaymath}

By examining chiral states among the UV theories above, we will give
nontrivial evidence that the different UV theories flow to the same IR
fixed point and have the IR affine symmetries indicated above.

\subsection{General remarks on chiral states}

\subsubsection{UV physics}

Let us focus on the first model in the last section, a heterotic
nonlinear sigma model described by the space and bundle
\begin{displaymath}
X \: = \: G(k,n), \: \: \:
{\cal E} \: = \: S^{A} \oplus (Q^*)^{2k+A-n} \oplus (\det S^*)^2 .
\end{displaymath}
To compute the chiral states, we first need to compute the bundle to which
the Fock vacuum couples.  To that end, in the model above,
\begin{displaymath}
\det {\cal E} \: = \: {\cal O}(-(A-2) \: - \: (2k+A-n)) \: = \:
{\cal O}(-2k -2A +n+2) ,
\end{displaymath}
and
\begin{displaymath}
K_X \: = \: {\cal O}(-n) ,
\end{displaymath}
hence
\begin{displaymath}
(\det {\cal E}) \otimes K_X \: = \: {\cal O}(-2k -2A +2), \: \: \:
(\det {\cal E})^{-1} \otimes K_X \: = \:
{\cal O}(2k+2A-2n-2) ,
\end{displaymath}
and so a square root always exists.  Furthermore, since the Grassmannian
is simply-connected, that square root is unique, and defines the bundle to
which the Fock vacuum couples.

Now, in this paper we are interested in more than merely counting states --
we also want to keep track of global symmetry representations.  To that end,
it will be useful to describe $X$ and ${\cal E}$ in terms of vector spaces
defining fundamental representations of global symmetry groups.  For example,
we will describe the model above as
\begin{displaymath}
X \: = \: G(k,\tilde{V}^*), \: \: \:
{\cal E} \: = \: U \otimes S \oplus V \otimes Q^* \oplus W \otimes
\det S^* ,
\end{displaymath}
where $U$, $V$, $W$, and $\tilde{V}$ are vector spaces of
dimensions $A$, $2k+A-n$, $2$, and $n$, respectively.

The original UV GLSM has a 
\begin{displaymath}
SU(A) \times SU(2k+A-n) \times SU(2) \times SU(n) \: = \:
SU(U) \times SU(V) \times SU(W) \times SU(\tilde{V})
\end{displaymath}
symmetry,
but in the sheaf cohomology groups, naively only the
\begin{displaymath}
SU(A) \times SU(2k+A-n) \times SU(2)
\: = \: SU(U) \times SU(V) \times SU(W)
\end{displaymath}
subgroup is explicit, in its action on the left-moving fermions.  
The remaining $SU(n)=SU(\tilde{V})$, which acts on the base, is made
manifest via the Bott-Borel-Weil theorem, reviewed in 
appendix~\ref{app:bbw},
which expresses sheaf cohomology of homogeneous vector bundles 
(including the ${\cal E}$ above) on Grassmannians $G(k,n)$ in terms
of representations of $U(n)$, making the relationship explicit.

To that end, when computing chiral rings, it will be important to keep
track of the difference between factors of {\it e.g.} $S$ and $Q^*$.
As holomorphic bundles, for example $\det S \cong \det Q^*$, but
they define different representations of the parabolic subgroup
$GL(k) \times GL(n-k)$ of $GL(n)$, and the difference will manifest
via Bott-Borel-Weil in terms of the precise representation of $U(n)$
appearing.  For example, if $n=2$, then on $G(1,\tilde{V}^*) = {\mathbb P}^1$,
$S^* \otimes Q^* \cong {\cal O}$.  However, applying Bott-Borel-Weil, we find
\begin{eqnarray*}
H^{\bullet}({\cal O}) & = & {\mathbb C} \delta^{\bullet,0}, \\
H^{\bullet}(S^* \otimes Q^*) & = & \wedge^2 \tilde{V} \delta^{\bullet,0} .
\end{eqnarray*}
The two sheaf cohomology groups have the same dimension -- as they should,
since the bundles are isomorphic as holomorphic bundles -- but encode
different representations of $GL(\tilde{V})$.
Thus, for example, when specifying Fock vacuum bundles, we must specify
not only a holomorphic line bundle, but in addition a precise representation
of the parabolic subgroup, {\it i.e.} a precise description as powers
of $S^*$ and $Q^*$.

\subsubsection{IR physics}

In principle, we would like to compare the UV sheaf cohomology groups
to the corresponding R sector states in the IR theory,
built from a right-moving part  of a Kazama-Suzuki
coset and a left-moving Kac-Moody algebra \cite{ggp-exact}.
Now, for a (2,2) Kazama-Suzuki coset $G/H$, it was argued in
\cite{lvw}[section 5] that part of the chiral ring is given as
the Lie algebra cohomology
\begin{displaymath}
H^{\bullet}\left( {\mathfrak t}_+, V_{\lambda} \right) ,
\end{displaymath}
where ${\mathfrak g} = {\mathfrak h} + {\mathfrak t}_+ + {\mathfrak t}_-$ is
a decomposition of the Lie algebra of $G$,
${\mathfrak h}$ the Lie algebra of $H$, and $V_{\lambda}$ is a 
${\mathfrak g}$-module corresponding to the representation of $G$ corresponding
to the ground state, {\it i.e.} an integrable representation.  
In the present case, for the IR limit in triality,
one would similarly expect 
that some of the chiral states can be expressed in the form
of Lie algebra cohomology
\begin{displaymath}
H^{\bullet}\left( {\mathfrak t}_+, M \right) ,
\end{displaymath}
where $M$ is a ${\mathfrak p}$-module
determined in part by the left-moving Kac-Moody algebra.  
Now, this description is incomplete -- even for (2,2) Kazama-Suzuki
cosets, it is not believed that the Lie algebra cohomology provides
a full description \cite{lvw,lerchepriv}, hence we cannot use this to try to
completely enumerate IR states to compare to UV states.

That said, one could still ask how the sheaf cohomology groups appearing
in the UV theory could be related, even in principle, to the Lie algebra
cohomology groups that appear, at least incompletely, in the IR.
The answer is provided by another version of the Bott-Borel-Weil
theorem \cite{kostant1,kostant2}.  As observed above, Bott-Borel-Weil
naturally says that for a homogeneous bundle ${\cal E}_{\xi}$ on 
$G/P$ defined by a representation $\xi$ of $P$, the sheaf cohomology
groups
\begin{displaymath}
H^{\bullet}(G/P, {\cal E}_{\xi})
\end{displaymath}
naturally come in representations of $G$.  It was observed in
\cite{kostant1,kostant2} that the Lie algebra cohomology groups
\begin{displaymath}
H^{\bullet}({\mathfrak t}_+, V_{\lambda})
\end{displaymath}
(for ${\mathfrak t}_+$ as above and $\lambda$ a representation
of $G$) naturally come in representations of $P$.  Moreover,
these sheaf cohomology and Lie algebra cohomology groups are
closely intertwined:  the multiplicity of representation
$\lambda$ of $G$ in $H^j(G/P,{\cal E}_{\xi})$ matches the
multiplicity of $\xi$ in $H^j({\mathfrak t}_+, V_{\lambda})$,
or more simply
\begin{displaymath}
H^j(G/P,{\cal E}_{\xi})_{\lambda} \: = \:
H^j({\mathfrak t}_+, V_{\lambda})_{\xi} ,
\end{displaymath}
where, for example,
\begin{displaymath}
H^j(G/P,{\cal E}_{\xi}) \: = \: \sum_{\lambda} V_{\lambda} \otimes
H^j(G/P,{\cal E}_{\xi})_{\lambda} .
\end{displaymath}
In other words, if $\lambda$ is a five-dimensional representation,
for example, and the sheaf cohomology group is $V_{\lambda}$,
then
\begin{displaymath}
H^j(G/P,{\cal E}_{\xi})_{\mu}
\: = \: \left\{ \begin{array}{cl}
{\mathbb C} & \mu = \lambda , \\
0 & \mu \neq \lambda ,
\end{array} \right.
\end{displaymath}
so that one copy of the five-dimensional $V_{\lambda}$ appears,
not five.

In these conventions,
\begin{displaymath}
H^{\bullet}(G/P, {\cal E}_{\xi}) \: = \:
\sum_{\mu} H^{\bullet}(G/P, {\cal E}_{\xi})_{\mu} \otimes V_{\mu} ,
\end{displaymath}
and
\begin{displaymath}
H^{\bullet}({\mathfrak n}, V_{\lambda}) \: = \: \sum_{\zeta}
H^{\bullet}({\mathfrak n}, V_{\lambda})_{\zeta} \otimes V_{\zeta} ,
\end{displaymath}
hence, for example,
\begin{displaymath}
H^{\bullet}(G/P, {\cal E}_{\xi}) \: = \:
\sum_{\mu} H^{\bullet}({\mathfrak t}_+, V_{\mu})_{\xi}  \otimes V_{\mu} .
\end{displaymath}
That said, in principle the ${\mathfrak t}_+$ pertinent to the IR
theory should be derived from a slightly more complicated Grassmannian,
so what we have outlined is not the complete story.  

In any event,
we will not try to   
compute chiral states in the IR theory, but,
we did think it important to demonstrate how in principle the
IR rings can be related to the sheaf cohomology groups we
discuss in this paper.

\subsection{Pseudo-topological twists}

Theories with (0,2) supersymmetry admit pseudo-topological twists,
resulting in theories known as the
A/2 and B/2 models \cite{ks,mm2}.  The A/2 model twist of a (0,2)
nonlinear sigma model on a space $X$ with bundle ${\cal E}$
is well-defined when,
in addition to the Green-Schwarz anomaly cancellation condition,
\begin{displaymath}
\det {\cal E}^* \: \cong \: K_X ,
\end{displaymath}
and the B/2 model twist is well-defined when, in addition to Green-Schwarz,
\begin{displaymath}
\det {\cal E} \: \cong \: K_X .
\end{displaymath}
Dualizing the bundle ${\cal E}$ yields an isomorphic quantum field theory,
in which the A/2 and B/2 twists are exchanged.

In principle, if the bundle ${\cal E}$ is reducible, then there are
further variants, further topological twists,
obtained by dualizing the various individual factors.  Dualizing those
bundle factors produces an isomorphic quantum field theory, but modifies
the twists.  If we think of the twists as twisting along a $U(1)$
symmetry of the theory, then the point here is that if the bundle is
reducible, then there are additional $U(1)$ symmetries (corresponding
to different phase factors on different factors)
which yield different pseudo-topological twists, or equivalently,
the A/2 and B/2 twist but for a bundle obtained by dualizing some of the
factors.  We will speak of a theory `admitting an A/2 or B/2 twist' when
the particular choice of ${\cal E}$ satisfies one of the conditions above.

Now, let us turn to the examples appearing in triality.
As a warm-up, consider
a (0,2) nonlinear sigma model on the Grassmannian $G(k,n)$ with
bundle
\begin{displaymath}
{\cal E} \: = \: S^{\oplus A} \oplus (Q^*)^{\oplus (2k+A-n)}\oplus 
(\det S^*)^{\oplus 2} ,
\end{displaymath}
from which we derive
\begin{eqnarray*}
\det {\cal E} & \cong & (\det S)^{A-2} \otimes (\det Q^*)^{2k+A-n}, \\
& \cong & (\det S)^{2A + 2k -n -2} ,
\end{eqnarray*}
using $\det Q^* \cong \det S$.
For the tangent bundle,
\begin{displaymath}
0 \: \longrightarrow \: S^* \otimes S \: \longrightarrow \:
S^* \otimes {\cal O}^n \: \longrightarrow \: T \: \longrightarrow \: 0 ,
\end{displaymath}
hence
$K_X \cong (\det S)^n$.
Putting this together, we find that this
model will admit an A/2 twist when
\begin{displaymath}
0 \: = \: A + k - 1,
\end{displaymath}
and the same model will admit a B/2 twist when
\begin{displaymath}
n \: = \: A + k - 1 .
\end{displaymath}
For bundles of this particular form, examples
admitting an A/2 twist will be rather rare, as it
requires $A+k=1$, but models admitting a B/2 twist are less uncommon.

Next, let us take advantage of the fact that ${\cal E}$ is reducible.
If we dualize the middle and last factors, we get the bundle
\begin{displaymath}
{\cal E}' \: = \: S^{\oplus A} \oplus (Q)^{\oplus (2k+A-n)}\oplus 
(\det S)^{\oplus 2} ,
\end{displaymath}
for which we compute
\begin{displaymath}
\det {\cal E}' \: \cong \: (\det S)^{A+2} \otimes (\det Q)^{2k+A-n}
\: \cong \: (\det Q)^{2k-n-2} ,
\end{displaymath}
and so
\begin{displaymath}
(\det {\cal E}')^{-1} \otimes K_X \: \cong \:
(\det S)^{2k-n-2} \otimes (\det S)^n \: \cong \: (\det S)^{2k-2} .
\end{displaymath}
Clearly, if $k=1$, then this presentation admits a B/2 twist.
However, if we are willing to make the global symmetry rotating
$Q^*$'s more obscure and dualize pairs of them individually, then we
can build an alternative bundle ${\cal E}''$ which admits a B/2 twist.

Thus, by suitably
dualizing gauge bundle factors, we can find a B/2 twist of 
any (0,2) theory related by triality.  That said, a given B/2 twist is not
invariant under triality, as even dualizing a bundle will replace the
original B/2 twist with something different.

\subsection{First example}

In this section we will compare chiral states in examples of
different UV NLSM's
that are related by triality -- some as different phases of the same
GLSM, others from different GLSM's.  We will find in the examples
we compute that all the different
presentations have some states in common, and a few states that differ
between presentations.  However, the states that are in common, all are
defined by integrable representations of the global symmetry groups,
integrable with respect to the proposed IR affine algebras.  Furthermore,
the mismatched states will not contribute to elliptic genera and are
defined by nonintegrable representations, strongly suggesting that they
become massive along the RG flow, indirectly verifying the triality proposal,
and also giving a very clean example of how non-protected operators can
change along RG flow.

\subsubsection{First GLSM}

We shall begin with a computation of chiral states in the two phases
of a (0,2) GLSM pertinent to triality. Let's take $k = 1$, $A = 3$, and $n = 3$, so $2k + A - n = 2$.  The two phases are defined by
\begin{displaymath}
{\cal E} \: \equiv \: U \otimes S \: + \: V \otimes Q^* \: + \:
W \otimes \det S^* \: \longrightarrow \: {\mathbb P}^2 \: = \: {\mathbb P} 
\tilde{V}^*
\end{displaymath}
for $r \gg 0$, and
\begin{displaymath}
{\cal F} \: \equiv \: U \otimes S^* \: + \: \tilde{V} \otimes Q^*
\: + \: W \otimes \det S \: \longrightarrow \: {\mathbb P}^1 \: = \:
{\mathbb P} V^*
\end{displaymath}
for $r \ll 0$, where in both phases,
\begin{displaymath}
U \: = \: {\mathbb C}^3, \: \: \:
V \: = \: {\mathbb C}^2, \: \: \:
W \: = \: {\mathbb C}^2, \: \: \:
\tilde{V} \: = \: {\mathbb C}^3.
\end{displaymath}
(For brevity, we only list the underlying geometries, rather than the
full matter content of each GLSM.  It may be worth observing that although
two of the triality phases are given by abelian GLSM's, the third is given
by a GLSM with gauge group $U(2)$.)

According to triality, these two nonlinear sigma models should flow in
the IR to the same point, hence, on the face of it, one would expect
them to have isomorphic chiral states.

It is straightforward to check that, in both cases, anomaly cancellation
holds, and furthermore each phase admits a B/2 twist, so that the Fock
vacuum line bundle is trivial.  As remarked earlier, for our purposes it is
important to give a precise presentation of the Fock vacuum line bundle
in terms of powers of $S^*$ and $Q^*$, and in both phases we will present 
it as the canonical trivial bundle, {\it i.e.}
$K_{(0)} S^* \otimes K_{(0,\cdots,0)} Q^*$ in the notation of
appendix~\ref{app:bbw}.  Hence we can
compute the chiral states in the form
\begin{displaymath}
H^{\bullet}(X, \wedge^{\bullet} {\cal E}) .
\end{displaymath}

Each phase has a set of global nonanomalous (chiral)
$U(1)^3$ symmetries, which are
given by
\begin{center}
\begin{tabular}{c|cccc}
& $\tilde{V}$ & $U$ & $V$ & $W$ \\ \hline
$U(1)_{(1)}$ & $0$ & $0$ & $-1$ & $-1$ \\
$U(1)_{(2)}$ & $1$ & $0$ & $0$ & $-3/2$ \\
$U(1)_{(3)}$ & $0$ & $1$ & $0$ & $+3/2$
\end{tabular}
\end{center}
In a slight variation from \cite{ggp1,ggp-exact}, 
we have assigned the same charge
to all elements of $W$, for simplicity in comparing states.

For the purposes of correctly comparing symmetries, it is useful
to distinguish ${\mathbb P} V$ from ${\mathbb P} V^*$, for example.
Briefly, on the space ${\mathbb P}V$,
the homogeneous coordinates naturally transform under $V^*$.  On that space,
\begin{displaymath}
0 \: \longrightarrow \: S \: \longrightarrow \: V \otimes {\cal O} \:
\longrightarrow \: Q \: \longrightarrow \: 0 ,
\end{displaymath}
so dualizing, taking the long exact sequence, and using
$H^{\bullet}(Q^*) =  0,$ we find that the homogeneous coordinates
are given by
\begin{displaymath}
H^0(S^*) \: = \: H^0(V^* \otimes {\cal O}) \: = \: V^*.
\end{displaymath}
Here, 
for $r \gg 0$, the homogeneous coordinates naturally transform under
$\tilde{V}$, and for $r \ll 0$, $V$, hence the two geometries are
naturally ${\mathbb P} \tilde{V}^*$ and ${\mathbb P}V^*$, respectively.

In the remaining tables in this section we list all of the states in the
two phases, beginning with all of the matching states.
Now, to match states in principle we need only match representations of
nonanomalous symmetries.  However, in this example, the bulk of the
matching states match full (anomalous) $GL(U) \times GL(V) \otimes GL(W)
\otimes GL(\tilde{V})$ representations.
In table~\ref{table:ex1:shared} we list all such states which match exactly,
as representations of the anomalous symmetry above, between the
two phases.
The state column lists the representation of
\begin{displaymath}
GL(U) \times GL(V) \times GL(W) \times GL(\tilde{V})
\end{displaymath}
obtained from the Bott-Borel-Weil computation.
As overall ${\mathbb C}^{\times}$ factors are individually anomalous,
we separately list the nonanomalous
\begin{displaymath}
SL(U) \times SL(V) \times SL(W) \times SL(\tilde{V})
\: = \: 
SU(3) \times SU(2) \times SU(2) \times SU(3)
\end{displaymath}
representations
and nonanomalous global
\begin{displaymath}
U(1)_{(1)} \times U(1)_{(2)} \times U(1)_{(3)}
\end{displaymath}
charges in the last two columns.
Immediately after the state listing, the
next two columns list in which wedge power of ${\cal E}$
the state was obtained, and the cohomological degree, respectively,
in the $r \gg 0$ phase, and the next two after that give the same information
for the $r \ll 0$ phase.

The $U(1)^3$ charges listed include the fractional charges from the Fock
vacua.  For the $r \gg 0$ phase, for example, the first and third $U(1)$'s
act linearly on the left-moving fermions, and so the fractional charges
can be computed directly using standard\footnote{
Alternatively, they can also be computed from our expression for the
Fock vacuum line bundle
\begin{displaymath}
(\det {\cal E})^{-1/2} \otimes K_X^{+1/2} .
\end{displaymath}
For the $r \gg 0$ phase, where
\begin{displaymath}
{\cal E} \: = \: U \otimes S \: + \: V \otimes Q^* \: + \: 
W \otimes S^* \: \longrightarrow \: G(1,\tilde{V}^*) ,
\end{displaymath}
it is straightforward to compute that
\begin{displaymath}
(\det {\cal E})^{-1/2} \otimes K_X^{+1/2} \: = \:
\left( \wedge^3 U \right)^{-1/2} \otimes \left( \wedge^2 V \right)^{-1}
\otimes \left( \wedge^2 W \right)^{-1/2} \otimes \left(
\wedge^3 \tilde{V}^* \right)^{-1/2} ,
\end{displaymath}
for which the first and third $U(1)$ charges are computed to match those
one would obtain by standard fractional fermion techniques.  We caution
against applying the same method for the fractional charge under the
second $U(1)$ in the $r \gg 0$ phase, as its action is not linear on the
right-moving fermions.  The method above will sometimes
give the correct result in that case, but will also often not.
} fractional fermion computations.
The second $U(1)$ does not act linearly on the right-moving fermions,
and so the corresponding vacuum charge is a bit more subtle to compute.
We computed it as $-1/2$ of the charge of the top-degree state, the
Serre dual to the state in $\wedge^0 {\cal E}$.  This guaranteed that the
first and last entries in table~\ref{table:ex1:shared} have opposite charges.
In any event, the result is that for both the $r \gg 0$ and $r \ll 0$ phases,
the fractional $U(1)^3$ charge of the vacuum was taken to be $(+3,0,-3)$.
It is a highly nontrivial consistency check that, with that choice,
all other states related by Serre duality also have opposite charges.

\begin{table}
\begin{tabular}{c|c|c|c|c|c|c}
& \multicolumn{2}{c}{$r \gg 0$} & \multicolumn{2}{c}{$r \ll 0$} & & \\
State & $\wedge^{\bullet} {\cal E}$ & $H^{\bullet}({\mathbb P}^2)$ &
$\wedge^{\bullet} {\cal F}$ & $H^{\bullet}({\mathbb P}^1)$ &
Rep' &
$U(1)^3$ \\ \hline
$1$ & $0$ & $0$ & $0$ & $0$ & $({\bf 1},{\bf 1},{\bf 1},{\bf 1})$ &
$(+3,0,-3)$ \\
$W \otimes \tilde{V}$ & $1$ & $0$ & $2$ & $1$ &
$({\bf 1},{\bf 1},{\bf 2}, {\bf 3})$ &
$(+2,-1/2,-3/2)$ \\
$\wedge^2 V \otimes \wedge^2 \tilde{V}$ & $2$ & $1$ & $2$ & $1$ &
$({\bf 1},{\bf 1},{\bf 1},{\bf \overline{3}})$ & $(+1,+2,-3)$ \\
$U \otimes V$ & $2$ & $1$ & $1$ & $0$ & $({\bf 3},{\bf 2},{\bf 1},{\bf 1})$
& $(+2,0,-2)$ \\
$U \otimes W$ & $2$ & $0$ & $2$ & $0$ & $({\bf 3},{\bf 1},{\bf 2},{\bf 1})$
& $(+2,-3/2,-1/2)$ \\
$V \otimes W \otimes \wedge^2 \tilde{V}$ & $2$ & $0$ & $3$ & $1$ &
$({\bf 1},{\bf 2},{\bf 2},{\bf \overline{3}})$ &
$(+1,+1/2,-3/2)$ \\
$U \otimes \wedge^2 V \otimes \tilde{V}$ & $3$ & $1$ & $2$ & $0$ &
$({\bf 3},{\bf 1},{\bf 1},{\bf 3})$ & $(+1,+1,-2)$ \\
$U \otimes \wedge^2 W \otimes \tilde{V}$ & $3$ & $0$ & $4$ & $1$ &
$({\bf 3},{\bf 1},{\bf 1},{\bf 3})$ &
$(+1,-2,+1)$ \\
$V \otimes \wedge^2 V \otimes \wedge^3 \tilde{V}$ & $3$ & $1$ & $3$ & $1$ &
$({\bf 1},{\bf 2},{\bf 1},{\bf 1})$ & $(0,+3,-3)$ \\
${\rm Sym}^2 V \otimes W \otimes \wedge^3 \tilde{V}$ & $3$ & $0$ & $4$ & $1$ &
$({\bf 1},{\bf 3},{\bf 2},{\bf 1})$ & $(0,+3/2,-3/2)$ \\
$\wedge^2 U \otimes {\rm Sym}^2 V$ & $4$ & $2$ & $2$ & $0$ &
$({\bf \overline{3}},{\bf 3},{\bf 1},{\bf 1})$ &
$(+1,0,-1)$ \\
$\wedge^2 U \otimes V \otimes W$ & $4$ & $1$ & $3$ & $0$ &
$({\bf \overline{3}},{\bf 2},{\bf 2},{\bf 1})$ &
$(+1,-3/2,+1/2)$ \\
$\wedge^2 U \otimes \wedge^2 W$ & $4$ & $0$ & $4$ & $0$ &
$({\bf \overline{3}},{\bf 1},{\bf 1},{\bf 1})$ &
$(+1,-3,+2)$ \\
$U \otimes \wedge^2 V \otimes W \otimes \wedge^2 \tilde{V}$ & $4$ & $1$ &
$4$ & $1$ & $({\bf 3},{\bf 1},{\bf 2},{\bf \overline{3}})$ &
$(0,+1/2,-1/2)$ \\
$U \otimes V \otimes \wedge^2 W \otimes \wedge^2 \tilde{V}$ & $4$ & $0$ &
$5$ & $1$ & $({\bf 3},{\bf 2},{\bf 1},{\bf \overline{3}})$ &
$(0,-1,+1)$ \\
$\wedge^2 U \otimes \wedge^2 V \otimes V \otimes \tilde{V}$ &
$5$ & $2$ & $3$ & $0$ & $({\bf \overline{3}},{\bf 2},{\bf 1}, {\bf 3})$ &
$(0,+1,-1)$ \\
$\wedge^2 U \otimes W \otimes \wedge^2 V \otimes \tilde{V}$ & $5$ & $1$ &
$4$ & $0$ & $({\bf \overline{3}},{\bf 1},{\bf 2},{\bf 3})$ &
$(0,-1/2,+1/2)$ \\
$U \otimes (\wedge^2 V)^2 \otimes \wedge^3 \tilde{V}$ & $5$ & $2$ &
$4$ & $1$ & $({\bf 3},{\bf 1},{\bf 1},{\bf 1})$ &
$(-1,+3,-2)$ \\
$U \otimes W \otimes \wedge^2 V \otimes V \otimes \wedge^3 \tilde{V}$ & $5$ &
$1$ & $5$ & $1$ & $({\bf 3},{\bf 2},{\bf 2},{\bf 1})$ &
$(-1,+3/2,-1/2)$ \\
$U \otimes {\rm Sym}^2 V \otimes \wedge^2 W \otimes \wedge^3 \tilde{V}$ &
$5$ & $0$ & $6$ & $1$ & $({\bf 3},{\bf 3},{\bf 1},{\bf 1})$ &
$(-1,0,+1)$ \\
$\wedge^3 U \otimes {\rm Sym}^2 V \otimes W$ & $6$ & $2$ & $4$ & $0$ &
$({\bf 1},{\bf 3},{\bf 2},{\bf 1})$ & $(0,-3/2,+3/2)$ \\
$\wedge^3 U \otimes V \otimes \wedge^2 W$ & $6$ & $1$ & $5$ & $0$ &
$({\bf 1},{\bf 2},{\bf 1},{\bf 1})$ & $(0,-3,+3)$ \\
$\wedge^2 U \otimes (\wedge^2 V)^2 \otimes \wedge^2 \tilde{V}$ & $6$ & $2$ &
$4$ & $0$ & $({\bf \overline{3}},{\bf 1},{\bf 1},{\bf \overline{3}})$ &
$(-1,+2,-1)$ \\
$\wedge^2 U \otimes \wedge^2 V \otimes \wedge^2 W \otimes \wedge^2
\tilde{V}$ & $6$ & $1$ & $6$ & $1$ &
$({\bf \overline{3}},{\bf 1},{\bf 1},{\bf \overline{3}})$ &
$(-1,-1,+2)$ \\
$\wedge^3 U \otimes \wedge^2 V \otimes V \otimes W \otimes \tilde{V}$ &
$7$ & $2$ & $5$ & $0$ &
$({\bf 1},{\bf 2},{\bf 2},{\bf 3})$ &
$(-1,-1/2,+3/2)$ \\
$\wedge^2 U \otimes (\wedge^2 V)^2 \otimes W \otimes \wedge^3 \tilde{V}$ &
$7$ & $2$ & $6$ & $1$ &
$({\bf \overline{3}},{\bf 1},{\bf 2},{\bf 1})$ &
$(-2,+3/2,+1/2)$ \\
$\wedge^2 U \otimes \wedge^2 V \otimes V \otimes \wedge^2 W \otimes 
\wedge^3 \tilde{V}$ & $7$ & $1$ & $7$ & $1$ &
$({\bf \overline{3}},{\bf 2},{\bf 1},{\bf 1})$ &
$(-2,0,+2)$ \\
$\wedge^3 U \otimes \wedge^2 V \otimes \wedge^2 W \otimes \tilde{V}$ &
$7$ & $1$ & $6$ & $0$ &
$({\bf 1},{\bf 1},{\bf 1},{\bf 3})$ &
$(-1,-2,+3)$ \\
$\wedge^3 U \otimes (\wedge^2 V)^2 \otimes W \otimes \wedge^2 \tilde{V}$ &
$8$ & $2$ & $6$ & $0$ &
$({\bf 1},{\bf 1},{\bf 2},{\bf \overline{3}})$ &
$(-2,+1/2,+3/2)$ \\
$\wedge^3 U \otimes (\wedge^2 V)^2 \otimes \wedge^2 W \otimes
\wedge^3 \tilde{V}$ & $9$ & $2$ & $8$ & $1$ &
$({\bf 1},{\bf 1},{\bf 1},{\bf 1})$ &
$(-3,0,+3)$
\end{tabular}
\caption{List of states shared between the two phases.}
\label{table:ex1:shared}
\end{table}

As a consistency check, note that all of the states in all of the
tables in this section come in Serre-dual pairs, exchanging not only
cohomology degrees but also dualizing representations and $U(1)^3$
charges.  (Recall that in $SU(2)$, ${\bf 2} = {\bf \overline{2}}$,
so the dualization only acts nontrivially on $SU(3)$ representations.)

A few additional states have matching representations of anomaly-free
global symmetries,
{\it i.e.} the same $SU(3)\times\cdots$ representations
and $U(1)^3$ charges, but are expressed differently in terms of
anomalous representations.  These are listed in
table~\ref{table:ex1:match-only-anom-free}.

\begin{table}
\begin{tabular}{c|c|c|c}
$r\gg 0$ & $r\ll 0$ & & \\
State, $\wedge^{\bullet} {\cal E}, H^{\bullet}({\mathbb P}^2)$ &
State,  $\wedge^{\bullet} {\cal F}, H^{\bullet}({\mathbb P}^1)$ &
Rep' & $U(1)^3$ \\ \hline
$\wedge^3 U \otimes \wedge^3 \tilde{V}^*$,  $3, 2$ &
$\wedge^2 W \otimes \wedge^2 V^*$, $2, 1$ &
$({\bf 1},{\bf 1},{\bf 1},{\bf 1})$ &
$(+3,-3,0)$ \\
$\wedge^3 U \otimes V \otimes \wedge^2 \tilde{V}^*$, $4, 2$ &
$\tilde{V} \otimes \wedge^2 W \otimes V^*$, $3, 1$ &
$({\bf 1},{\bf 2},{\bf 1},{\bf 3})$ &
$(+2,-2,0)$ \\
${\rm Sym}^2 V \otimes \wedge^2 W \otimes \tilde{V} \otimes
\wedge^3 \tilde{V}$, $4, 0$ &
$\wedge^3 U \otimes \tilde{V} \otimes \wedge^2 V \otimes 
{\rm Sym}^2 V$, $4, 0$ &
$({\bf 1},{\bf 3},{\bf 1},{\bf 3})$ &
$(-1,+1,0)$ \\
$\wedge^3 U \otimes {\rm Sym}^2 V \otimes \wedge^3 \tilde{V}^* \otimes
\wedge^2 \tilde{V}$, $5, 2$ &
$\wedge^2 \tilde{V} \otimes \wedge^2 W \otimes \wedge^2 V^* \otimes
{\rm Sym}^2 V$, $4, 1$ &
$({\bf 1},{\bf 3},{\bf 1},{\bf \overline{3}})$ &
$(+1,-1,0)$ \\
$\wedge^2 V \otimes V \otimes \wedge^2 W \otimes \wedge^3 \tilde{V}
\otimes \wedge^2 \tilde{V}$, $5, 0$ &
$\wedge^3 U \otimes \wedge^2 \tilde{V} \otimes (\wedge^2 V)^2 \otimes V$,
$5, 0$ & $({\bf 1},{\bf 2},{\bf 1},{\bf \overline{3}})$ &
$(-2,+2,0)$ \\
$(\wedge^2 V)^2 \otimes \wedge^2 W \otimes (\wedge^3 \tilde{V})^2$, $6, 0$ &
$\wedge^3 U \otimes \wedge^3 \tilde{V} \otimes (\wedge^2 V)^3$,
$6, 0$ & $({\bf 1},{\bf 1},{\bf 1},{\bf 1})$ &
$(-3,+3,0)$
\end{tabular}
\caption{Additional shared states
defined by matching representations of anomaly-free global
symmetries.}
\label{table:ex1:match-only-anom-free}
\end{table}

Finally, there are a few remaining states in each phase
that do not match any states in the other phase at all,
listed in table~\ref{table:ex1:mismatch}.
Note in particular that the states in neither phase are a proper
subset of the states in the other:  both phases have states not in
the other.

\begin{table}
\begin{tabular}{c|c|c|c|c|c|c}
& \multicolumn{2}{c}{$r \gg 0$} & \multicolumn{2}{c}{$r \ll 0$} & & \\
State & $\wedge^{\bullet} {\cal E}$ & $H^{\bullet}({\mathbb P}^2)$ &
$\wedge^{\bullet} {\cal F}$ & $H^{\bullet}({\mathbb P}^1)$ &
Rep' &
$U(1)^3$ \\ \hline
$\wedge^2 W \otimes {\rm Sym}^2 \tilde{V}$ & $2$ & $0$ & $-$ & $-$ &
$({\bf 1},{\bf 1},{\bf 1},{\bf 6})$ & $(+1,-1,0)$ \\
$V \otimes \wedge^2 W \otimes K_{(2,1,0)} \tilde{V}$ & $3$ & $0$ &
$-$ & $-$ &
$({\bf 1},{\bf 2},{\bf 1},{\bf 8})$ &
$(0,0,0)$ \\
$\wedge^2 V \otimes \wedge^2 W \otimes K_{(2,2,0)} \tilde{V}$ & $4$ & $0$ &
$-$ & $-$ &
$({\bf 1},{\bf 1},{\bf 1},{\bf 6})$ & $(-1,+1,0)$ \\
$\wedge^3 U \otimes \wedge^2 V \otimes \wedge^3 \tilde{V}^* \otimes
{\rm Sym}^2 \tilde{V}$ & $5$ & $2$ & $-$ & $-$ &
$({\bf 1},{\bf 1},{\bf 1},{\bf 6})$ & $(+1,-1,0)$ \\
$\wedge^3 U \otimes K_{(2,1)} V \otimes K_{(1,0,-1)} \tilde{V}$ & $6$ & $2$ &
$-$ & $-$ & $({\bf 1},{\bf 2},{\bf 1},{\bf 8})$ &
$(0,0,0)$ \\
$\wedge^3 U \otimes (\wedge^2 V)^2 \otimes \wedge^3 \tilde{V}
\otimes {\rm Sym}^2 \tilde{V}^*$ & $7$ & $2$ & $-$ & $-$ &
$({\bf 1},{\bf 1},{\bf 1},{\bf 6})$ &
$(-1,+1,0)$ \\
$\wedge^3 U \otimes {\rm Sym}^3 V$ & $-$ & $-$ & $3$ & $0$ &
$({\bf 1},{\bf 4},{\bf 1},{\bf 1})$ & $(0,0,0)$ \\
$\wedge^3 \tilde{V} \otimes \wedge^2 W \otimes \wedge^2 V^* \otimes
{\rm Sym}^3 V$ & $-$ & $-$ & $5$ & $1$ &
$({\bf 1},{\bf 4},{\bf 1},{\bf 1})$ & $(0,0,0)$
\end{tabular}
\caption{List of all states which are not shared between the two phases.}
\label{table:ex1:mismatch}
\end{table}

To aid the reader, as the methods are not commonly used in the
physics community, let us take a moment to illustrate how, for example,
the next-to-last entry in table~\ref{table:ex1:shared} was computed, in the
$r\gg 0$ phase.  This entry arose as the only nonzero contribution to
$H^{\bullet}({\mathbb P}^2, \wedge^8 {\cal E})$.  Now,
\begin{eqnarray*}
\wedge^8 {\cal E} & = & \wedge^8 \left( U \otimes S \: + \: 
V \otimes Q^* \: + \: W \otimes S^* \right) , \\
& = & \wedge^2(U \otimes S) \otimes \wedge^4 (V \otimes Q^*)
\otimes \wedge^2 (W \otimes S^*)
\\
& & \hspace*{0.5in} \: + \:
\wedge^3(U \otimes S) \otimes \wedge^3(V \otimes Q^*) \otimes
\wedge^2(W \otimes S^*)
\\
& &  \hspace*{0.5in} \: + \: 
\wedge^3(U \otimes S) \otimes \wedge^4(V \otimes Q^*) \otimes 
\wedge^1(W \otimes S^*) ,
\end{eqnarray*}
since $U\otimes S$ has rank 3, $V \otimes Q^*$ has rank 4, and
$W \otimes S^*$ has rank 2.  (More generally,
we sum over all combinations of
wedge powers adding up to the given power, which in this case is eight.)
To compute each of the wedge powers appearing, we use the identity
\begin{displaymath}
\wedge^r(A \otimes B) \: = \: \sum_{|\lambda|=r} K_{\lambda} A \otimes
K_{\lambda^T} B ,
\end{displaymath}
where the sum is over Young diagrams $\lambda$ with $r$ boxes,
and $K_{\lambda} A$ denotes a tensor product of $A$'s determined by
the Young diagram, for example
\begin{displaymath}
K_{\tiny{\yng(2)}} A \: = \: {\rm Sym}^2 A, \: \: \:
K_{\tiny{\yng(1,1)}} A \: = \: \wedge^2 A .
\end{displaymath}
Thus, for example,
\begin{eqnarray*}
\wedge^2(W \otimes S^*) & = & K_{\tiny{\yng(2)}} W \otimes 
K_{\tiny{\yng(1,1)}} S^*
\: + \: K_{\tiny{\yng(1,1)}} W \otimes K_{\tiny{\yng(2)}} S^* ,
\\
& = &
{\rm Sym}^2 W \otimes \wedge^2 S^* \: + \: \wedge^2 W \otimes {\rm Sym}^2 S^* .
\end{eqnarray*}
However, since $S^*$ has rank one, $\wedge^2 S^* = 0$, and so
\begin{displaymath}
\wedge^2(W \otimes S^*) \: = \: \wedge^2 W \otimes
{\rm Sym}^2 S^* .
\end{displaymath}
Similarly,
\begin{displaymath}
\wedge^3(V \otimes Q^*) \: = \: K_{\tiny{\yng(2,1)}} V \otimes
K_{\tiny{\yng(2,1)}} Q^*, \: \: \:
\wedge^4(V \otimes Q^*) \: = \: K_{\tiny{\yng(2,2)}} V \otimes
K_{\tiny{\yng(2,2)}} Q^* ,
\end{displaymath}
since both $V$ and $Q^*$ have rank 2, eliminating most possible
contributions, and
\begin{displaymath}
\wedge^2(U \otimes S) \: = \: \wedge^2 U \otimes {\rm Sym}^2 S, \: \: \:
\wedge^3(U \otimes S) \: = \: \wedge^3 U \otimes {\rm Sym}^3 S ,
\end{displaymath}
since $S$ has rank 1, eliminating most possible contributions.
Putting this together, we find
\begin{eqnarray*}
\wedge^8 {\cal E} & = & \wedge^2 U \otimes K_{\tiny{\yng(2,2)}} V
\otimes \wedge^2 W \otimes {\rm Sym}^2 S \otimes K_{\tiny{\yng(2,2)}}
Q^* \otimes {\rm Sym}^2 S^*\\
& & \hspace*{0.5in} \: + \:
\wedge^3 U \otimes K_{\tiny{\yng(2,1)}} V \otimes \wedge^2 W \otimes
{\rm Sym}^3 S \otimes K_{\tiny{\yng(2,1)}} Q^* \otimes {\rm Sym}^2 S^*
\\
& & \hspace*{0.5in} \: + \:
\wedge^3 U \otimes K_{\tiny{\yng(2,2)}} V \otimes W \otimes
{\rm Sym}^3 S \otimes K_{\tiny{\yng(2,2)}} Q^* \otimes S^* .
\end{eqnarray*}
Thus,
\begin{eqnarray*}
H^{\bullet}({\mathbb P}^2,\wedge^8 {\cal E}) & = &
\wedge^2 U \otimes K_{\tiny{\yng(2,2)}} V
\otimes \wedge^2 W \otimes H^{\bullet}\left({\mathbb P}^2,
K_{(2,2)} Q^* \right) \\
& & \hspace*{0.5in} \: + \:
\wedge^3 U \otimes K_{\tiny{\yng(2,1)}} V \otimes \wedge^2 W \otimes
H^{\bullet}\left( {\mathbb P}^2, K_{(-1)} S^* \otimes K_{(2,1)} Q^*
\right) \\
& & \hspace*{0.5in} \: + \:
\wedge^3 U \otimes K_{\tiny{\yng(2,2)}} V \otimes W \otimes
H^{\bullet}\left( {\mathbb P}^2, K_{(-2)} S^* \otimes K_{(2,2)} Q^*
\right) ,
\end{eqnarray*}
and from Bott-Borel-Weil,
\begin{displaymath}
H^{\bullet}\left({\mathbb P}^2,
K_{(2,2)} Q^* \right) \: = \: 0 \: = \:
H^{\bullet}\left( {\mathbb P}^2, K_{(-1)} S^* \otimes K_{(2,1)} Q^*
\right), 
\end{displaymath}
\begin{displaymath}
H^{\bullet}\left( {\mathbb P}^2, K_{(-2)} S^* \otimes K_{(2,2)} Q^*
\right)
\: = \: K_{(1,1,0)} \tilde{V} \, \delta^{\bullet,2}.
\end{displaymath}
The next-to-last entry in table~\ref{table:ex1:shared} follows.

Note that the states in table~\ref{table:ex1:mismatch} might not make a net
contribution to elliptic genera.  From \cite{jsw}[section 2.1], 
since this theory is B/2-twistable, the leading term in any refined
NLSM elliptic genus will be of the form
\begin{equation}    \label{eq:b2-ellgen-lead}
(-)^{r/2} q^{+(r-n)/12} \sum_{s=0}^r (-)^s \chi_y(\wedge^s {\cal E}) ,
\end{equation}
where $y$ represents the refinement by any nonanomalous symmetry,
$r$ is the rank of ${\cal E}$, and $n$ is the dimension of the base space.
Since in our example the rank is greater than the dimension of the
base, the unrefined elliptic genus will vanish \cite{jsw}.  However, since
there are a number of nonanomalous symmetries, the elliptic genus can be
refined, and it is straightforward to check that by adding suitable
refinements, the elliptic genus can certainly be made nonzero.  

Now, the contribution of any state to the weighted 
sum~(\ref{eq:b2-ellgen-lead}) is weighted in part by a sign determined by the
wedge power of the bundle and the degree of the
cohomology group in which the state appears.  However, all of the mismatched
states in table~\ref{table:ex1:mismatch} come in pairs with matching 
representations of nonanomalous symmetries but different signs, and so
cancel out.  Consider as a prototypical example the two states in the
$r \ll 0$ phase.  They both live in the same representation $({\bf 1},
{\bf 4}, {\bf 1}, {\bf 1})$, and have the same $U(1)^3$ charges, but one
enters with a $+$ and the other with a $-$, and so their net contribution to
the leading term of any elliptic genus refined by any of the symmetries
listed is necessarily zero.  Thus, the mismatched states
listed in table~\ref{table:ex1:mismatch} make no net contribution to the
leading term of any refined elliptic genus we shall consider.  
Of course,
that guarantees neither that higher order contributions will also vanish,
nor that there are no other nonanomalous symmetries whose refinements
might receive a contribution.  However, we do find it to be a suggestive
observation, supporting the idea that these (mismatched) states might all pair
up and become massive along the RG flow.

Now, let us examine these states from the perspective of 
the exact IR limit proposed in \cite{ggp-exact}.  In the notation
of that reference, in the IR there is an affine\footnote{
The $SU(W)$ affine contribution is not mentioned explicitly in
\cite{ggp-exact}.  The level can be computed from the trace anomaly
\begin{displaymath}
k_{\Omega} \: = \: {\rm tr}\, \gamma^3 J J \: = \: - 1/2
\end{displaymath}
(in their conventions).  The corresponding level of the IR current algebra
is
\begin{displaymath}
2 | k_{\Omega} | \: = \: 1
\end{displaymath}
and the sign indicates that it is left-moving.  
}
\begin{displaymath}
SU(U)_{1} \times SU(V)_{2} \times SU(W)_1 \times SU(\tilde{V})_{1}
\: = \:
SU(3)_1 \times SU(2)_2 \times SU(2)_1 \times SU(3)_1
\end{displaymath}
symmetry.  The corresponding left-chiral states in the IR should be
determined in part by integrable representations of the affine symmetry
above.

In terms of Young diagrams, a given representation of $SU(N)_k$ is
integrable if the Young diagram has no row with more than $k$ boxes
\cite{difranc}[section 16.6].
For example, for $k=1$, there are $N$ integrable representations,
given by
\begin{displaymath}
{\bf 1}, \: \: \:
{\bf N}, \: \: \:
\wedge^2 {\bf N}, \: \: \: \cdots, \: \: \: \wedge^{N-1}
{\bf N}.
\end{displaymath}
corresponding to Young diagrams
\begin{displaymath}
\Yvcentermath1
1, \: \: \:
\yng(1), \: \: \: \yng(1,1), \: \: \:
\yng(1,1,1), \: \: \: \cdots .
\end{displaymath}
In the case $k=2$, in addition to the $N$ diagrams above,
there are $N-1$ diagrams of the form
\begin{displaymath}
\Yvcentermath1
\yng(2), \: \: \:
\yng(2,1), \: \: \:
\yng(2,1,1), \: \: \:
\yng(2,1,1,1), \: \: \: \cdots ,
\end{displaymath}
plus $N-2$ diagrams of the form
\begin{displaymath}
\Yvcentermath1
\yng(2,2), \: \: \:
\yng(2,2,1), \: \: \:
\yng(2,2,1,1), \: \: \: \cdots ,
\end{displaymath}
and so forth, for a total of
\begin{displaymath}
1\: + \: 2 \: + \: \cdots \: + \: N-1 \: + \: N \: = \: \frac{1}{2} N (N+1)
\end{displaymath}
integrable representations of $su(N)$ at level $k=2$.

In the present case, $SU(3)_1$ has integrable representations
\begin{displaymath}
{\bf 1}, \: \: \:
{\bf 3}, \: \: \:
{\bf \overline{3}},
\end{displaymath}
$SU(2)_1$ has integrable representations
\begin{displaymath}
{\bf 1}, {\bf 2},
\end{displaymath}
and $SU(2)_2$ has integrable representations
\begin{displaymath}
{\bf 1}, \: \: \:
{\bf 2}, \: \: \:
{\bf 3}.
\end{displaymath}

Note that all of the states in tables~\ref{table:ex1:shared} and
\ref{table:ex1:match-only-anom-free} have integrable representations of
the nonabelian UV global symmetry groups.  On the other hand, all of the
mismatched states in table~\ref{table:ex1:mismatch} have at least
one non-integrable representation.  If any of the mismatched states
survived to the IR, they would then appear to contradict the assertion of
\cite{ggp-exact} that these theories have a nontrivial IR fixed point
of the form described there.  However, given that their contributions to
refined elliptic genera vanish, we find it much more likely that they become
massive, leaving only states which are both common across presentations
and defined by suitable representations.  In this fashion, we have an
indirect test of triality.

\subsubsection{Other GLSMs}  \label{sect:1st:other}

In the previous subsection we analyzed the chiral states in the two geometric
phases of one of three GLSMs that is believed to flow to a single
fixed point.  Next we shall repeat the same analysis for another
GLSM related to the first by triality.  We shall find an analogous 
structure -- states in integrable representations match between
the phases, and moreover, we shall see that the states that match
between phases also match between GLSMs.

The other two GLSMs can be obtained by cyclically permuting
\begin{displaymath}
U \: \longrightarrow \: V \: \longrightarrow \: \tilde{V}^* \: \longrightarrow
\:  U \: \longrightarrow \: \cdots .
\end{displaymath}
The three large-radius phases correspond to the bundles and spaces given by
\begin{eqnarray*}
(1): & U \otimes S \: + \: V \otimes Q^* \: + \: W \otimes \det S^* &
\longrightarrow \: G(1,\tilde{V}^*) ,\\
(2): & V \otimes S \: + \: \tilde{V}^* \otimes Q^* \: + \: W \otimes \det S^*
& \longrightarrow \: G(2,U) , \\
(3): & \tilde{V}^* \otimes S \: + \: U \otimes Q^* \: + \: W \otimes \det S^*
& \longrightarrow \: G(1,V) ,\\
\end{eqnarray*}
and the three $r \ll 0$ phases are described by
\begin{eqnarray*}
(1): & U \otimes S^* \: + \: \tilde{V} \otimes Q^* \: + \: W \otimes \det S
& \longrightarrow \: G(1,V^*) ,\\
(2): & V \otimes S^* \: + \: U^* \otimes Q^* \: + \: W \otimes \det S 
& \longrightarrow \: G(2,\tilde{V}) ,\\
(3): & \tilde{V}^* \otimes S^* \: + \: V^* \otimes Q^* \: + \: W \otimes \det S
& \longrightarrow \: G(1,U^*) .
\end{eqnarray*}
(As a consistency check, the $U(1)^3$ global symmetry is nonanomalous
in each of the six phases above.)
In the previous section, we computed the chiral rings in both phases of
GLSM (1), and note that the $r \ll 0$ phase of (1) is closely related by
mathematical duality to the $r \gg 0$ phase of (3).  The remaining geometry
can be described as either the $r \gg 0$ phase of (2) or the $r \ll 0$ phase
of (3).  Let us therefore focus on GLSM (3), and compute the chiral rings
in the two NLSM phases.

The reader should note that the phases of GLSM (3) are closely related to
those of (1).  Specifically, we can get the $r \ll 0$ phase of (3) from
the $r \gg 0$ phase of (1), and the $r \gg 0$ phase of (3) from the
$r \ll 0$ phase of (1), by making the substitutions
\begin{displaymath}
U \: \leftrightarrow \: \tilde{V}, \: \: \:
V \: \leftrightarrow \: V^*, \: \: \:
W \: \leftrightarrow \: W^*,
\end{displaymath}
so rather than re-compute spectra from scratch, we can simply re-use the
existing tables of states by making the replacements above.

The $U(1)^3$ charges are computed from a slightly different action than
in the first GLSM,
given by
\begin{center}
\begin{tabular}{c|cccc}
& $\tilde{V}$ & $U$ & $V$ & $W$ \\ \hline
$U(1)_{(1)}$ & $0$ & $0$ & $+1$ & $+1$ \\
$U(1)_{(2)}$ & $1$ & $0$ & $0$ & $-3/2$ \\
$U(1)_{(3)}$ & $0$ & $1$ & $0$ & $+3/2$
\end{tabular}
\end{center}
(These are almost the same as in the first GLSM, except that there is
a sign flip on the charges of the first $U(1)$.)
With these (nonanomalous) charge assignments, we shall see that the
states we compute in this GLSM which are shared between the two geometric
phases, are also shared with the first GLSM.
In any event, following the same procedure as in the last
section, we compute $(+3,-3,0)$ for both the $r \gg 0$ and
$r \ll 0$ phases.

By making the substitutions described above, we can immediately
write down the chiral states in the two phases of this GLSM.
These states are encoded in tables~\ref{table:ex1:var3:shared},
\ref{table:ex1:var3:match-only-anom-free}, and \ref{table:ex1:var3:mismatch},
which are precise analogues of the corresponding 
tables~\ref{table:ex1:shared}, 
\ref{table:ex1:match-only-anom-free}, \ref{table:ex1:mismatch}
for the previous GLSM related by triality.
As a consistency check, it is straightforward to check that all states
come in Serre dual pairs in which representations are dualized.
Furthermore, all of the states in table~\ref{table:ex1:var3:mismatch}
cancel out of the leading term in any elliptic genus refined by any of
the displayed nonanomalous global symmetries, as before.

\begin{table}
\begin{tabular}{c|c|c|c|c|c|c}
& \multicolumn{2}{c}{$r \ll 0$} & \multicolumn{2}{c}{$r \gg 0$} & & \\
State & $\wedge^{\bullet} {\cal E}$ & $H^{\bullet}({\mathbb P}^2)$ &
$\wedge^{\bullet} {\cal F}$ & $H^{\bullet}({\mathbb P}^1)$ &
Rep' &
$U(1)^3$ \\ \hline
$1$ & $0$ & $0$ & $0$ & $0$ & $({\bf 1},{\bf 1},{\bf 1},{\bf 1})$ &
$(+3,-3,0)$ \\
$W^* \otimes U$ & $1$ & $0$ & $2$ & $1$ & $({\bf 3},{\bf 1},{\bf 2},{\bf 1})$
& $(+2,-3/2,-1/2)$ \\
$\wedge^2 V^* \otimes \wedge^2 U$ & $2$ & $1$ & $2$ & $1$ &
$({\bf \overline{3}},{\bf 1},{\bf 1},{\bf 1})$ &
$(+1,-3,+2)$ \\
$\tilde{V} \otimes V^*$ & $2$ & $1$ & $1$ & $0$ & 
$({\bf 1},{\bf 2},{\bf 1},{\bf 3})$ &
$(+2,-2,0)$ \\
$\tilde{V} \otimes W^*$ & $2$ & $0$ & $2$ & $0$ & 
$({\bf 1},{\bf 1},{\bf 2},{\bf 3})$ &
$(+2,-1/2,-3/2)$ \\
$V^* \otimes W^* \otimes \wedge^2 U$ & $2$ & $0$ & $3$ & $1$ &
$({\bf \overline{3}},{\bf 2},{\bf 2},{\bf 1})$ &
$(+1,-3/2,+1/2)$ \\
$\tilde{V} \otimes \wedge^2 V^* \otimes U$ & $3$ & $1$ & $2$ & $0$ &
$({\bf 3},{\bf 1},{\bf 1},{\bf 3})$ & $(+1,-2,+1)$ \\
$\tilde{V} \otimes \wedge^2 W^* \otimes U$ & $3$ & $0$ & $4$ & $1$ &
$({\bf 3},{\bf 1},{\bf 1},{\bf 3})$ & $(+1,+1,-2)$ \\
$V^* \otimes \wedge^2 V^* \otimes \wedge^3 U$ & $3$ & $1$ & $3$ & $1$ &
$({\bf 1},{\bf 2},{\bf 1},{\bf 1})$ & $(0,-3,+3)$ \\
${\rm Sym}^2 V^* \otimes W^* \otimes \wedge^3 U$ & $3$ & $0$ & $4$ & $1$ &
$({\bf 1},{\bf 3},{\bf 2},{\bf 1})$ & $(0,-3/2,+3/2)$ \\
$\wedge^2 \tilde{V} \otimes {\rm Sym}^2 V^*$ & $4$ & $2$ & $2$ & $0$ &
$({\bf 1},{\bf 3},{\bf 1},{\bf \overline{3}})$ & $(+1,-1,0)$  \\
$\wedge^2 \tilde{V} \otimes V^* \otimes W^*$ & $4$ & $1$ & $3$ & $0$ &
$({\bf 1},{\bf 2},{\bf 2},{\bf \overline{3}})$ & $(+1,+1/2,-3/2)$ \\
$\wedge^2 \tilde{V} \otimes \wedge^2 W^*$ & $4$ & $0$ & $4$ & $0$ &
$({\bf 1},{\bf 1},{\bf 1},{\bf \overline{3}})$ & $(+1,+2,-3)$ \\
$\tilde{V} \otimes \wedge^2 V^* \otimes W^* \otimes \wedge^2 U$ &
$4$ & $1$ & $4$ & $1$ & $({\bf \overline{3}},{\bf 1},{\bf 2},{\bf 3})$ &
$(0,-1/2,+1/2)$ \\
$\tilde{V} \otimes V^* \otimes \wedge^2 W^* \otimes \wedge^2 U$ &
$4$ & $0$ & $5$ & $1$ & 
$({\bf \overline{3}},{\bf 2},{\bf 1},{\bf 3})$ &
$(0,+1,-1)$ \\
$\wedge^2 \tilde{V} \otimes \wedge^2 V^* \otimes V^* \otimes U$ &
$5$ & $2$ & $3$ & $0$ & $({\bf 3},{\bf 2},{\bf 1},{\bf \overline{3}})$ &
$(0,-1,+1)$ \\
$\wedge^2 \tilde{V} \otimes W^* \otimes \wedge^2 V^* \otimes U$ &
$5$ & $1$ & $4$ & $0$ & $({\bf 3},{\bf 1},{\bf 2},{\bf \overline{3}})$
& $(0,+1/2,-1/2)$ \\
$\tilde{V} \otimes (\wedge^2 V^*)^2 \otimes \wedge^3 U$ & 
$5$ & $2$ & $4$ & $1$ & 
$({\bf 1},{\bf 1},{\bf 1},{\bf 3})$ &
$(-1,-2,+3)$ \\
$\tilde{V} \otimes W^* \otimes \wedge^2 V^* \otimes V^* \otimes \wedge^3 U$ &
$5$ & $1$ & $5$ & $1$ & $({\bf 1},{\bf 2},{\bf 2},{\bf 3})$ &
$(-1,-1/2,+3/2)$ \\
$\tilde{V} \otimes {\rm Sym}^2 V^* \otimes \wedge^2 W^* \otimes 
\wedge^3 U$ & $5$ & $0$ & $6$ & $1$ &
$({\bf 1},{\bf 3},{\bf 1},{\bf 3})$ &
$(-1,+1,0)$ \\
$\wedge^3 \tilde{V} \otimes {\rm Sym}^2 V^* \otimes W^*$ &
$6$ & $2$ & $4$ & $0$ &
$({\bf 1},{\bf 3},{\bf 2},{\bf 1})$ &
$(0,+3/2,-3/2)$ \\
$\wedge^3 \tilde{V} \otimes V^* \otimes \wedge^2 W^*$ & $6$ & $1$ &
$5$ & $0$ & $({\bf 1},{\bf 2},{\bf 1},{\bf 1})$ &
$(0,+3,-3)$ \\
$\wedge^2 \tilde{V} \otimes (\wedge^2 V^*)^2 \otimes \wedge^2 U$ &
$6$ & $2$ & $4$ & $0$ & 
$({\bf \overline{3}},{\bf 1},{\bf 1},{\bf \overline{3}})$ &
$(-1,-1,+2)$ \\
$\wedge^2 \tilde{V} \otimes \wedge^2 V^* \otimes \wedge^2 W^* \otimes
\wedge^2 U$ & $6$ & $1$ & $6$ & $1$ &
$({\bf \overline{3}},{\bf 1},{\bf 1},{\bf \overline{3}})$ &
$(-1,+2,-1)$ \\
$\wedge^3 \tilde{V} \otimes \wedge^2 V^* \otimes V^* \otimes W^* \otimes U$ &
$7$ & $2$ & $5$ & $0$ &
$({\bf 3},{\bf 2},{\bf 2},{\bf 1})$ & $(-1,+3/2,-1/2)$ \\
$\wedge^2 \tilde{V} \otimes ( \wedge^2 V)^2 \otimes W^* \otimes \wedge^3 U$
& $7$ & $2$ & $6$ & $1$ & 
$({\bf 1},{\bf 1},{\bf 2},{\bf \overline{3}})$ &
$(-2,+1/2,+3/2)$ \\
$\wedge^2 \tilde{V} \otimes \wedge^2 V^* \otimes V^* \otimes \wedge^2 W^*
\otimes \wedge^3 U$ & $7$ & $1$ & $7$ & $1$ &
$({\bf 1},{\bf 2},{\bf 1},{\bf \overline{3}})$ &
$(-2,+2,0)$ \\
$\wedge^3 \tilde{V} \otimes \wedge^2 V^* \otimes \wedge^2 W^* \otimes U$ &
$7$ & $1$ & $6$ & $0$ &
$({\bf 3},{\bf 1},{\bf 1},{\bf 1})$ &
$(-1,+3,-2)$ \\
$\wedge^3 \tilde{V} \otimes (\wedge^2 V^*)^2 \otimes W^* \otimes 
\wedge^2 U$ & $8$ & $2$ & $6$ & $0$ &
$({\bf \overline{3}},{\bf 1},{\bf 2},{\bf 1})$ &
$(-2,+3/2,+1/2)$ \\
$\wedge^3 \tilde{V} \otimes (\wedge^2 V^*)^2 \otimes \wedge^2 W^*
\otimes \wedge^3 U$ & $9$ & $2$ & $8$ & $1$ &
$({\bf 1},{\bf 1},{\bf 1},{\bf 1})$ &
$(-3,+3,0)$
\end{tabular}
\caption{List of states shared between the two phases.}
\label{table:ex1:var3:shared}
\end{table}

\begin{table}
\begin{tabular}{c|c|c|c}
$r\ll 0$ & $r\gg 0$ & & \\
State, $\wedge^{\bullet} {\cal E}, H^{\bullet}({\mathbb P}^2)$ &
State,  $\wedge^{\bullet} {\cal F}, H^{\bullet}({\mathbb P}^1)$ &
Rep' & $U(1)^3$ \\ \hline
$\wedge^3 \tilde{V} \otimes \wedge^3 U^*$, $3, 2$ &
$\wedge^2 W^* \otimes \wedge^2 V$, $2, 1$ &
$({\bf 1},{\bf 1},{\bf 1},{\bf 1})$ &
$(+3,0,-3)$ \\
$\wedge^3 \tilde{V} \otimes V^* \otimes \wedge^2 U^*$, $4, 2$ &
$U \otimes \wedge^2 W^* \otimes V$, $3, 1$ &
$({\bf 3},{\bf 2},{\bf 1},{\bf 1})$ &
$(+2,0,-2)$ \\
${\rm Sym}^2 V^* \otimes \wedge^2 W^* \otimes 
K_{(2,1,1)} U$,
$4, 0$ &
$\wedge^3 \tilde{V} \otimes U \otimes 
K_{(3,1)} V^*$,
$4, 0$ &
$({\bf 3},{\bf 3},{\bf 1},{\bf 1})$ &
$(-1,0,+1)$ \\
$\wedge^3 \tilde{V} \otimes {\rm Sym}^2 V^* \otimes 
K_{(0,0,-1)} U$,
$5, 2$ &
$\wedge^2 U \otimes \wedge^2 W^* \otimes K_{(1,-1)} V$,
$4, 1$ &
$({\bf \overline{3}},{\bf 3},{\bf 1},{\bf 1})$ &
$(+1,0,-1)$ \\
$K_{(2,1)} V^* 
\otimes \wedge^2 W^* \otimes K_{(2,2,1)} U$,
$5, 0$ &
$\wedge^3 \tilde{V} \otimes \wedge^2 U \otimes K_{(3,2)} V^*$,
$5, 0$ &
$({\bf \overline{3}},{\bf 2},{\bf 1},{\bf 1})$ &
$(-2,0,+2)$ \\
$(\wedge^2 V^*)^2 \otimes \wedge^2 W^* \otimes (\wedge^3 U)^2$, $6, 0$ &
$\wedge^3 \tilde{V} \otimes \wedge^3 U \otimes (\wedge^2 V^*)^3$, $6, 0$ &
$({\bf 1},{\bf 1},{\bf 1},{\bf 1})$ &
$(-3,0,+3)$
\end{tabular}
\caption{Additional shared states
defined by matching representations of anomaly-free global
symmetries.}
\label{table:ex1:var3:match-only-anom-free}
\end{table}

\begin{table}
\begin{tabular}{c|c|c|c|c|c|c}
& \multicolumn{2}{c}{$r \ll 0$} & \multicolumn{2}{c}{$r \gg 0$} & & \\
State & $\wedge^{\bullet} {\cal E}$ & $H^{\bullet}({\mathbb P}^2)$ &
$\wedge^{\bullet} {\cal F}$ & $H^{\bullet}({\mathbb P}^1)$ &
Rep' &
$U(1)^3$ \\ \hline
$\wedge^2 W^* \otimes {\rm Sym}^2 U$ & $2$ & $0$ & $-$ & $-$ & 
$({\bf 6},{\bf 1},{\bf 1},{\bf 1})$ & $(+1,0,-1)$ \\
$V^* \otimes \wedge^2 W^* \otimes K_{(2,1,0)} U$ & $3$ & $0$ & $-$ & $-$ &
$({\bf 8},{\bf 2},{\bf 1},{\bf 1})$ & $(0,0,0)$ \\
$\wedge^2 V^* \otimes \wedge^2 W^* \otimes K_{(2,2,0)} U$ & $4$ & $0$ &
$-$ & $-$ & $({\bf 6},{\bf 1},{\bf 1},{\bf 1})$ &
$(-1,0,+1)$ \\
$\wedge^3 \tilde{V} \otimes \wedge^2 V^* \otimes \wedge^3 U^* \otimes
{\rm Sym}^2 U$ & $5$ & $2$ & $-$ & $-$ &
$({\bf 6},{\bf 1},{\bf 1},{\bf 1})$ &
$(+1,0,-1)$ \\
$\wedge^3 \tilde{V} \otimes K_{(2,1)} V^* \otimes K_{(1,0,-1)} U$ &
$6$ & $2$ & $-$ & $-$ & 
$({\bf 8},{\bf 2},{\bf 1},{\bf 1})$ & $(0,0,0)$ \\
$\wedge^3 \tilde{V} \otimes (\wedge^2 V^*)^2 \otimes \wedge^3 U
\otimes {\rm Sym}^2 U^*$ & $7$ & $2$ & $-$ & $-$ &
$({\bf 6},{\bf 1},{\bf 1},{\bf 1})$ &
$(-1,0,+1)$ \\
$\wedge^3 \tilde{V} \otimes {\rm Sym}^3 V^*$ & $-$ & $-$ & $3$ & $0$ &
$({\bf 1},{\bf 4},{\bf 1},{\bf 1})$ & $(0,0,0)$ \\
$\wedge^3 U \otimes \wedge^2 W^* \otimes \wedge^2 V \otimes
{\rm Sym}^3 V^*$ & $-$ & $-$ & $5$ & $1$ &
$({\bf 1},{\bf 4},{\bf 1},{\bf 1})$ & $(0,0,0)$ 
\end{tabular}
\caption{List of all states which are not shared between the two phases.}
\label{table:ex1:var3:mismatch}
\end{table}

Finally, we compare the chiral states from the first GLSM,
in tables~\ref{table:ex1:shared} and \ref{table:ex1:match-only-anom-free},
with those from the second GLSM, in tables~\ref{table:ex1:var3:shared}
and \ref{table:ex1:var3:match-only-anom-free}.
It is straightforward to check that they match -- the states which are believed
to flow to the IR in the first GLSM, are isomorphic to states in the second
GLSM which are believed to flow to the IR.
(States that mismatch between two phases of the GLSM, we do not consider,
as we do not believe they flow to the IR.)
This supports the triality proposal, in that it is a check that not only
phases of a single GLSM, but phases of multiple GLSMs, all have the
same IR limit.

In passing, we have seen how non-protected operators can pair up and
become massive along RG flow, but in principle the opposite can also happen
-- pairs of massive operators can become massless and enter the RG flow.
We do not seem to observe this in any of the examples discussed in this paper,
and we leave open the question of whether more general triality examples
exhibit that phenomenon, or whether for some reason it does not happen
in UV presentation sof triality.

\subsection{Second example}

\subsubsection{First GLSM}

Next, we shall consider a (0,2) GLSM with $k = 1$, $A = 4$, and $n = 2$, so $2k + A - n =4$. Its two phases are defined by
\begin{displaymath}
{\cal E} \: \equiv \: U \otimes S \: + \: V \otimes Q^* \: + \: 
W \otimes \det S^*
\: \longrightarrow \: {\mathbb P}^1 \: = \: {\mathbb P} \tilde{V}^*
\end{displaymath}
for $r \gg 0$, and
\begin{displaymath}
{\cal F} \: \equiv \: U \otimes S^* \: + \: \tilde{V} \otimes Q^* \: + \:
W \otimes \det S \: \longrightarrow \: {\mathbb P}^3 \: = \:
{\mathbb P} V^*
\end{displaymath}
for $r \ll 0$, where in both phases,
\begin{displaymath}
U \: = \: {\mathbb C}^4, \: \: \:
V \: = \: {\mathbb C}^4, \: \: \:
W \: = \: {\mathbb C}^2, \: \: \:
\tilde{V} \: = \: {\mathbb C}^2.
\end{displaymath}

Each phase has a set of global nonanomalous $U(1)^3$ symmetries,
which are given by
\begin{center}
\begin{tabular}{c|cccc}
& $\tilde{V}$ & $U$ & $V$ & $W$ \\ \hline
$U(1)_{(1)}$ & $0$ & $0$ & $-1$ & $-2$ \\
$U(1)_{(2)}$ & $1$ & $0$ & $0$ & $-1$ \\
$U(1)_{(3)}$ & $0$ & $1$ & $0$ & $+2$
\end{tabular}
\end{center}
In a slight variation from \cite{ggp1,ggp-exact}, 
we have assigned the same charge
to all elements of $W$, for simplicity in comparing states.

As in the previous example, the ${\mathbb P}^1$ of the $r \gg 0$ phase
is identified with ${\mathbb P} \tilde{V}^*$ rather than
${\mathbb P} \tilde{V}$.

These nonlinear sigma models are neither A/2 nor B/2-twistable; the
Fock vacuum line bundle is nontrivial.
In the $r \gg 0$ phase, we take the Fock vacuum line bundle to be
\begin{displaymath}
{\cal L} \: = \: (Q^*)^{-2} .
\end{displaymath}
In the $r \ll 0$ phase, we take the Fock vacuum line bundle to be
\begin{displaymath}
{\cal L} \: = \: K_{(-3)} S^* \otimes K_{(-1,-1,-1)} Q^* .
\end{displaymath}
This vacuum is somewhat more interesting, as it has no cohomology, hence
there are no chiral states in $H^{\bullet}(\wedge^0 {\cal F} \otimes 
{\cal L})$.  The theory still has a Fock vacuum, but it does not define
a nontrivial element of BRST cohomology.

In our previous examples, many of the states matched `on the nose,'
in representations of not only anomaly-free symmetries but also anomalous
symmetries.  In this example, there are no states that match in
representations of anomalous symmetries, only in anomaly-free
symmetries.

Matching states are listed in tables~\ref{table:ex2:match-only-anom-free-1}
and \ref{table:ex2:match-only-anom-free-2}.  (Because of the sheer
number of states, they could not all be listed in a single table,
so they were broken into two sets related by Serre duality.)
Displayed are states, the wedge power of the gauge bundle involved,
the cohomology degree in which they appear, the representation of
\begin{displaymath}
SL(U) \times SL(V) \times SL(W) \times SL(\tilde{V}) \: = \:
SU(4) \times SU(4) \times SU(2) \times SU(2)
\end{displaymath}
and charges under the nonanomalous global
\begin{displaymath}
U(1)_{(1)} \times U(1)_{(2)} \times U(1)_{(3)}
\end{displaymath}
symmetries, in the same format as for the previous example.
The $U(1)^3$ charges listed include the fractional contributions from
the Fock vacua.  For $r \gg 0$, we compute that the Fock vacuum has
charge $(+4,+2,-4)$, whereas for $r \ll 0$, we compute that the Fock vacuum
has charge $(0,-2,-4)$.

Furthermore, as before, each geometric phase has a few states not
possessed by the other phase.  These are listed in
table~\ref{table:ex2:mismatch}.

\begin{table}
\begin{tabular}{c|c|c|c}
$r\gg 0$ & $r\ll 0$ & & \\
State, $\wedge^{\bullet} {\cal E}, H^{\bullet}({\mathbb P}^1)$ &
State,  $\wedge^{\bullet} {\cal F}, H^{\bullet}({\mathbb P}^3)$ &
Rep' & $U(1)^3$ \\ \hline
${\rm Sym}^2 \tilde{V}^*$, $0, 0$ &
${\rm Sym}^2 \tilde{V} \otimes (\wedge^4 V)^{-1}$, $2, 2$ &
$({\bf 1}, {\bf 1}, {\bf 1}, {\bf 3})$ &
$(+4,0,-4)$ \\
$U \otimes (\wedge^2 \tilde{V})^{-1} \otimes \tilde{V}^*$, $1, 0$ &
$U \otimes \tilde{V} \otimes (\wedge^4 V)^{-1}$, $2, 1$ &
$({\bf 4},{\bf 1},{\bf 1},{\bf 2})$ & $(+4,-1,-3)$ \\
$V \otimes \tilde{V}^*$, $1, 0$ &
$(\wedge^2 \tilde{V}) \otimes \tilde{V} \otimes (\wedge^4 V)^{-1} \otimes V$,
$3, 2$ &
$({\bf 1},{\bf 4},{\bf 1},{\bf 2})$ &
$(+3,+1,-4)$ \\
$W \otimes K_{(1,-2)} \tilde{V}$, $1, 0$ &
$W \otimes {\rm Sym}^3 \tilde{V} \otimes (\wedge^4 V)^{-1}$, $4, 3$ &
$({\bf 1},{\bf 1},{\bf 2},{\bf 4})$ & $(+2,0,-2)$ \\
$\wedge^2 U \otimes (\wedge^2 \tilde{V})^{-2}$, $2, 0$ &
$\wedge^2 U \otimes (\wedge^4 V)^{-1}$, $2, 0$ &
$({\bf 6},{\bf 1},{\bf 1},{\bf 1})$ &
$(+4,-2,-2)$ \\
$U \otimes V \otimes (\wedge^2 \tilde{V})^{-1}$, $2, 0$ &
$U \otimes \wedge^2 \tilde{V} \otimes (\wedge^4 V)^{-1} \otimes V$, $3, 1$ &
$({\bf 4},{\bf 4},{\bf 1},{\bf 1})$ & $(+3,0,-3)$ \\
$U \otimes W \otimes {\rm Sym}^2 \tilde{V}^*$, $2, 0$ &
$U \otimes W \otimes {\rm Sym}^2 \tilde{V} \otimes (\wedge^4 V)^{-1}$, $4, 2$ &
$({\bf 4},{\bf 1},{\bf 2},{\bf 3})$ & $(+2,-1,-1)$ \\
$\wedge^2 V$, $2, 0$ &
$(\wedge^2 \tilde{V})^2 \otimes (\wedge^4 V)^{-1} \otimes \wedge^2 V$, $4, 2$ &
$({\bf 1},{\bf 6},{\bf 1},{\bf 1})$ & $(+2,+2,-4)$ \\
$V \otimes W \otimes K_{(1,-1)} \tilde{V}$, $2, 0$ &
$W \otimes K_{(3,1)} \tilde{V} \otimes K_{(0,-1,-1,-1)} V$, $5, 3$ &
$({\bf 1},{\bf 4},{\bf 2},{\bf 3})$ & $(+1,+1,-2)$ \\
$\wedge^2 U \otimes W \otimes K_{(-1,-2)} \tilde{V}$, $3, 0$ &
$\wedge^2 U \otimes \tilde{V} \otimes W \otimes (\wedge^4 V)^{-1}$, $4, 1$ &
$({\bf 6},{\bf 1},{\bf 2},{\bf 2})$ & $(+2,-2,0)$ \\
$U \otimes V \otimes W \otimes \tilde{V}^*$, $3, 0$ &
$U \otimes W \otimes K_{(2,1)} \tilde{V} \otimes K_{(0,-1,-1,-1)} V$, $5, 2$ &
$({\bf 4},{\bf 4},{\bf 2},{\bf 2})$ & $(+1,0,-1)$ \\
$U \otimes \wedge^2 W \otimes K_{(+1,-2)} \tilde{V}$, $3,0$ &
$U \otimes \wedge^2 W \otimes {\rm Sym}^3 \tilde{V} \otimes (\wedge^4 V)^{-1}$,
$6, 3$ &
$({\bf 4},{\bf 1},{\bf 1},{\bf 4})$ & $(0,-1,+1)$ \\
$\wedge^2 V \otimes W \otimes \tilde{V}$, $3, 0$ &
$W \otimes K_{(3,2)} \tilde{V} \otimes K_{(0,0,-1,-1)} V$, $6, 3$ &
$({\bf 1},{\bf 6},{\bf 2},{\bf 2})$ & $(0,+2,-2)$ \\
$V \otimes \wedge^2 W \otimes K_{(2,-1)} \tilde{V}$, $3, 0$ &
$\wedge^4 U \otimes {\rm Sym}^3 \tilde{V} \otimes V$, $7, 0$ &
$({\bf 1},{\bf 4},{\bf 1},{\bf 4})$ & $(-1,+1,0)$ \\
$\wedge^4 U \otimes K_{(-3,-3)} \tilde{V}$, $4, 1$ &
$\wedge^2 W \otimes K_{(-2,-2,-2,-2)} V$, $2, 3$ &
$({\bf 1},{\bf 1},{\bf 1},{\bf 1})$ & $(+4,-4,0)$ \\
$\wedge^3 U \otimes V \otimes K_{(-2,-2)} \tilde{V}$, $4, 1$ &
$\wedge^3 U \otimes K_{(0,-1,-1,-1)} V$, $3, 0$ &
$({\bf \overline{4}},{\bf 4},{\bf 1},{\bf 1})$ & $(+3,-2,-1)$ \\
$\wedge^3 U \otimes W \otimes K_{(-2,-2)} \tilde{V}$, $4, 0$ &
$\wedge^3 U \otimes W \otimes (\wedge^4 V)^{-1}$, $4, 0$ &
$({\bf \overline{4}},{\bf 1},{\bf 2},{\bf 1})$ & $(+2,-3,+1)$ \\
$\wedge^2 U \otimes \wedge^2 V \otimes \wedge^2 \tilde{V}^*$, $4, 1$ &
$\wedge^2 U \otimes \wedge^2 \tilde{V} \otimes \wedge^2 V^*$, $4, 1$ &
$({\bf 6},{\bf 6},{\bf 1},{\bf 1})$ & $(+2,0,-2)$ \\
$\wedge^2 U \otimes V \otimes W \otimes \wedge^2 \tilde{V}^*$, $4, 0$ &
$\wedge^2 U \otimes W \otimes \wedge^2 \tilde{V} \otimes \wedge^3 V^*$, $5, 1$ &
$({\bf 6},{\bf 4},{\bf 2},{\bf 1})$ & $(+1,-1,0)$ \\
$\wedge^2 U \otimes \wedge^2 W \otimes {\rm Sym}^2 \tilde{V}^*$, $4, 0$ &
$\wedge^2 U \otimes \wedge^2 W \otimes K_{(2,0)} \tilde{V} \otimes 
(\wedge^4 V)^{-1}$, $6, 2$ &
$({\bf 6},{\bf 1},{\bf 1},{\bf 3})$ & $(0,-2,+2)$ \\
$U \otimes \wedge^3 V$, $4, 1$ &
$U \otimes K_{(2,2)} \tilde{V} \otimes V^*$, $5, 2$ &
$({\bf 4},{\bf \overline{4}},{\bf 1},{\bf 1})$ & $(+1,+2,-3)$ \\
$U \otimes \wedge^2 V \otimes W$, $4, 0$ &
$U \otimes W \otimes K_{(2,2)} \tilde{V} \otimes \wedge^2 V^*$, $6, 2$ &
$({\bf 4},{\bf 6},{\bf 2},{\bf 1})$ & $(0,+1,-1)$ \\
$U \otimes V \otimes \wedge^2 W \otimes K_{(1,-1)} \tilde{V}$, $4, 0$ &
$U \otimes \wedge^2 W \otimes K_{(3,1)} \tilde{V} \otimes \wedge^3 V^*$, $7, 3$
&
$({\bf 4},{\bf 4},{\bf 1},{\bf 3})$ & $(-1,0,+1)$ \\
$\wedge^4 V \otimes \wedge^2 \tilde{V}$, $4, 1$ &
$(\wedge^2 \tilde{V})^3$, $6, 3$ &
$({\bf 1},{\bf 1},{\bf 1},{\bf 1})$ & $(0,+4,-4)$ \\
$\wedge^3 V \otimes W \otimes \wedge^2 \tilde{V}$, $4, 0$ &
$W \otimes K_{(3,3)} \tilde{V} \otimes V^*$, $7, 3$ &
$({\bf 1},{\bf \overline{4}},{\bf 2},{\bf 1})$ & $(-1,+3,-2)$ \\
$\wedge^2 V \otimes \wedge^2 W \otimes {\rm Sym}^2 \tilde{V}$, $4, 0$ &
$\wedge^4 U \otimes K_{(3,1)} \tilde{V} \otimes \wedge^2 V$, $8, 0$ &
$({\bf 1},{\bf 6},{\bf 1},{\bf 3})$ & $(-2,+2,0)$ \\
$\wedge^4 U \otimes V \otimes K_{(-2,-3)} \tilde{V}$, $5, 1$ &
$\wedge^2 W \otimes \tilde{V} \otimes K_{(-1,-2,-2,-2)} V$, $3, 3$ &
$({\bf 1},{\bf 4},{\bf 1},{\bf 2})$ & $(+3,-3,0)$ \\
$\wedge^3 U \otimes \wedge^2 V \otimes K_{(-1,-2)} \tilde{V}$, $5, 1$ &
$\wedge^3 U \otimes \tilde{V} \otimes \wedge^2 V^*$, $4, 0$ &
$({\bf \overline{4}},{\bf 6},{\bf 1},{\bf 2})$ & $(+2,-1,-1)$ \\
$\wedge^3 U \otimes \wedge^2 W \otimes K_{(-1,-2)} \tilde{V}$, $5, 0$ &
$\wedge^3 U \otimes \wedge^2 W \otimes \tilde{V} \otimes \wedge^4 V^*$, $6, 1$ &
$({\bf \overline{4}},{\bf 1},{\bf 1},{\bf 2})$ & $(0,-3,+3)$ \\
$\wedge^2 U \otimes \wedge^3 V \otimes \tilde{V}^*$, $5, 1$ &
$\wedge^2 U \otimes K_{(2,1)} \tilde{V} \otimes V^*$, $5, 1$ &
$({\bf 6},{\bf \overline{4}},{\bf 1},{\bf 2})$ & $(+1,+1,-2)$ \\
$\wedge^2 U \otimes V \otimes \wedge^2 W \otimes \tilde{V}^*$, $5, 0$ &
$\wedge^2 U \otimes \wedge^2 W \otimes K_{(2,1)} \tilde{V} \otimes
\wedge^3 V^*$, $7, 2$ &
$({\bf 6},{\bf 4},{\bf 1},{\bf 2})$ & $(-1,-1,+2)$ \\
$U \otimes \wedge^4 V \otimes \tilde{V}$, $5, 1$ &
$U \otimes K_{(3,2)} \tilde{V}$, $6, 2$ &
$({\bf 4},{\bf 1},{\bf 1},{\bf 2})$ & $(0,+3,-3)$ \\
$U \otimes \wedge^2 V \otimes \wedge^2 W \otimes \tilde{V}$, $5, 0$ &
$U \otimes \wedge^2 W \otimes K_{(3,2)} \tilde{V} \otimes \wedge^2 V^*$, $8, 3$ &
$({\bf 4},{\bf 6},{\bf 1},{\bf 2})$ & $(-2, +1,+1)$\\
$\wedge^3 V \otimes \wedge^2 W \otimes K_{(2,1)} \tilde{V}$, $5, 0$ &
$\wedge^4 U \otimes K_{(3,2)} \tilde{V} \otimes \wedge^3 V$, $9, 0$ &
$({\bf 1},{\bf \overline{4}},{\bf 1},{\bf 2})$ & $(-3,+3,0)$ \\
\end{tabular}
\caption{First half of shared states
defined by matching representations of anomaly-free global
symmetries.}
\label{table:ex2:match-only-anom-free-1}
\end{table}

\begin{table}
\begin{tabular}{c|c|c|c}
$r\gg 0$ & $r\ll 0$ & & \\
State, $\wedge^{\bullet} {\cal E}, H^{\bullet}({\mathbb P}^1)$ &
State,  $\wedge^{\bullet} {\cal F}, H^{\bullet}({\mathbb P}^3)$ &
Rep' & $U(1)^3$ \\ \hline
$\wedge^4 U \otimes \wedge^2 V \otimes K_{(-1,-3)} \tilde{V}$, $6, 1$ &
$\wedge^2 W \otimes {\rm S}^2 \tilde{V} \otimes K_{(2,2,1,1)} V^*$, $4, 3$
&
$({\bf 1},{\bf 6},{\bf 1},{\bf 3})$ & $(+2,-2,0)$ \\
$\wedge^4 U \otimes V \otimes W \otimes K_{(-2,-2)} \tilde{V}$, $6, 1$ &
$\wedge^4 U \otimes W \otimes \wedge^3 V^*$, $5, 0$ &
$({\bf 1},{\bf 4},{\bf 2},{\bf 1})$ & $(+1,-3,+2)$ \\
$\wedge^4 U \otimes \wedge^2 W \otimes K_{(-2,-2)} \tilde{V}$, $6, 0$ &
$\wedge^4 U \otimes \wedge^2 W \otimes \wedge^4 V^*$, $6, 0$ &
$({\bf 1},{\bf 1},{\bf 1},{\bf 1})$ & $(0,-4,+4)$\\
$\wedge^3 U \otimes \wedge^3 V \otimes {\rm Sym}^2 \tilde{V}^*$, $6, 1$ &
$\wedge^3 U \otimes {\rm Sym}^2 \tilde{V} \otimes V^*$, $5, 0$ &
$({\bf \overline{4}},{\bf \overline{4}},{\bf 1},{\bf 3})$ &
$(+1,0,-1)$ \\
$\wedge^3 U \otimes \wedge^2 V \otimes W \otimes \wedge^2 \tilde{V}^*$, $6, 1$ &
$\wedge^3 U \otimes W \otimes \wedge^2 \tilde{V} \otimes \wedge^2 V^*$, $6, 1$ &
$({\bf \overline{4}},{\bf 6},{\bf 2},{\bf 1})$ & $(0,-1,+1)$ \\
$\wedge^3 U \otimes V \otimes \wedge^2 W \otimes \wedge^2 \tilde{V}^*$, $6, 0$ &
$\wedge^3 U \otimes \wedge^2 W \otimes \wedge^2 \tilde{V} \otimes \wedge^3 V^*$,
$7, 1$ &
$({\bf \overline{4}},{\bf 4},{\bf 1},{\bf 1})$ & $(-1,-2,+3)$ \\
$\wedge^2 U \otimes \wedge^4 V \otimes K_{(1,-1)} \tilde{V}$, $6, 1$ &
$\wedge^2 U \otimes K_{(3,1)} \tilde{V}$, $6, 1$ &
$({\bf 6},{\bf 1},{\bf 1},{\bf 3})$ & $(0,+2,-2)$ \\
$\wedge^2 U \otimes \wedge^3 V \otimes W$, $6, 1$ &
$\wedge^2 U \otimes W \otimes K_{(2,2)} \tilde{V} \otimes V^*$, $7, 2$ &
$({\bf 6},{\bf \overline{4}},{\bf 2},{\bf 1})$ & $(-1,+1,0)$ \\
$\wedge^2 U \otimes \wedge^2 V \otimes \wedge^2 W$, $6, 0$ &
$\wedge^2 U \otimes \wedge^2 W \otimes (\wedge^2 \tilde{V})^2 \otimes 
\wedge^2 V^*$, $8, 2$ &
$({\bf 6},{\bf 6},{\bf 1},{\bf 1})$ & $(-2,0,+2)$ \\
$U \otimes \wedge^4 V \otimes W \otimes \wedge^2 \tilde{V}$, $6, 1$ &
$U \otimes W \otimes K_{(3,3)} \tilde{V}$, $8, 3$ &
$({\bf 4},{\bf 1},{\bf 2},{\bf 1})$ & $(-2,+3,-1)$ \\
$U \otimes \wedge^3 V \otimes \wedge^2 W \otimes \wedge^2 \tilde{V}$, $6, 0$ &
$U \otimes K_{(3,3)} \tilde{V} \otimes \wedge^2 W \otimes V^*$, $9, 3$ &
$({\bf 4},{\bf \overline{4}},{\bf 1},{\bf 1})$ & $(-3,+2,+1)$ \\
$\wedge^4 V \otimes \wedge^2 W \otimes K_{(2,2)} \tilde{V}$, $6, 0$ &
$\wedge^4 U \otimes K_{(3,3)} \tilde{V} \otimes \wedge^4 V$, $10, 0$ &
$({\bf 1},{\bf 1},{\bf 1},{\bf 1})$ & $(-4,+4,0)$ \\
$\wedge^4 U \otimes \wedge^3 V \otimes {\rm Sym}^3 \tilde{V}^*$, $7, 1$ &
$\wedge^2 W \otimes {\rm S}^3 \tilde{V} \otimes K_{(2,1,1,1)} V^*$, $5, 3$
&
$({\bf 1},{\bf \overline{4}},{\bf 1},{\bf 4})$ & $(+1,-1,0)$ \\
$\wedge^4 U \otimes \wedge^2 V \otimes W \otimes K_{(2,1)} \tilde{V}^*$, $7, 1$
&
$\wedge^4 U \otimes \tilde{V} \otimes W \otimes \wedge^2 V^*$, $6, 0$ &
$({\bf 1},{\bf 6},{\bf 2},{\bf 2})$ & $(0,-2,+2)$ \\
$\wedge^3 U \otimes \wedge^4 V \otimes K_{(1,-2)} \tilde{V}$, $7, 1$ &
$\wedge^3 U \otimes {\rm Sym}^3 \tilde{V}$, $6, 0$ &
$({\bf \overline{4}},{\bf 1},{\bf 1},{\bf 4})$ & $(0,+1,-1)$ \\
$\wedge^3 U \otimes \wedge^3 V \otimes W \otimes \tilde{V}^*$, $7, 1$ &
$\wedge^3 U \otimes W \otimes K_{(2,1)} \tilde{V} \otimes V^*$, $7, 1$ &
$({\bf \overline{4}},{\bf \overline{4}},{\bf 2},{\bf 2})$ &
$(-1,0,+1)$ \\
$\wedge^2 U \otimes \wedge^4 V \otimes W \otimes \tilde{V}$, $7, 1$ &
$\wedge^2 U \otimes W \otimes K_{(3,2)} \tilde{V}$, $8, 2$ &
$({\bf 6},{\bf 1},{\bf 2},{\bf 2})$ & $(-2,+2,0)$ \\
$\wedge^4 U \otimes \wedge^3 V \otimes W \otimes {\rm Sym}^2 \tilde{V}^*$,
$8, 1$ &
$\wedge^4 U \otimes W \otimes {\rm Sym}^2 \tilde{V} \otimes V^*$, $7, 0$ &
$({\bf 1},{\bf \overline{4}},{\bf 2},{\bf 3})$ & $(-1,-1,+2)$ \\
$\wedge^4 U \otimes \wedge^2 V \otimes \wedge^2 W \otimes \wedge^2 \tilde{V}^*$,
$8, 1$ &
$\wedge^4 U \otimes \wedge^2 W \otimes \wedge^2 \tilde{V} \otimes \wedge^2
V^*$, $8, 1$ &
$({\bf 1},{\bf 6},{\bf 1},{\bf 1})$ & $(-2,-2,+4)$ \\
$\wedge^3 U \otimes \wedge^4 V \otimes W \otimes K_{(1,-1)} \tilde{V}$, $8, 1$ &
$\wedge^3 U \otimes W \otimes K_{(3,1)} \tilde{V}$, $8, 1$ &
$({\bf \overline{4}},{\bf 1},{\bf 2},{\bf 3})$ &
$(-2,+1,+1)$ \\
$\wedge^3 U \otimes \wedge^3 V \otimes \wedge^2 W$, $8, 1$ &
$\wedge^3 U \otimes \wedge^2 W \otimes K_{(2,2)} \tilde{V} \otimes V^*$, $9, 2$
& $({\bf \overline{4}},{\bf \overline{4}},{\bf 1},{\bf 1})$ &
$(-3,0,+3)$ \\
$\wedge^2 U \otimes \wedge^4 V \otimes \wedge^2 W \otimes \wedge^2 \tilde{V}$,
$8, 1$ &
$\wedge^2 U \otimes K_{(3,3)} \tilde{V} \otimes \wedge^2 W$, $10, 3$ &
$({\bf 6},{\bf 1},{\bf 1},{\bf 1})$ & $(-4,+2,+2)$ \\
$\wedge^4 U \otimes \wedge^4 V \otimes W \otimes K_{(1,-2)} \tilde{V}$, $9, 1$ &
$\wedge^4 U \otimes {\rm Sym}^3 \tilde{V} \otimes W$, $8, 0$ &
$({\bf 1},{\bf 1},{\bf 2},{\bf 4})$ & $(-2,0,+2)$ \\
$\wedge^4 U \otimes \wedge^3 V \otimes \wedge^2 W \otimes \tilde{V}^*$, $9, 1$ &
$\wedge^4 U \otimes \wedge^2 W \otimes K_{(2,1)} \tilde{V} \otimes V^*$, $9, 1$
& $({\bf 1},{\bf \overline{4}},{\bf 1},{\bf 2})$ & $(-3,-1,+4)$ \\
$\wedge^3 U \otimes \wedge^4 V \otimes \wedge^2 W \otimes \tilde{V}$, $9, 1$ &
$\wedge^3 U \otimes \wedge^2 W \otimes K_{(3,2)} \tilde{V}$, $10, 2$ &
$({\bf \overline{4}},{\bf 1},{\bf 1},{\bf 2})$ &
$(-4,+1,+3)$ \\
$\wedge^4 U \otimes \wedge^4 V \otimes \wedge^2 W \otimes K_{(1,-1)} \tilde{V}$,
$10, 1$ &
$\wedge^4 U \otimes \wedge^2 W \otimes K_{(3,1)} \tilde{V}$, $10, 1$ &
$({\bf 1},{\bf 1},{\bf 1},{\bf 3})$ & $(-4,0,+4)$
\end{tabular}
\caption{Second half of shared states
defined by matching representations of anomaly-free global
symmetries.  We have occasionally used the symbol ${\rm S}$ as
an abbreviation for ${\rm Sym}$.}
\label{table:ex2:match-only-anom-free-2}
\end{table}

\begin{table}
\begin{tabular}{c|c|c|c|c|c|c}
& \multicolumn{2}{c}{$r \gg 0$} & \multicolumn{2}{c}{$r \ll 0$} & & \\
State & $\wedge^{\bullet} {\cal E}$ & $H^{\bullet}({\mathbb P}^2)$ &
$\wedge^{\bullet} {\cal F}$ & $H^{\bullet}({\mathbb P}^1)$ &
Rep' &
$U(1)^3$ \\ \hline
$\wedge^2 W \otimes K_{(2,-2)} \tilde{V}$ & $2$ & $0$ & $-$ & $-$ &
$({\bf 1},{\bf 1},{\bf 1},{\bf 5})$ & $(0,0,0)$ \\
$\wedge^4 U \otimes \wedge^4 V \otimes K_{(1,-3)} \tilde{V}$ & $8$ & $1$ &
$-$ & $-$ & $({\bf 1},{\bf 1},{\bf 1},{\bf 5})$ & $(0,0,0)$\\
$\wedge^4 U \otimes K_{(+1,-1,-1,-1)} V$ & $-$ & $-$ & $4$ & $0$ &
$({\bf 1},{\bf 10},{\bf 1},{\bf 1})$ & $(+2,-2,0)$ \\
$\wedge^2 W \otimes \wedge^2 \tilde{V} \otimes K_{(0,-2,-2,-2)} V$ & $-$ & $-$
& $4$ & $3$ & $({\bf 1},{\bf \overline{10}},{\bf 1},{\bf 1})$ & $(+2,-2,0)$ \\
$\wedge^4 U \otimes \tilde{V} \otimes K_{(+1,0,-1,-1}) V$ & $-$ & $-$ &
$5$ & $0$ & $({\bf 1},{\bf 20},{\bf 1},{\bf 2})$ & $(+1,-1,0)$ \\
$\wedge^2 W \otimes K_{(2,1)} \tilde{V} \otimes K_{(0,-1,-2,-2)} V$ & $-$ & $-$
& $5$ & $3$ & $({\bf 1},{\bf 20},{\bf 1},{\bf 2})$ & $(+1,-1,0)$ \\
$\wedge^4 U \otimes \wedge^2 \tilde{V} \otimes K_{(+1,+1,-1,-1)} V$ & $-$ & $-$
& $6$ & $0$ & $({\bf 1},{\bf 20},{\bf 1},{\bf 1})$ & $(0,0,0)$ \\
$\wedge^4 U \otimes {\rm Sym}^2 \tilde{V} \otimes K_{(+1,0,0,-1)} V$ & $-$ & $-$
& $6$ & $0$ & $({\bf 1},{\bf 15},{\bf 1},{\bf 3})$ & $(0,0,0)$ \\
$\wedge^2 W \otimes K_{(3,1)} \tilde{V} \otimes K_{(0,-1,-1,-2)} V$ & $-$ & $-$
& $6$ & $3$ & $({\bf 1},{\bf 15},{\bf 1},{\bf 3})$ & $(0,0,0)$ \\
$\wedge^2 W \otimes K_{(2,2)} \tilde{V} \otimes K_{(0,0,-2,-2)} V$ & $-$ & $-$
& $6$ & $3$ & $({\bf 1},{\bf 20},{\bf 1},{\bf 1})$ & $(0,0,0)$ \\
$\wedge^4 U \otimes K_{(2,1)} \tilde{V} \otimes K_{(1,1,0,-1)} V$ & $-$ & $-$
& $7$ & $0$ & $({\bf 1},{\bf 20},{\bf 1},{\bf 2})$ & $(-1,+1,0)$ \\
$\wedge^2 W \otimes K_{(3,2)} \tilde{V} \otimes K_{(0,0,-1,-2)} V$ & $-$ & $-$
& $7$ & $3$ & ${\bf 1},{\bf 20},{\bf 1},{\bf 2})$ & $(-1,+1,0)$ \\
$\wedge^4 U \otimes K_{(2,2)} \tilde{V} \otimes K_{(1,1,1,-1)} V$ & $-$ & $-$
& $8$ & $0$ & $({\bf 1},{\bf 10},{\bf 1},{\bf 1})$ & $(-2,+2,0)$ \\
$K_{(3,3)} \tilde{V} \otimes \wedge^2 W \otimes {\rm Sym}^2 V^*$ & $-$ & $-$
& $8$ & $3$ & $({\bf 1},{\bf \overline{10}},{\bf 1},{\bf 1})$
& $(-2,+2,0)$
\end{tabular}
\caption{List of all states which are not shared between the two phases.}
\label{table:ex2:mismatch}
\end{table}

As a consistency check, note that all of the states come in Serre dual
pairs, just as in the previous example.  In addition, the mismatched states 
come in pairs that cancel out of refined elliptic genus computations,
as before, which means they can plausibly become massive along the RG
flow.

In the IR, these theories are believed to flow to an SCFT in which the
nonanomalous global symmetry
\begin{displaymath}
SU(U) \times SU(V) \times SU(W) \times SU(\tilde{V})
\end{displaymath}
is enhanced to an affine symmetry
\begin{displaymath}
SU(U)_1 \times SU(V)_1 \times SU(W)_1 \times SU(\tilde{V})_3
\: = \:
SU(4)_1 \times SU(4)_1 \times SU(2)_1 \times SU(2)_3 ,
\end{displaymath}
and the states in the IR should all be associated with integrable
representations of the affine algebra above.
Using the fact that the integrable representations of $SU(2)_3$ are given
by
\begin{displaymath}
{\bf 1}, \: \: \:
{\bf 2}, \: \: \:
{\bf 3}, \: \: \:
{\bf 4} ,
\end{displaymath}
and the integrable representations of $SU(4)_1$ are given by
\begin{displaymath}
{\bf 1}, \: \: \:
{\bf 4}, \: \: \:
\wedge^2 {\bf 4} \: = \: {\bf 6}, \: \: \:
\wedge^3 {\bf 4} \: = \: {\bf \overline{4}} ,
\end{displaymath}
it is straightforward to check explicitly that all of the matching states
listed in tables~\ref{table:ex2:match-only-anom-free-1} 
and \ref{table:ex2:match-only-anom-free-2} 
are indeed associated with
integrable representations, whereas by contrast all of the non-matching
states in table~\ref{table:ex2:mismatch} 
are associated with non-integrable representations.
As before, it is our expectation that the non-matching states listed
in table~\ref{table:ex2:mismatch} do not survive to the IR.

\subsubsection{Other GLSMs}

Now, let us compute the chiral states in another GLSM related to the previous
one by triality.  We shall see, as before,
that states in integrable representations
match between phases, and also that those same states match between GLSMs.
 
Just as in section~\ref{sect:1st:other}, we can obtain the other GLSMs 
related by triality by cyclically permuting
\begin{displaymath}
U \: \longrightarrow \: V \: \longrightarrow \: \tilde{V}^* \: \longrightarrow
\:  U \: \longrightarrow \: \cdots
\end{displaymath}
(and changing the gauge group).
The three large-radius phases correspond to the bundles and spaces given by
\begin{eqnarray*}
(1): & U \otimes S \: + \: V \otimes Q^* \: + \: W \otimes \det S^* &
\longrightarrow \: G(1,\tilde{V}^*) ,\\ 
(2): & V \otimes S \: + \: \tilde{V}^* \otimes Q^* \: + \: W \otimes \det S^*
& \longrightarrow \: G(1,U) , \\
(3): & \tilde{V}^* \otimes S \: + \: U \otimes Q^* \: + \: W \otimes \det S^*
& \longrightarrow \: G(3,V) ,\\
\end{eqnarray*} 
and the three $r \ll 0$ phases are described by
\begin{eqnarray*}
(1): & U \otimes S^* \: + \: \tilde{V} \otimes Q^* \: + \: W \otimes \det S
& \longrightarrow \: G(1,V^*) ,\\
(2): & V \otimes S^* \: + \: U^* \otimes Q^* \: + \: W \otimes \det S 
& \longrightarrow \: G(1,\tilde{V}) ,\\
(3): & \tilde{V}^* \otimes S^* \: + \: V^* \otimes Q^* \: + \: W \otimes \det S
& \longrightarrow \: G(3,U^*) .
\end{eqnarray*}
(As a consistency check, the $U(1)^3$ global symmetry is nonanomalous
in each of the six phases above.)

Note that the $r \gg 0$ phase of the first GLSM and the $r \ll 0$ phase
of the second GLSM are closely related:  if we exchange
\begin{displaymath}
U \: \leftrightarrow \: V^*, \: \: \:
W \: \leftrightarrow \: W^*, \: \: \:
\tilde{V} \: \leftrightarrow \: \tilde{V}^* ,
\end{displaymath}
then we can map one theory into the other.  Similarly, the $r \ll 0$ phase
of the first GLSM and the $r \gg 0$ phase of the second.  Therefore, rather
than compute a new set of chiral states from scratch, we can re-use our
existing computations to derive the states for the second GLSM.

The $U(1)^3$ action that matches that of the previous GLSM is given by
\begin{center}
\begin{tabular}{c|cccc}
& $\tilde{V}$ & $U$ & $V$ & $W$ \\ \hline
$U(1)_{(1)}$ & $0$ & $0$ & $-1$ & $-2$ \\
$U(1)_{(2)}$ & $-1$ & $0$ & $0$ & $+1$ \\
$U(1)_{(3)}$ & $0$ & $1$ & $0$ & $+2$
\end{tabular}
\end{center}
The vacuum charges in the $r \ll 0$ and $r \gg 0$ phases are given by,
respectively, $(-4,+2,+4)$ and $(-4,-2,0)$.

Our results for the second GLSM are listed in 
tables~\ref{table:ex2:var2:match-only-anom-free-1}, 
\ref{table:ex2:var2:match-only-anom-free-2}, 
\ref{table:ex2:var2:match-only-anom-free-3}, and \ref{table:ex2:var2:mismatch}.
Tables~\ref{table:ex2:var2:match-only-anom-free-1}, 
\ref{table:ex2:var2:match-only-anom-free-2}, and 
\ref{table:ex2:var2:match-only-anom-free-3} 
list states that match between the $r \gg 0$
and $r \ll 0$ phases; table~\ref{table:ex2:var2:mismatch} lists the remainder.
It is straightforward to check that all states come in Serre dual pairs,
that all of the matching states lie in integrable representations,
and that the mismatched states do not lie in integrable representations.
In addition, as before, the mismatched states naturally come in pairs such that
they can make no net contribution to the (leading term of the) elliptic
genus, refined by any of the listed nonanomalous symmetries.

In addition, it is straightforward to check that all of the matching
states in this GLSM, are in one-to-one correspondence with matching states
in the previous GLSM related by triality -- for any matching state in
this GLSM, one can find a matching state in the previous GLSM in the
same representation of nonanomalous symmetries.

\begin{table}
\begin{tabular}{c|c|c|c}
$r\ll 0$ & $r\gg 0$ & & \\
State, $\wedge^{\bullet} {\cal E}, H^{\bullet}({\mathbb P}^1)$ &
State,  $\wedge^{\bullet} {\cal F}, H^{\bullet}({\mathbb P}^3)$ &
Rep' & $U(1)^3$ \\ \hline
${\rm Sym}^2 \tilde{V}$, $0, 0$ &
${\rm Sym}^2 \tilde{V}^* \otimes \wedge^4 U$, $2, 2$ &
$({\bf 1},{\bf 1},{\bf 1},{\bf 3})$ & $(-4,0,+4)$ \\
$V^* \otimes \wedge^2 \tilde{V} \otimes \tilde{V}$, $1, 0$ &
$V^* \otimes \tilde{V}^* \otimes \wedge^4 U$, $2, 1$ &
$({\bf 1},{\bf \overline{4}},{\bf 1},{\bf 2})$ & $(-3,-1,+4)$ \\
$U^* \otimes \tilde{V}$, $1, 0$ &
$\wedge^2 \tilde{V}^* \otimes \tilde{V}^* \otimes \wedge^4 U \otimes U^*$, 
$3, 2$ &
$({\bf \overline{4}},{\bf 1},{\bf 1},{\bf 2})$ & $(-4,+1,+3)$ \\
$W^* \otimes K_{(1,-2)} \tilde{V}^*$, $1, 0$ &
$W^* \otimes {\rm Sym}^3 \tilde{V}^* \otimes \wedge^4 U$, $4, 3$ &
$({\bf 1},{\bf 1},{\bf 2},{\bf 4})$ & $(-2,0,+2)$ \\
$\wedge^2 V^* \otimes (\wedge^2 \tilde{V})^2$, $2, 0$ &
$\wedge^2 V^* \otimes \wedge^4 U$, $2, 0$ &
$({\bf 1},{\bf 6},{\bf 1},{\bf 1})$ & $(-2,-2,+4)$ \\
$V^* \otimes U^* \otimes \wedge^2 \tilde{V}$, $2, 0$ &
$V^* \otimes \wedge^2 \tilde{V}^* \otimes \wedge^4 U \otimes U^*$, $3, 1$ &
$({\bf \overline{4}},{\bf \overline{4}},{\bf 1},{\bf 1})$ & $(-3,0,+3)$ \\
$V^* \otimes W^* \otimes {\rm Sym}^2 \tilde{V}$, $2, 0$ &
$V^* \otimes W^* \otimes {\rm Sym}^2 \tilde{V}^* \otimes \wedge^4 U$, $4, 2$ &
$({\bf 1},{\bf \overline{4}},{\bf 2},{\bf 3})$ & $(-1,-1,+2)$ \\
$\wedge^2 U^*$, $2, 0$ &
$(\wedge^2 \tilde{V}^*)^2 \otimes \wedge^4 U \otimes \wedge^2 U^*$, $4, 2$ &
$({\bf 6},{\bf 1},{\bf 1},{\bf 1})$ & $(-4,+2,+2)$ \\
$U^* \otimes W^* \otimes K_{(1,-1)} \tilde{V}^*$, $2, 0$ &
$W^* \otimes K_{(3,1)} \tilde{V}^* \otimes \wedge^3 U$, $5, 3$ &
$({\bf \overline{4}},{\bf 1},{\bf 2},{\bf 3})$ & $(-2,+1,+1)$ \\
$\wedge^2 V^* \otimes W^* \otimes K_{(2,1)} \tilde{V}$, $3, 0$ &
$\wedge^2 V^* \otimes \tilde{V}^* \otimes W^* \otimes \wedge^4 U$, $4, 1$ &
$({\bf 1},{\bf 6},{\bf 2},{\bf 2})$ & $(0,-2,+2)$ \\
$V^* \otimes U^* \otimes W^* \otimes \tilde{V}$, $3, 0$ &
$V^* \otimes W^* \otimes K_{(2,1)} \tilde{V}^* \otimes \wedge^3 U$, $5, 2$ &
$({\bf \overline{4}},{\bf \overline{4}},{\bf 2},{\bf 2})$ & $(-1,0,+1)$ \\
$V^* \otimes \wedge^2 W^* \otimes K_{(1,-2)} \tilde{V}^*$, $3, 0$ &
$V^* \otimes \wedge^2 W^* \otimes {\rm Sym}^3 \tilde{V}^* \otimes 
\wedge^4 U$, $6, 3$ &
$({\bf 1},{\bf \overline{4}},{\bf 1},{\bf 4})$ & $(+1,-1,0)$ \\
$\wedge^2 U^* \otimes W^* \otimes \tilde{V}^*$, $3, 0$ &
$W^* \otimes K_{(3,2)} \tilde{V}^* \otimes \wedge^2 U$, $6, 3$ &
$({\bf 6},{\bf 1},{\bf 2},{\bf 2})$ & $(-2,+2,0)$ \\
$U^* \otimes \wedge^2 W^* \otimes K_{(2,-1)} \tilde{V}^*$, $3, 0$ &
$\wedge^4 V^* \otimes {\rm Sym}^3 \tilde{V}^* \otimes U^*$, $7, 0$ &
$({\bf \overline{4}},{\bf 1},{\bf 1},{\bf 4})$ & $(0,+1,-1)$ \\
$\wedge^4 V^* \otimes K_{(3,3)} \tilde{V}$, $4, 1$ &
$\wedge^2 W^* \otimes K_{(2,2,2,2)} U$, $2, 3$ &
$({\bf 1},{\bf 1},{\bf 1},{\bf 1})$ & $(0,-4,+4)$ \\
$\wedge^3 V^* \otimes U^* \otimes K_{(2,2)} \tilde{V}$, $4, 1$ &
$\wedge^3 V^* \otimes \wedge^3 U$, $3, 0$ &
$({\bf \overline{4}},{\bf 4},{\bf 1},{\bf 1})$ & $(-1,-2,+3)$ \\
$\wedge^3 V^* \otimes W^* \otimes K_{(2,2)} \tilde{V}$, $4, 0$ &
$\wedge^3 V^* \otimes W^* \otimes \wedge^4 U$, $4, 0$ &
$({\bf 1},{\bf 4},{\bf 2},{\bf 1})$ & $(+1,-3,+2)$ \\
$\wedge^2 V^* \otimes \wedge^2 U^* \otimes \wedge^2 \tilde{V}$, $4, 1$ &
$\wedge^2 V^* \otimes \wedge^2 \tilde{V}^* \otimes \wedge^2 U$, $4, 1$ &
$({\bf 6},{\bf 6},{\bf 1},{\bf 1})$ & $(-2,0,+2)$ \\
$\wedge^2 V^* \otimes U^* \otimes W^* \otimes \wedge^2 \tilde{V}$, $4, 0$ &
$\wedge^2 V^* \otimes W^* \otimes \wedge^2 \tilde{V}^* \otimes \wedge^3 U$,
$5, 1$ &
$({\bf \overline{4}},{\bf 6},{\bf 2},{\bf 1})$ & $(0,-1,+1)$ \\
$\wedge^2 V^* \otimes \wedge^2 W^* \otimes {\rm Sym}^2 \tilde{V}$, $4, 0$ &
$\wedge^2 V^* \otimes \wedge^2 W^* \otimes K_{(2,0)} \tilde{V}^* \otimes
\wedge^4 U$, $6, 2$ &
$({\bf 1},{\bf 6},{\bf 1},{\bf 3})$ & $(+2,-2,0)$ \\
$V^* \otimes \wedge^3 U^*$, $4, 1$ &
$V^* \otimes K_{(2,2)} \tilde{V}^* \otimes U$, $5, 2$ &
$({\bf 4},{\bf \overline{4}},{\bf 1},{\bf 1})$ & $(-3,+2,+1)$ \\
$V^* \otimes \wedge^2 U^* \otimes W^*$, $4, 0$ &
$V^* \otimes W^* \otimes K_{(2,2)} \tilde{V}^* \otimes \wedge^2 U$, $6, 2$ &
$({\bf 6},{\bf \overline{4}},{\bf 2},{\bf 1})$ & $(-1,+1,0)$ \\
$V^* \otimes U^* \otimes \wedge^2 W^* \otimes K_{(1,-1)} \tilde{V}^*$, $4, 0$ &
$V^* \otimes \wedge^2 W^* \otimes K_{(3,1)} \tilde{V}^* \otimes \wedge^3 U$,
$7, 3$ &
$({\bf \overline{4}},{\bf \overline{4}},{\bf 1},{\bf 3})$ & $(+1,0,-1)$ \\
$\wedge^4 U^* \otimes \wedge^2 \tilde{V}^*$, $4, 1$ &
$(\wedge^2 \tilde{V}^*)^3$, $6, 3$ &
$({\bf 1},{\bf 1},{\bf 1},{\bf 1})$ & $(-4,+4,0)$ \\
$\wedge^3 U^* \otimes W^* \otimes \wedge^2 \tilde{V}^*$, $4, 0$ &
$W^* \otimes K_{(3,3)} \tilde{V}^* \otimes U$, $7, 3$ &
$({\bf 4},{\bf 1},{\bf 2},{\bf 1})$ & $(-2,+3,-1)$ \\
$\wedge^2 U^* \otimes \wedge^2 W^* \otimes {\rm Sym}^2 \tilde{V}^*$, $4, 0$ &
$\wedge^4 V^* \otimes K_{(3,1)} \tilde{V}^* \otimes \wedge^2 U^*$, $8, 0$ &
$({\bf 6},{\bf 1},{\bf 1},{\bf 3})$ & $(0,+2,-2)$ \\
$\wedge^4 V^* \otimes U^* \otimes K_{(3,2)} \tilde{V}$, $5, 1$ &
$\wedge^2 W^* \otimes \tilde{V}^* \otimes K_{(2,2,2,1)} U$, $3, 3$ &
$({\bf \overline{4}},{\bf 1},{\bf 1},{\bf 2})$ & $(0,-3,+3)$ \\
$\wedge^3 V^* \otimes \wedge^2 U^* \otimes K_{(2,1)} \tilde{V}$, $5, 1$ &
$\wedge^3 V^* \otimes \tilde{V}^* \otimes \wedge^2 U$, $4, 0$ &
$({\bf 6},{\bf 4},{\bf 1},{\bf 2})$ & $(-1,-1,+2)$ \\
$\wedge^3 V^* \otimes \wedge^2 W^* \otimes K_{(2,1)} \tilde{V}$, $5, 0$ &
$\wedge^3 V^* \otimes \wedge^2 W^* \otimes \tilde{V}^* \otimes 
\wedge^4 U$, $6, 1$ &
$({\bf 1},{\bf 4},{\bf 1},{\bf 2})$ & $(+3,-3,0)$ \\
$\wedge^2 V^* \otimes \wedge^3 U^* \otimes \tilde{V}$, $5, 1$ &
$\wedge^2 V^* \otimes K_{(2,1)} \tilde{V}^* \otimes U$, $5, 1$ &
$({\bf 4},{\bf 6},{\bf 1},{\bf 2})$ & $(-2,+1,+1)$ \\
$\wedge^2 V^* \otimes U^* \otimes \wedge^2 W^* \otimes \tilde{V}$, $5, 0$ &
$\wedge^2 V^* \otimes \wedge^2 W^* \otimes K_{(2,1)} \tilde{V}^* \otimes 
\wedge^3 U$, $7, 2$ &
$({\bf \overline{4}},{\bf 6},{\bf 1},{\bf 2})$ & $(+2,-1,-1)$ \\
$V^* \otimes \wedge^4 U^* \otimes \tilde{V}^*$, $5, 1$ & 
$V^* \otimes K_{(3,2)} \tilde{V}^*$, $6, 2$ &
$({\bf 1},{\bf \overline{4}},{\bf 1},{\bf 2})$ & $(-3,+3,0)$ \\
$V^* \otimes \wedge^2 U^* \otimes \wedge^2 W^* \otimes \tilde{V}^*$, $5, 0$ &
$V^* \otimes \wedge^2 W^* \otimes K_{(3,2)} \tilde{V}^* \otimes \wedge^2 U$,
$8, 3$ &
$({\bf 6},{\bf \overline{4}},{\bf 1},{\bf 2})$ & $(+1,+1,-2)$ \\
$\wedge^3 U^* \otimes \wedge^2 W^* \otimes K_{(2,1)} \tilde{V}^*$, $5, 0$ &
$\wedge^4 V^* \otimes K_{(3,2)} \tilde{V}^* \otimes \wedge^3 U^*$, $9, 0$ &
$({\bf 4},{\bf 1},{\bf 1},{\bf 2})$ & $(0,+3,-3)$
\end{tabular}
\caption{First half of shared states
defined by matching representations of anomaly-free global
symmetries.}
\label{table:ex2:var2:match-only-anom-free-1}
\end{table}

\begin{table}
\begin{tabular}{c|c|c|c}
$r\ll 0$ & $r\gg 0$ & & \\
State, $\wedge^{\bullet} {\cal E}, H^{\bullet}({\mathbb P}^1)$ &
State,  $\wedge^{\bullet} {\cal F}, H^{\bullet}({\mathbb P}^3)$ &
Rep' & $U(1)^3$ \\ \hline
$\wedge^4 V^* \otimes \wedge^2 U^* \otimes K_{(3,1)} \tilde{V}$, $6, 1$ &
$\wedge^2 W^* \otimes S^2 \tilde{V}^* \otimes K_{(2,2,1,1)} U$, $4, 3$ &
$({\bf 6},{\bf 1},{\bf 1},{\bf 3})$ & $(0,-2,+2)$ \\
$\wedge^4 V^* \otimes U^* \otimes W^* \otimes K_{(2,2)} \tilde{V}$, $6, 1$ &
$\wedge^4 V^* \otimes W^* \otimes \wedge^3 U$, $5, 0$ &
$({\bf \overline{4}},{\bf 1},{\bf 2},{\bf 1})$ & $(+2,-3,+1)$ \\
$\wedge^4 V^* \otimes \wedge^2 W^* \otimes K_{(2,2)} \tilde{V}$, $6, 0$ &
$\wedge^4 V^* \otimes \wedge^2 W^* \otimes \wedge^4 U$, $6, 0$ &
$({\bf 1},{\bf 1},{\bf 1},{\bf 1})$ & $(+4,-4,0)$ \\
$\wedge^3 V^* \otimes \wedge^3 U^* \otimes {\rm Sym}^2 \tilde{V}$, $6, 1$ &
$\wedge^3 V^* \otimes {\rm Sym}^2 \tilde{V}^* \otimes U$, $5, 0$ &
$({\bf 4},{\bf 4},{\bf 1},{\bf 3})$ & $(-1,0,+1)$ \\
$\wedge^3 V^* \otimes \wedge^2 U^* \otimes W^* \otimes \wedge^2 \tilde{V}$, 
$6, 1$ &
$\wedge^3 V^* \otimes W^* \otimes \wedge^2 \tilde{V}^* \otimes \wedge^2 U$,
$6, 1$ &
$({\bf 6},{\bf 4},{\bf 2},{\bf 1})$ & $(+1,-1,0)$ \\
$\wedge^3 V^* \otimes U^* \otimes \wedge^2 W^* \otimes \wedge^2 \tilde{V}$,
$6, 0$ &
$\wedge^3 V^* \otimes \wedge^2 W^* \otimes \wedge^2 \tilde{V}^* \otimes
\wedge^3 U$, $7, 1$ &
$({\bf \overline{4}},{\bf 4},{\bf 1},{\bf 1})$ & $(+3,-2,-1)$ \\
$\wedge^2 V^* \otimes \wedge^4 U^* \otimes K_{(1,-1)} \tilde{V}^*$, $6, 1$ &
$\wedge^2 V^* \otimes K_{(3,1)} \tilde{V}^*$, $6, 1$ &
$({\bf 1},{\bf 6},{\bf 1},{\bf 3})$ & $(-2,+2,0)$ \\
$\wedge^2 V^* \otimes \wedge^3 U^* \otimes W^*$, $6, 1$ &
$\wedge^2 V^* \otimes W^* \otimes K_{(2,2)} \tilde{V}^* \otimes U$, $7, 2$ &
$({\bf 4},{\bf 6},{\bf 2},{\bf 1})$ & $(0,+1,-1)$ \\
$\wedge^2 V^* \otimes \wedge^2 U^* \otimes \wedge^2 W^*$, $6, 0$ &
$(\wedge^2 V \otimes \wedge^2 W \otimes (\wedge^2 \tilde{V})^2)^* \otimes
\wedge^2 U$, $8, 2$ &
$({\bf 6},{\bf 6},{\bf 1},{\bf 1})$ & $(+2,0,-2)$ \\
$V^* \otimes \wedge^4 U^* \otimes W^* \otimes \wedge^2 \tilde{V}^*$, $6, 1$ &
$V^* \otimes W^* \otimes K_{(3,3)} \tilde{V}^*$, $8, 3$ &
$({\bf 1},{\bf \overline{4}},{\bf 2},{\bf 1})$ & $(-1,+3,-2)$ \\
$V^* \otimes \wedge^3 U^* \otimes \wedge^2 W^* \otimes \wedge^2 \tilde{V}^*$,
$6, 0$ &
$V^* \otimes K_{(3,3)} \tilde{V}^* \otimes \wedge^2 W^* \otimes U$, $9, 3$ &
$({\bf 4},{\bf \overline{4}},{\bf 1},{\bf 1})$ & $(+1,+2,-3)$ \\
$\wedge^4 U^* \otimes \wedge^2 W^* \otimes K_{(2,2)} \tilde{V}^*$, $6, 0$ &
$\wedge^4 V^* \otimes K_{(3,3)} \tilde{V}^* \otimes \wedge^4 U^*$, $10, 0$ &
$({\bf 1},{\bf 1},{\bf 1},{\bf 1})$ & $(0,+4,-4)$ \\
\end{tabular}
\caption{Second portion of shared states
defined by matching representations of anomaly-free global
symmetries.  We have occasionally used the symbol ${\rm S}$ as
an abbreviation for ${\rm Sym}$.}
\label{table:ex2:var2:match-only-anom-free-2}
\end{table}

\begin{table}
\begin{tabular}{c|c|c|c}
$r\ll 0$ & $r\gg 0$ & & \\
State, $\wedge^{\bullet} {\cal E}, H^{\bullet}({\mathbb P}^1)$ &
State,  $\wedge^{\bullet} {\cal F}, H^{\bullet}({\mathbb P}^3)$ &
Rep' & $U(1)^3$ \\ \hline
$\wedge^4 V^* \otimes \wedge^3 U^* \otimes {\rm S}^3 \tilde{V}$, $7, 1$ &
$\wedge^2 W^* \otimes {\rm S}^3 \tilde{V}^* \otimes K_{(2,1,1,1)} U$,
$5, 3$ &
$({\bf 4},{\bf 1},{\bf 1},{\bf 4})$ & $(0,-1,+1)$ \\
$\wedge^4 V^* \otimes \wedge^2 U^* \otimes W^* \otimes K_{(2,1)} \tilde{V}$,
$7, 1$ &
$\wedge^4 V^* \otimes \tilde{V}^* \otimes W^* \otimes \wedge^2 U$, $6, 0$ &
$({\bf 6},{\bf 1},{\bf 2},{\bf 2})$ & $(+2,-2,0)$ \\
$\wedge^3 V^* \otimes \wedge^4 U^* \otimes K_{(2,-1)} \tilde{V}$, $7, 1$ &
$\wedge^3 V^* \otimes {\rm Sym}^3 \tilde{V}^*$, $6, 0$ &
$({\bf 1},{\bf 4},{\bf 1},{\bf 4})$ & $(-1,+1,0)$ \\
$\wedge^3 V^* \otimes \wedge^3 U^* \otimes W^* \otimes \tilde{V}$, $7, 1$ &
$\wedge^3 V^* \otimes W^* \otimes K_{(2,1)} \tilde{V}^* \otimes U$, $7, 1$ &
$({\bf 4},{\bf 4},{\bf 2},{\bf 2})$ & $(+1,0,-1)$ \\
$\wedge^2 V^* \otimes \wedge^4 U^* \otimes W^* \otimes \tilde{V}^*$, $7, 1$ &
$\wedge^2 V^* \otimes W^* \otimes K_{(3,2)} \tilde{V}^*$, $8, 2$ &
$({\bf 1},{\bf 6},{\bf 2},{\bf 2})$ & $(0,+2,-2)$ \\
$\wedge^4 V^* \otimes \wedge^3 U^* \otimes W^* \otimes {\rm Sym}^2 \tilde{V}$,
$8, 1$ &
$\wedge^4 V^* \otimes W^* \otimes {\rm Sym}^2 \tilde{V}^* \otimes U$, $7, 0$ &
$({\bf 4},{\bf 1},{\bf 2},{\bf 3})$ & $(+2,-1,-1)$ \\
$\wedge^4 V^* \otimes \wedge^2 U^* \otimes \wedge^2 W^* \otimes \wedge^2
\tilde{V}$, $8, 1$ &
$\wedge^4 V^* \otimes \wedge^2 W^* \otimes \wedge^2 \tilde{V}^* \otimes
\wedge^2 U$, $8, 1$ &
$({\bf 6},{\bf 1},{\bf 1},{\bf 1})$ & $(+4,-2,-2)$ \\
$(\wedge^3 V \otimes \wedge^4 U \otimes W \otimes K_{(1,-1)} \tilde{V})^*$,
$8, 1$ &
$\wedge^3 V^* \otimes W^* \otimes K_{(3,1)} \tilde{V}^*$, $8, 1$ &
$({\bf 1},{\bf 4},{\bf 2},{\bf 3})$ & $(+1,+1,-2)$ \\
$\wedge^3 V^* \otimes \wedge^3 U^* \otimes \wedge^2 W^*$, $8, 1$ &
$(\wedge^3 V \otimes \wedge^2 W \otimes K_{(2,2)} \tilde{V})^* \otimes U$,
$9, 2$ &
$({\bf 4},{\bf 4},{\bf 1},{\bf 1})$ & $(+3,0,-3)$ \\
$\wedge^2 V^* \otimes \wedge^4 U^* \otimes \wedge^2 W^* \otimes \wedge^2
\tilde{V}^*$, $8, 1$ &
$\wedge^2 V^* \otimes K_{(3,3)} \tilde{V}^* \otimes \wedge^2 W^*$, $10, 3$ &
$({\bf 1},{\bf 6},{\bf 1},{\bf 1})$ & $(+2,+2,-4)$ \\
$\wedge^4 V^* \otimes \wedge^4 U^* \otimes W^* \otimes K_{(2,-1)} \tilde{V}$,
$9, 1$ &
$\wedge^4 V^* \otimes {\rm Sym}^3 \tilde{V}^* \otimes W^*$, $8, 0$ &
$({\bf 1},{\bf 1},{\bf 2},{\bf 4})$ & $(+2,0,-2)$ \\
$\wedge^4 V^* \otimes \wedge^3 U^* \otimes \wedge^2 W^* \otimes \tilde{V}$,
$9, 1$ &
$(\wedge^4 V \otimes \wedge^2 W \otimes K_{(2,1)} \tilde{V})^* \otimes U$,
$9, 1$ &
$({\bf 4},{\bf 1},{\bf 1},{\bf 2})$ & $(+4,-1,-3)$ \\
$\wedge^3 V^* \otimes \wedge^4 U^* \otimes \wedge^2 W^* \otimes \tilde{V}^*$,
$9, 1$ &
$\wedge^3 V^* \otimes \wedge^2 W^* \otimes K_{(3,2)} \tilde{V}^*$, 
$10, 2$ &
$({\bf 1},{\bf 2},{\bf 1},{\bf 2})$ & $(+3,+1,-4)$ \\
$(\wedge^4 V \otimes \wedge^4 U \otimes \wedge^2 W \otimes K_{(1,-1)} 
\tilde{V})^*$, $10, 1$ &
$\wedge^4 V^* \otimes \wedge^2 W^* \otimes K_{(3,1)} \tilde{V}^*$, $10, 1$ &
$({\bf 1},{\bf 1},{\bf 1},{\bf 3})$ & $(+4,0,-4)$
\end{tabular}
\caption{Third portion of shared states
defined by matching representations of anomaly-free global
symmetries.  We have occasionally used the symbol ${\rm S}$ as
an abbreviation for ${\rm Sym}$.}
\label{table:ex2:var2:match-only-anom-free-3}
\end{table}

\begin{table}
\begin{tabular}{c|c|c|c|c|c|c}
& \multicolumn{2}{c}{$r \ll 0$} & \multicolumn{2}{c}{$r \gg 0$} & & \\
State & $\wedge^{\bullet} {\cal E}$ & $H^{\bullet}({\mathbb P}^2)$ &
$\wedge^{\bullet} {\cal F}$ & $H^{\bullet}({\mathbb P}^1)$ &
Rep' &
$U(1)^3$ \\ \hline
$\wedge^2 W^* \otimes K_{(2,-2)} \tilde{V}^*$ & $2$ & $0$ & $-$ & $-$ &
$({\bf 1},{\bf 1},{\bf 1},{\bf 5})$ & $(0,0,0)$ \\
$\wedge^4 V^* \otimes \wedge^4 U^* \otimes K_{(3,-1)} \tilde{V}$ & $8$ & $1$ &
$-$ & $-$ & $({\bf 1},{\bf 1},{\bf 1},{\bf 5})$ & $(0,0,0)$  \\
$\wedge^4 V^* \otimes K_{(1,1,1,-1)} U$ & $-$ & $-$ & $4$ & $0$ &
$({\bf 10},{\bf 1},{\bf 1},{\bf 1})$ & $(0,-2,+2)$ \\
$\wedge^2 W^* \otimes \wedge^2 \tilde{V}^* \otimes K_{(2,2,2,0)} U$ & $-$ & $-$
& $4$ & $3$ & $({\bf 10},{\bf 1},{\bf 1},{\bf 1})$ & $(0,-2,+2)$ \\
$\wedge^4 V^* \otimes \tilde{V}^* \otimes K_{(1,1,0,-1)} U$ & $-$ & $-$ &
$5$ & $0$ & $({\bf 20},{\bf 1},{\bf 1},{\bf 2})$ & $(0,-1,+1)$ \\
$\wedge^2 W^* \otimes K_{(2,1)} \tilde{V}^* \otimes K_{(2,2,1,0)} U$ &
$-$ & $-$ & $5$ & $3$ & $({\bf 20},{\bf 1},{\bf 1},{\bf 2})$ & $(0,-1,+1)$ \\
$\wedge^4 V^* \otimes \wedge^2 \tilde{V}^* \otimes K_{(1,1,-1,-1)} U$ &
$-$ & $-$ & $6$ & $0$ & $({\bf 20},{\bf 1},{\bf 1},{\bf 1})$ & $(0,0,0)$ \\
$\wedge^4 V^* \otimes {\rm Sym}^2 \tilde{V}^* \otimes K_{(1,0,0,-1)} U$ &
$-$ & $-$ & $6$ & $0$ & $({\bf 15},{\bf 1},{\bf 1},{\bf 3})$ & $(0,0,0)$ \\
$\wedge^2 W^* \otimes K_{(3,1)} \tilde{V}^* \otimes K_{(2,1,1,0)} U$ &
$-$ & $-$ & $6$ & $3$ & $({\bf 15},{\bf 1},{\bf 1},{\bf 3})$ & $(0,0,0)$ \\
$\wedge^2 W^* \otimes K_{(2,2)} \tilde{V}^* \otimes K_{(2,2,0,0)} U$ &
$-$ & $-$ & $6$ & $3$ & $({\bf 20},{\bf 1},{\bf 1},{\bf 1})$ & $(0,0,0)$ \\
$\wedge^4 V^* \otimes K_{(2,1)} \tilde{V}^* \otimes K_{(1,1,0,-1)} U^*$ &
$-$ & $-$ & $7$ & $0$ & $({\bf 20},{\bf 1},{\bf 1},{\bf 2})$ &
$(0,+1,-1)$ \\
$\wedge^2 W^* \otimes K_{(3,2)} \tilde{V}^* \otimes K_{(2,1,0,0)} U$ & 
$-$ & $-$ & $7$ & $3$ & $({\bf 20},{\bf 1},{\bf 1},{\bf 2})$ & $(0,+1,-1)$ \\
$\wedge^4 V^* \otimes K_{(2,2)} \tilde{V}^* \otimes K_{(1,-1,-1,-1)} U$ &
$-$ & $-$ & $8$ & $0$ & $({\bf 10},{\bf 1},{\bf 1},{\bf 1})$ & $(0,+2,-2)$ \\
$K_{(3,3)} \tilde{V}^* \otimes \wedge^2 W^* \otimes {\rm Sym}^2 U$ & $-$ & $-$
& $8$ & $3$ & $({\bf 10},{\bf 1},{\bf 1},{\bf 1})$ &
$(0,+2,-2)$
\end{tabular}
\caption{List of all states which are not shared between the two phases.}
\label{table:ex2:var2:mismatch}
\end{table}

\subsection{Third example:  $T_{222}$}

\subsubsection{First GLSM}

This example is the case in which $k=1$, $A=2$, and $n=2$, so
$2k+A-n=2$.  There is only one geometry appearing in this example,
namely the space $G(1,2) = {\mathbb P}^1$, with bundle
\begin{displaymath}
{\cal E} \: = \: S^2 \oplus (Q^*)^2 \oplus (\det S^*)^2 .
\end{displaymath}
In the notations above, we will consider the GLSM which for
$r \gg 0$ describes
\begin{displaymath}
{\cal E} \: = \: U \otimes S \: + \: V \otimes Q^* \: + \: W \otimes 
\det S^* \: \longrightarrow \: G(1,\tilde{V}^*),
\end{displaymath}
for
\begin{displaymath}
U \: = \: {\mathbb C}^2 \: = \: V \: = \: W \: = \: \tilde{V},
\end{displaymath}
and which for $r \ll 0$ is described by
\begin{displaymath}
{\cal F} \: = \: U \otimes S^* \: + \: \tilde{V} \otimes Q^* \: + \:
W \otimes \det S \: \longrightarrow \: G(1,V^*) .
\end{displaymath}

Both of the phases above are B/2-twistable without further dualization; 
the Fock vacuum line bundle
is trivial.  As noted earlier, we need to make a choice of presentation
as powers of $S^*$ and $Q^*$ -- although the choice does not enter
the final nonanomalous representations, we must make a choice in order
to initially compute the chiral states.  We choose the canonical
trivial presentation, as $K_{(0)}S^* \otimes K_{(0)} Q^*$ in both
phases.

We will take the nonanomalous (chiral) global $U(1)^3$ to be defined by
\begin{center}
\begin{tabular}{c|cccc}
& $\tilde{V}$ & $U$ & $V$ & $W$ \\ \hline
$U(1)_{(1)}$ & $0$ & $0$ & $-1$ & $-1$ \\
$U(1)_{(2)}$ & $1$ & $0$ & $0$ & $-1$ \\
$U(1)_{(3)}$ & $0$ & $1$ & $0$ & $1$ 
\end{tabular}
\end{center}
For both the $r \gg 0$ and $r \ll 0$ phases, we compute that the Fock
vacuum has charge $(+2,0,-2)$.

Table~\ref{table:t222:match-only-anom-free} lists all the states which
match between the two phases of this GLSM.  A few states do not match;
these are listed in table~\ref{table:t222:mismatch}.  As a consistency
check, note that both tables are invariant under Serre duality as expected.

It was predicted in \cite{ggp-exact}[equ'n (3.1)] 
that the $SU(2)^4$ global flavor symmetries
of this model should be promoted to an
\begin{displaymath}
SU(2)_1 \times SU(2)_1 \times SU(2)_1 \times SU(2)_1
\end{displaymath}
affine symmetry in the IR SCFT, and indeed, note that all of the
matching representations in table~\ref{table:t222:match-only-anom-free}
are integrable, whereas all of the mismatched representations in
table~\ref{table:t222:mismatch} are non-integrable, suggesting that the
mismatched representations do not survive to the IR.
In addition, it is also easy to check that the mismatched states in
table~\ref{table:t222:mismatch} cancel out of the leading term in
elliptic genera, refined by any listed nonanomalous symmetry, which is
consistent with the prediction that they do not survive to the IR.

\begin{center}
\begin{table}
\begin{tabular}{c|c|c|c}
$r\gg 0$ & $r\ll 0$ & & \\
State, $\wedge^{\bullet} {\cal E}, H^{\bullet}({\mathbb P}^1)$ &
State,  $\wedge^{\bullet} {\cal F}, H^{\bullet}({\mathbb P}^1)$ &
Rep' & $U(1)^3$ \\ \hline
$1$, $0$, $0$ & $1$, $0$, $0$ & 
$({\bf 1},{\bf 1},{\bf 1},{\bf 1})$ & $(+2,0,-2)$ \\
$W \otimes \tilde{V}$, $1$, $0$ &
$W \otimes \tilde{V}$, $2$, $1$ &
$({\bf 1},{\bf 1},{\bf 2},{\bf 2})$ & $(+1,0,-1)$ \\
$\wedge^2 U \otimes K_{(-1,-1)} \tilde{V}$, $2$, $1$ &
$\wedge^2 W \otimes K_{(-1,-1)} V$, $2$, $1$ & 
$({\bf 1},{\bf 1},{\bf 1},{\bf 1})$ & $(+2,-2,0)$ \\
$U \otimes V$, $2$, $1$ &
$U \otimes V$, $1$, $0$ &
$({\bf 2},{\bf 2},{\bf 1},{\bf 1})$ & $(+1,0,-1)$ \\
$U \otimes W$, $2$, $0$ & 
$U \otimes W$, $2$, $0$ &
$({\bf 2},{\bf 1},{\bf 2},{\bf 1})$ & $(+1,-1,0)$ \\
$\wedge^2 V \otimes \wedge^2 \tilde{V}$, $2$, $1$ &
$\wedge^2 V \otimes \wedge^2 \tilde{V}$, $2$, $1$ &
$({\bf 1},{\bf 1},{\bf 1},{\bf 1})$ & $(0,+2,-2)$ \\
$V \otimes W \otimes \wedge^2 \tilde{V}$, $2$, $0$ &
$V \otimes W \otimes \wedge^2 \tilde{V}$, $3$, $1$ &
$({\bf 1},{\bf 2},{\bf 2},{\bf 1})$ & $(0,+1,-1)$ \\
$\wedge^2 U \otimes V \otimes \tilde{V}^*$, $3$, $1$ &
$\tilde{V} \otimes \wedge^2 W \otimes V^*$, $3$, $1$ &
$({\bf 1},{\bf 2},{\bf 1},{\bf 2})$ & $(+1,-1,0)$ \\
$U \otimes \wedge^2 V \otimes \tilde{V}$, $3$, $1$ &
$U \otimes \tilde{V} \otimes \wedge^2 V$, $2$, $0$ &
$({\bf 2},{\bf 1},{\bf 1},{\bf 2})$ & $(0,+1,-1)$ \\
$U \otimes \wedge^2 W \otimes \tilde{V}$, $3$, $0$ &
$U \otimes \wedge^2 W \otimes \tilde{V}$, $4$, $1$ &
$({\bf 2},{\bf 1},{\bf 1},{\bf 2})$ & $(0,-1,+1)$ \\
$V \otimes \wedge^2 W \otimes K_{(2,1)} \tilde{V}$, $3$, $0$ &
$\wedge^2 U \otimes \tilde{V} \otimes K_{(2,1)} V$, $3$, $0$ &
$({\bf 1},{\bf 2},{\bf 1},{\bf 2})$ & $(-1,+1,0)$ \\
$\wedge^2 U \otimes V \otimes W$, $4$, $1$ &
$\wedge^2 U \otimes V \otimes W$, $3$, $0$ &
$({\bf 1},{\bf 2},{\bf 2},{\bf 1})$ & $(0,-1,+1)$ \\
$\wedge^2 U \otimes \wedge^2 W$, $4$, $0$ & 
$\wedge^2 U \otimes \wedge^2 W$, $4$, $0$ &
$({\bf 1},{\bf 1},{\bf 1},{\bf 1})$ & $(0,-2,+2)$ \\
$U \otimes \wedge^2 V \otimes W \otimes \wedge^2 \tilde{V}$, $4$, $1$ &
$U \otimes \wedge^2 V \otimes W \otimes \wedge^2 \tilde{V}$, $4$, $1$ &
$({\bf 2},{\bf 1},{\bf 2},{\bf 1})$ & $(-1,+1,0)$ \\
$U \otimes V \otimes \wedge^2 W \otimes \wedge^2 \tilde{V}$, $4$, $0$ &
$U \otimes V \otimes \wedge^2 W \otimes \wedge^2 \tilde{V}$, $5$, $1$ &
$({\bf 2},{\bf 2},{\bf 1},{\bf 1})$ & $(-1,0,+1)$ \\
$\wedge^2 V \otimes \wedge^2 W \otimes K_{(2,2)} \tilde{V}$, $4$, $0$ &
$\wedge^2 U \otimes \wedge^2 \tilde{V} \otimes K_{(2,2)} V$, $4$, $0$ &
$({\bf 1},{\bf 1},{\bf 1},{\bf 1})$ & $(-2,+2,0)$ \\
$\wedge^2 U \otimes \wedge^2 V \otimes W \otimes \tilde{V}$, $5$, $1$ &
$\wedge^2 U \otimes \tilde{V} \otimes W \otimes \wedge^2 V$, $4$, $0$ &
$({\bf 1},{\bf 1},{\bf 2},{\bf 2})$ & $(-1,0,+1)$ \\
$\wedge^2 U \otimes \wedge^2 V \otimes \wedge^2 W \otimes \wedge^2 \tilde{V}$,
$6$, $1$ &
$\wedge^2 U \otimes \wedge^2 V \otimes \wedge^2 W \otimes \wedge^2 \tilde{V}$,
$6$, $1$ &
$({\bf 1},{\bf 1},{\bf 1},{\bf 1})$ & $(-2,0,+2)$
\end{tabular}
\caption{Shared states defined by matching representations of anomaly-free
global symmetries.}
\label{table:t222:match-only-anom-free}
\end{table}
\end{center}

\begin{center}
\begin{table}
\begin{tabular}{c|c|c|c|c|c|c}
& \multicolumn{2}{c}{$r \gg 0$} & \multicolumn{2}{c}{$r \ll 0$} & & \\
State & $\wedge^{\bullet} {\cal E}$ & $H^{\bullet}({\mathbb P}^1)$ &
$\wedge^{\bullet} {\cal F}$ & $H^{\bullet}({\mathbb P}^1)$ &
Rep' &
$U(1)^3$ \\ \hline
$\wedge^2 W \otimes K_{(2,0)} \tilde{V}$ & $2$ & $0$ & $-$ & $-$ &
$({\bf 1},{\bf 1},{\bf 1},{\bf 3})$ & $(0,0,0)$ \\
$\wedge^2 U \otimes \wedge^2 V \otimes K_{(1,-1)} \tilde{V}$ & $4$ & $1$ &
$-$ & $-$ &
$({\bf 1},{\bf 1},{\bf 1},{\bf 3})$ & $(0,0,0)$ \\
$\wedge^2 U \otimes K_{(2,0)} V$ & $-$ & $-$ & $2$ & $0$ &
$({\bf 1},{\bf 3},{\bf 1},{\bf 1})$ & $(0,0,0)$ \\
$\wedge^2 \tilde{V} \otimes \wedge^2 W \otimes K_{(1,-1)} V$ & $-$ & $-$ &
$4$ & $1$ & $({\bf 1},{\bf 3},{\bf 1},{\bf 1})$ & $(0,0,0)$
\end{tabular}
\caption{List of all states which are not shared between the two phases.}
\label{table:t222:mismatch}
\end{table}
\end{center}

Another prediction of \cite{ggp-exact} is that in this particular
model, the $SU(2)_1^4$ affine symmetry of the IR SCFT should be enhanced
to an $E_6$ symmetry at level 1.  Briefly, the $SU(W)$ of the two fermi
fields can combine with the $SU(U)$ to form $SU(4)$, and by triality,
any one of the other $SU(2)$'s at a time.  The resulting structure can
be naturally interpreted as a subgroup of $E_6$.

We can see some evidence for this in the
states above.  In particular, the total number of matching states is
54, twice the dimension of the ${\bf 27}$ representation, suggesting that
the states above flow to copies of either the ${\bf 27}$ or
${\bf \overline{27}}$ representations.  Indeed, the only integrable
representations of $E_6$ at level 1 are ${\bf 1}$, ${\bf 27}$,
and ${\bf \overline{27}}$.

Now that we have verified the dimensions are consistent, let us see if the
precise representations appearing above match those in the decomposition
of a ${\bf 27}$ or ${\bf \overline{27}}$ under an $su(2)^4$ 
subalgebra\footnote{
We would like to thank A.~Knutson for a discussion of the corresponding
group theory.
}.  We can think of the $su(2)^4$ subalgebra of $E_6$ as obtained
from a maximal $( SU(2) \times SU(6) )/{\mathbb Z}_2$ subgroup,
and the $SU(6)$ naturally contains an $SU(2)^3$ from omitting two of the
nodes in its Dynkin diagram.  The $SU(W)$, which can recombine with any
one other $SU(2)$, can be understood as the middle node in the
$E_6$ Dynkin diagram, or equivalently the middle node in the
$SU(6)$ Dynkin diagram.  Under the $( SU(2) \times SU(6) )/{\mathbb Z}_2$
subgroup, the ${\bf 27}$ of $E_6$ decomposes as \cite{slansky}[table 15]
\begin{displaymath}
{\bf 27} \: = \: ({\bf 2},{\bf \overline{6}}) \: + \: ({\bf 1},{\bf 15}) .
\end{displaymath}
Furthermore, under the $SU(2)^3$ subgroup of $SU(6)$, 
\begin{displaymath}
{\bf \overline{6}} \: = \: ({\bf 2},{\bf 1},{\bf 1}) \: + \:
({\bf 1},{\bf 2},{\bf 1}) \: + \:
({\bf 1},{\bf 1},{\bf 2}) .
\end{displaymath}
(The ${\bf 6}$ has the same decomposition -- the $su(2)^4$ subalgebra
cannot distinguish the ${\bf 6}$ from ${\bf \overline{6}}$.)
Similarly, the ${\bf 15} = \wedge^2 {\bf 6}$ should decompose as
\begin{eqnarray*}
{\bf 15} & = &
\wedge^2 ({\bf 2},{\bf 1},{\bf 1}) \: + \:
\wedge^2 ({\bf 1},{\bf 2},{\bf 1}) \: + \:
\wedge^2 ({\bf 1},{\bf 1},{\bf 2})
\: + \:
({\bf 2},{\bf 2},{\bf 1}) \: + \:
({\bf 2},{\bf 1},{\bf 2}) \: + \:
({\bf 1},{\bf 2},{\bf 2}) ,
\\
& = &
3 ({\bf 1},{\bf 1},{\bf 1}) \: + \:
({\bf 2},{\bf 2},{\bf 1}) \: + \:
({\bf 2},{\bf 1},{\bf 2}) \: + \:
({\bf 1},{\bf 2},{\bf 2}) .
\end{eqnarray*}
Thus, under the $( SU(2)^4 )/{\bf Z}_2$ subgroup of $E_6$,
\begin{eqnarray*}
{\bf 27} & = &
({\bf 2},{\bf 2},{\bf 1},{\bf 1}) \: + \:
({\bf 2},{\bf 1},{\bf 2},{\bf 1}) \: + \:
({\bf 2},{\bf 1},{\bf 1},{\bf 2}) \: + \:
({\bf 1},{\bf 2},{\bf 2},{\bf 1}) \: + \:
({\bf 1},{\bf 2},{\bf 1},{\bf 2}) \: + \:
({\bf 1},{\bf 1},{\bf 2},{\bf 2})
\\
& &  \: + \:
3 ({\bf 1},{\bf 1},{\bf 1},{\bf 1}) .
\end{eqnarray*}
(The ${\bf \overline{27}}$ has the same decomposition -- the
$su(2)^4$ subalgebra cannot distinguish ${\bf 27}$ from
${\bf \overline{27}}$.)
It is straightforward to check that the states in
table~\ref{table:t222:match-only-anom-free} do indeed form two copies of the
decomposition above, partially verifying that in the IR, the states 
transform (two) ${\bf 27}$'s or ${\bf \overline{27}}$'s.  

It may well be possible to say more.  For example,
\cite{ggp-exact}[equ'n (3.28)] contains a prediction for the left NS
partition function, from which one could extract a counting of left NS
states.  However, our chiral ring computations above are in the
left R sector, and for reasons discussed earlier, it is not entirely
clear that those two sectors should match in these examples, so we will
not pursue that direction further.

\subsubsection{Other GLSMs}

Other GLSMs related by triality can be constructed by cyclically exchanging the
vector spaces
\begin{displaymath}
U \: \mapsto \: V \: \mapsto \: \tilde{V}^* \: \mapsto \: U \: \mapsto \:
\cdots .
\end{displaymath}
It is straightforward to check that the same $U(1)^3$ is nonanomalous for each 
GLSM so obtained; however, to match charges, we need to pick a slightly
different $U(1)^3$ action for the different GLSMs.
Moreover, since the dimensions of these
vector spaces all match, we can immediately derive the chiral states
from tables~\ref{table:t222:match-only-anom-free} and \ref{table:t222:mismatch} 
for the first GLSM for $T_{2,2,2}$.

For completeness, we list in tables~\ref{table:t222:var2:match-only-anom-free}
and \ref{table:t222:var2:mismatch} the corresponding chiral
states for one cyclic rotation, {\it i.e.} $U$ replaced with $V$ and so forth.
The global $U(1)^3$ charges listed are those for the action defined by
\begin{center}
\begin{tabular}{c|cccc}
& $\tilde{V}$ & $U$ & $V$ & $W$ \\ \hline
$U(1)_{(1)}$ & $0$ & $0$ & $-1$ & $-1$ \\
$U(1)_{(2)}$ & $-1$ & $0$ & $0$ & $+1$ \\
$U(1)_{(3)}$ & $0$ & $1$ & $0$ & $1$
\end{tabular}
\end{center}
For both geometric phases, we compute that the Fock vacuum has charge
$(+2,-2,0)$.
As before, all states come in Serre-dual pairs which dualize representations.
Also as before, all the representations appearing amongst matched states
are integrable, whereas the unmatched states in 
table~\ref{table:t222:var2:mismatch} have nonintegrable representations.
Furthermore, the unmatched states cancel out of leading terms in refined
elliptic
genera, because they come in pairs with opposite chirality and matching
representations of global symmetries.

With the $U(1)^3$ charges above, it is straightforwrad to check that
all of the 
states below in table~\ref{table:t222:var2:match-only-anom-free},
which match between the two phases of the second GLSM and are expected to
survive to the IR, also appear in table~\ref{table:t222:match-only-anom-free},
which listed the states of the first GLSM that are expected to survive
to the IR.  Thus, these rings provide evidence not only that different
phases of each GLSM flow to the same IR fixed point, but that in addition,
phases of related distinct GLSMs also flow to the same IR fixed point.

\begin{center}
\begin{table}
\begin{tabular}{c|c|c|c}
$r\gg 0$ & $r\ll 0$ & & \\
State, $\wedge^{\bullet} {\cal E}, H^{\bullet}({\mathbb P}^1)$ &
State,  $\wedge^{\bullet} {\cal F}, H^{\bullet}({\mathbb P}^1)$ &
Rep' & $U(1)^3$ \\ \hline
$1$, $0, 0$ & $1$, $0, 0$ &
$({\bf 1},{\bf 1},{\bf 1},{\bf 1})$ & $(+2,-2,0)$ \\
$W \otimes U^*$, $1, 0$ & 
$W \otimes U^*$, $2, 1$ & $({\bf 2},{\bf 1},{\bf 2},{\bf 1})$ &
$(+1,-1,0)$ \\
$\wedge^2 V \otimes K_{(-1,-1)} U^*$, $2, 1$ &
$\wedge^2 W \otimes K_{(-1,-1)} \tilde{V}^*$, $2, 1$ &
$({\bf 1},{\bf 1},{\bf 1},{\bf 1})$ &
$(0,-2,+2)$ \\
$V \otimes \tilde{V}^*$, $2, 1$ &
$V \otimes \tilde{V}^*$, $1, 0$ &
$({\bf 1},{\bf 2},{\bf 1},{\bf 2})$ &
$(+1,-1,0)$ \\
$V \otimes W$, $2, 0$ &
$V \otimes W$, $2, 0$ &
$({\bf 1},{\bf 2},{\bf 2},{\bf 1})$ &
$(0,-1,+1)$ \\
$\wedge^2 \tilde{V}^* \otimes \wedge^2 U^*$, $2, 1$ &
$\wedge^2 \tilde{V}^* \otimes \wedge^2 U^*$, $2, 1$ &
$({\bf 1},{\bf 1},{\bf 1},{\bf 1})$ & $(+2,0,-2)$ \\
$\tilde{V}^* \otimes W \otimes \wedge^2 U^*$, $2, 0$ &
$\tilde{V}^* \otimes W \otimes \wedge^2 U^*$, $3, 1$ &
$({\bf 1},{\bf 1},{\bf 2},{\bf 2})$ & $(+1,0,-1)$ \\
$\wedge^2 V \otimes \tilde{V}^* \otimes U$, $3, 1$ &
$U^* \otimes \wedge^2 W \otimes \tilde{V}$, $3, 1$ &
$({\bf 2},{\bf 1},{\bf 1},{\bf 2})$ & $(0,-1,+1)$ \\
$V \otimes \wedge^2 \tilde{V}^* \otimes U^*$, $3, 1$ &
$V \otimes U^* \otimes \wedge^2 \tilde{V}^*$, $2, 0$ &
$({\bf 2},{\bf 2},{\bf 1},{\bf 1})$ & $(+1,0,-1)$ \\
$V \otimes \wedge^2 W \otimes U^*$, $3, 0$ &
$V \otimes \wedge^2 W \otimes U^*$, $4, 1$ &
$({\bf 2},{\bf 2},{\bf 1},{\bf 1})$ & $(-1,0,+1)$ \\
$\tilde{V}^* \otimes \wedge^2 W \otimes K_{(2,1)} U^*$, $3, 0$ &
$\wedge^2 V \otimes U^* \otimes K_{(2,1)} \tilde{V}^*$, $3, 0$ &
$({\bf 2},{\bf 1},{\bf 1},{\bf 2})$ & $(0,+1,-1)$ \\
$\wedge^2 V \otimes \tilde{V}^* \otimes W$, $4, 1$ &
$\wedge^2 V \otimes \tilde{V}^* \otimes W$, $3, 0$ &
$({\bf 1},{\bf 1},{\bf 2},{\bf 2})$ & $(-1,0,+1)$ \\
$\wedge^2 V \otimes \wedge^2 W$, $4, 0$ &
$\wedge^2 V \otimes \wedge^2 W$, $4, 0$ &
$({\bf 1},{\bf 1},{\bf 1},{\bf 1})$ & $(-2,0,+2)$ \\
$V \otimes \wedge^2 \tilde{V}^* \otimes W \otimes \wedge^2 U^*$, $4, 1$ &
$V \otimes \wedge^2 \tilde{V}^* \otimes W \otimes \wedge^2 U^*$, $4, 1$ &
$({\bf 1},{\bf 2},{\bf 2},{\bf 1})$ & $(0,+1,-1)$ \\
$V \otimes \tilde{V}^* \otimes \wedge^2 W \otimes \wedge^2 U^*$, $4, 0$ &
$V \otimes \tilde{V}^* \otimes \wedge^2 W \otimes \wedge^2 U^*$, $5, 1$ &
$({\bf 1},{\bf 2},{\bf 1},{\bf 2})$ & $(-1,+1,0)$ \\
$\wedge^2 \tilde{V}^* \otimes \wedge^2 W \otimes K_{(2,2)} U^*$, $4, 0$ &
$\wedge^2 V \otimes \wedge^2 U^* \otimes K_{(2,2)} \tilde{V}^*$, $4, 0$ &
$({\bf 1},{\bf 1},{\bf 1},{\bf 1})$ & $(0,+2,-2)$ \\
$\wedge^2 V \otimes \wedge^2 \tilde{V}^* \otimes W \otimes U^*$, $5, 1$ &
$\wedge^2 V \otimes U^* \otimes W \otimes \wedge^2 \tilde{V}^*$, $4, 0$ &
$({\bf 2},{\bf 1},{\bf 2},{\bf 1})$ & $(-1,+1,0)$ \\
$\wedge^2 V \otimes \wedge^2 \tilde{V}^* \otimes \wedge^2 W \otimes
\wedge^2 U^*$, $6, 1$ &
$\wedge^2 V \otimes \wedge^2 \tilde{V}^* \otimes \wedge^2 W \otimes 
\wedge^2 U^*$, $6, 1$ &
$({\bf 1},{\bf 1},{\bf 1},{\bf 1})$ & $(-2,+2,0)$ 
\end{tabular}
\caption{Shared states defined by matching representations of anomaly-free
global symmetries.}
\label{table:t222:var2:match-only-anom-free}
\end{table}
\end{center}

\begin{center}
\begin{table}
\begin{tabular}{c|c|c|c|c|c|c}
& \multicolumn{2}{c}{$r \gg 0$} & \multicolumn{2}{c}{$r \ll 0$} & & \\
State & $\wedge^{\bullet} {\cal E}$ & $H^{\bullet}({\mathbb P}^1)$ &
$\wedge^{\bullet} {\cal F}$ & $H^{\bullet}({\mathbb P}^1)$ &
Rep' &
$U(1)^3$ \\ \hline
$\wedge^2 W \otimes K_{(2,0)} U^*$ & $2$ & $0$ & $-$ & $-$ &
$({\bf 3},{\bf 1},{\bf 1},{\bf 1})$ & $(0,0,0)$ \\
$\wedge^2 V \otimes \wedge^2 \tilde{V}^* \otimes K_{(1,-1)} U^*$ & $4$ & $1$ &
$-$ & $-$ & $({\bf 3},{\bf 1},{\bf 1},{\bf 1})$ & $(0,0,0)$ \\
$\wedge^2 V \otimes K_{(2,0)} \tilde{V}^*$ & $-$ & $-$ & $2$ & $0$ &
$({\bf 1},{\bf 1},{\bf 1},{\bf 3})$ & $(0,0,0)$ \\
$\wedge^2 U^* \otimes \wedge^2 W \otimes K_{(1,-1)} \tilde{V}^*$ & $-$ & $-$
& $4$ & $1$ & $({\bf 1},{\bf 1},{\bf 1},{\bf 3})$ & $(0,0,0)$
\end{tabular}
\caption{List of all states which are not shared between the two phases.}
\label{table:t222:var2:mismatch}
\end{table}
\end{center}

\section{Conclusions}

In this paper we have used chiral state computations to confirm the
triality conjecture of \cite{ggp1,ggp-exact}, which says that triples of
two-dimensional (0,2) theories RG flow to the same IR fixed point.

We began with a review of chiral states and rings in (2,2) and (0,2)
theories.  The fact that the Fock vacuum in a nonlinear sigma model
transforms as a section of a line bundle over the target played
an important role in our computations, and we elaborated on this property,
and also explained how even in (2,2) theories, this arises in describing choices
of spin structures on the target space.

We then turned to chiral states in the theories arising in triality.
As (0,2) chiral states are not protected against nonperturbative
corrections, and the theories in question RG flow from weakly-coupled
nonlinear sigma models to strongly coupled regimes, the chiral states
need not match identically everywhere along RG flow.  (That said, the opposite
has often been implicitly assumed, so triality provides clean examples 
demonstrating that (0,2) chiral rings are not protected.)  In several examples,
we computed chiral states in various phases and different GLSMs related
by triality, and indeed discovered that the states did not all match.
However, the states that did match, were all consistent with the proposed
affine symmetry algebras of the IR fixed point.  Furthermore, the mismatched
states were all in non-integrable representations, and appear in pairs
which cancel out of refined elliptic genera, suggesting that they are lifted
in RG flow.  In this fashion, we were able to confirm triality.  
(In principle, it could also happen that pairs of massive states become
massless, but we did not observe this in any of the computed
examples.)  Finally,
we were also able to support predictions for enhanced IR symmetries
in certain theories, such as the enhanced $E_6$ in the $T_{2,2,2}$ theory,
by confirming that the matching UV states fall into the decomposition of
integrable representations of the $E_6$ algebra.

\section{Acknowledgements}

We would like to thank A.~C\u ald\u araru, R.~Donagi,
A.~Gadde, H.~Jockers,
S.~Katz, A.~Knutson, W.~Lerche, V.~Lu, I.~Melnikov, and T.~Pantev 
for useful discussions.
Bei Jia was partially supported by NSF grant PHY-1316033.
Eric Sharpe was partially supported by NSF grants PHY-1068725, PHY-1417410.

\appendix

\section{Bott-Borel-Weil}  \label{app:bbw}

As the Bott-Borel-Weil theorem plays a crucial role in our results,
but is perhaps not well known in the physics community, this appendix
will provide a short self-contained introduction.

First, we will use the notation $K_{\lambda} A$ for a vector space or
bundle $A$ to denote a tensor product of copies of $A$ determined by
a $U(n)$ representation
$\lambda$.  For example, for the special case of
$SU(n)$ representations, we can associate representations with
Young diagrams, and
\begin{displaymath}
K_{\tiny{\yng(2)}} A \: = \: {\rm Sym}^2 A, \: \: \:
K_{\tiny{\yng(1,1)}} A \: = \: \wedge^2 A .
\end{displaymath}
More properly, a $U(n)$ representation is determined by a 
non-increasing sequence of integers, as described in
{\it e.g.} \cite{jsw}.  In the special case of $SU(n)$ representations,
those integers count the number of boxes in rows of Young diagrams.
For example, for $U(2)$,
\begin{displaymath}
K_{(2,0)} A \: = \: {\rm Sym}^2 A, \: \: \:
K_{(1,1)} A \: = \: \wedge^2 A, \: \: \:
K_{(1,-1)} A \: = \: ( \wedge^{\rm top} A )^{-1} \otimes 
{\rm Sym}^2 A .
\end{displaymath}

The Bott-Borel-Weil theorem says that the
only nonvanishing cohomology of $K_{\beta} S^* \otimes K_{\gamma}
Q^*$ over $G(k,V)$
lives in $\ell(\alpha)$, where $\ell(\alpha)$ is the number of
`mutations' required to transform $\alpha = (\beta,\gamma)$
into a dominant weight $\tilde{\alpha}$ of $GL(V)$.  Furthermore,
\begin{displaymath}
H^{\ell(\alpha)}(K_{\beta} S^* \otimes K_{\gamma} Q^*) \: = \:
K_{\tilde{\alpha}} V^* .
\end{displaymath}
Mutations are defined as follows:
For
\begin{displaymath}
\alpha \: = \: \left( \alpha_1, \alpha_2, \cdots, \alpha_n \right)
\end{displaymath}
a mutation is \cite{weyman}[remark 4.1.5]
\begin{displaymath}
\sigma_i \cdot \alpha \: = \: \left( \alpha_1, \cdots, \alpha_{i-1},
\alpha_{i+1}-1, \alpha_i+1, \alpha_{i+2}, \cdots, \alpha_n \right) .
\end{displaymath}
In principle, we count the number of mutations needed to turn $(\beta,\gamma)$
into a nonincreasing sequence.  If at any point there exists a mutation that
leaves the sequence invariant, then all the cohomology vanishes.

For a simple example, consider the cohomology of $S = {\cal O}(-1)$ on
${\mathbb P}^1 = G(1,2)$.  Here, $\beta = (-1)$, $\gamma=(0)$.  Thus,
we need to mutate $(-1,0)$ to a nonincreasing sequence.  However, this
sequence is invariant under mutation, hence, all the cohomology vanishes:
\begin{displaymath}
H^{\bullet}({\cal O}(-1)) \: = \: 0 ,
\end{displaymath}
which is a standard result.
More generally, the cohomology of ${\cal O}(k)$ on ${\mathbb P}^1$ can
be computed as the cohomology of $K_{(k)} S^*$.  To compute this cohomology,
we mutate $(n,0)$ into a nonincreasing sequence.  If $k \geq 0$,
no mutations are required, all the cohomology lives in degree zero,
and so
\begin{displaymath}
H^0({\cal O}(k)) \: = \: K_{(k,0)} V^* \: = \: {\rm Sym}^k V^*
\end{displaymath}
for $V$ a two-dimensional vector space.  Note that
\begin{displaymath}
{\rm dim}\, {\rm Sym}^k V^* \: = \: k+1 ,
\end{displaymath}
and so the methods above predict that
\begin{displaymath}
h^0({\cal O}(k)) \: = \: k+1
\end{displaymath}
for $k \geq 0$, in agreement with standard results.
The case $k=-1$ we have already discussed.  For $k < -1$, we can try the
mutation
\begin{displaymath}
(k,0) \: \rightsquigarrow \: (-1,k+1) .
\end{displaymath}
If $n<-1$, then this is a nonincreasing sequence, and so we have
\begin{displaymath}
H^1( {\cal O}(k)) \: = \: K_{(-1,k+1)} V^* .
\end{displaymath}

\end{document}